\documentclass[11pt]{article}
\usepackage{amsmath,amssymb,color,graphics,epsfig,cite}
\UseRawInputEncoding
\usepackage{amsfonts}
\newcommand{\be}{\begin{equation}}
\newcommand{\ee}{\end{equation}}
\newcommand{\bea}{\setlength\arraycolsep{2pt} \begin{eqnarray}}
\newcommand{\eea}{\end{eqnarray}}
\newcommand{\nn}{\nonumber}

\def\ft#1#2{{\textstyle{\frac{\scriptstyle #1}{\scriptstyle #2} } }}
\def\fft#1#2{{\frac{#1}{#2}}}

\def\0{{\sst{(0)}}}
\def\1{{\sst{(1)}}}
\def\2{{\sst{(2)}}}
\def\3{{\sst{(3)}}}
\def\4{{\sst{(4)}}}
\def\5{{\sst{(5)}}}
\def\6{{\sst{(6)}}}
\def\7{{\sst{(7)}}}
\def\8{{\sst{(8)}}}
\def\sst#1{{\scriptscriptstyle #1}}

\begin{document}

\begin{center}
{\Large {\bf Revisiting the AdS Boson Stars: the Mass-Charge Relations}}

\vspace{20pt}

{\large Hai-Shan Liu, H. L\"u and Yi Pang}

\vspace{10pt}

{\it  Center for Joint Quantum Studies and Department of Physics,\\
School of Science, Tianjin University, Tianjin 300350, China}

\vspace{40pt}

\underline{ABSTRACT}
\end{center}

Motivated by the recent progress in solving the large charge sector of conformal field theories, we revisit the mass-charge relation of boson stars asymptotic to global AdS. We construct and classify a large number of electrically charged boson star solutions in a toy model and two supergravity models arising from the $SU(3)$ and  $U(1)^4$ truncation of $D=4$ $SO(8)$ gauged maximal supergravity. We find a simple ansatz for the chemical potential that can fit the numerical data in striking accuracy for the full range of charge. Combining with the first law of thermodynamics, we can then evaluate the mass as a function of the charge and obtain the free energy in the fixed charge ensemble. We show that in the toy model, the ground state can be either the extremal RN black hole or the boson stars depending on the parameter region. For the $SU(3)$ truncation, there always exists a boson star that has smaller free energy than the extremal RN black hole, in contrast to the $U(1)^4$ model where the extremal RN black hole is always the ground state. In all models, for boson star solutions with arbitrarily large charge, we show that the large charge expansion of the mass reproduces the same structure exhibited in the CFT side.

\vfill{\footnotesize  hsliu.zju@gmail.com \ \ \ mrhonglu@gmail.com \ \ \ pangyi1@tju.edu.cn }
%\vfill {\footnotesize mrhonglu@gmail.com}

%{\footnotesize \hoch{*}Corresponding author}

\thispagestyle{empty}
\pagebreak

\tableofcontents
\addtocontents{toc}{\protect\setcounter{tocdepth}{2}}

\newpage

\section{Introduction}

When it comes to solving a physical system, it is very useful to look for a small parameter about which a systematic perturbative expansion can be carried out. This approach led to the discovery of soluble sectors of strongly coupled conformal field theory (CFT) at large spin \cite{Fitzpatrick:2012yx, Komargodski:2012ek,Alday1,Alday2,Alday3} or large global $U(1)$ charge \cite{Orlando1,Orlando2,Orlando3,Monin:2016jmo,delaFuente:2018qwv,Badel:2019khk,Jafferis:2017zna} where the spectrum of the theory simplifies. For instance, in the $O(2)$ model describing the superfluid Helium, the lowest scaling dimension in the sector with large global $U(1)$ charge is carried by a scalar operator and takes the form \cite{Orlando1}
\be
\label{mq1}
\Delta+0.0937256=c_{\fft32}\, Q^{\fft32}+c_{\fft12}\, Q^{\fft12}+ c_{-\fft12}\, Q^{-\fft12} + {\cal O}(Q^{-\fft32})\,,
\ee
where $c_{\ft32}$, $c_{\ft12}$, {\it etc.}~are model dependent calculable coefficients. The universal constant $c_0=-0.0937256$ comes from the one-loop contribution of the relativistic Goldstone boson associated with the breaking of the global $U(1)$ symmetry due to the non-vanishing charge $Q$ present in the vacuum. Similar behavior also appeared in three-dimensional $\Phi^3$ Wess-Zumino model, $O(N)$ vector model \cite{Orlando2} and $SU(N)$ matrix model \cite{Orlando3} and was conjectured to be universal for a large class of CFTs with spontaneously broken global symmetry.

When the CFT is placed on $R\times S^2$, the spectrum of the dilation operator coincides with that of the Hamiltonian. Therefore, the large charge behavior for the lowest scaling dimension directly translates into that of the ground state energy of the CFT on $R\times S^2$. Furthermore, if the three dimensional CFT admits a dual gravitational description, this implies that within the fixed charge ensemble, the solution of the lowest energy is non-rotating with a mass-charge relation belonging to the same asymptotic class in the large charge limit.
Initial attempt has been made relating the ground state of the large charge sector of a strongly coupled CFT on $R\times S^2$ to an extremal Reissner-Nordstr\"om (RN) black hole that is asymptotic to the anti-de Sitter spacetime (AdS) \cite{Loukas:2018zjh}. However, doubts about this correspondence has been raised in \cite{Nakayama:2020dle}. It was pointed out recently that the extremal RN-AdS black hole is not the appropriate gravity dual as it carries a large amount of entropy absent from the CFT side \cite{delaFuente:2020yua}. Instead, the ground state of the large charge sector of the CFT on $R\times S^2$ should be dual to a global AdS boson star solution (a charged soliton that is asymptotic to global AdS) that possesses no entropy while exhibiting similar mass-charge relation in the large charge limit \cite{delaFuente:2020yua}.

The gravity model considered by  \cite{delaFuente:2020yua} is simply the four dimensional Einstein-Maxwell coupled to a charged massive scalar field without self interaction. Different from previous work  \cite{Astefanesei:2003qy,Hartmann:2012gw,Buchel:2013uba} on AdS boson stars where the Maxwell field is absent, here the $U(1)$ gauge field coupled to the scalar field is indispensable as it implies the presence of a global $U(1)$ symmetry in the dual CFT. Boson star solutions in the same model have already been studied in \cite{Gentle:2011kv} which also observed that $M\propto Q^{3/2}$ in the strictly $Q\rightarrow\infty$ limit. The reason behind such scaling is quite simple. Numerical evidence \cite{Gentle:2011kv} suggests that, for boson stars admitting arbitrarily large charge, the $Q\rightarrow\infty$ limit amounts to the planar limit under which the mass parameter $M$ should be interpreted as the energy density of the CFT on two-torus while the charge parameter $Q$ corresponds to the charge density. Thus, $M\propto Q^{3/2}$ simply follows from the fact that
in three-dimensions, the energy density has mass dimension 3 while the charge density has mass dimension 2. This dimension analysis also explains the  universal large $Q$ behavior of the boson star mass even for bulk theories that are less likely to have CFT duals, such as the one considered in \cite{delaFuente:2020yua}.

Motivated by the recent progress in solving the large charge sector of conformal field theories, we carry out a quantitatively more detailed study on the mass-charge relation of AdS boson stars than \cite{Gentle:2011kv}.  We construct and classify more AdS boson star solutions not only in the phenomenological toy model, but also in two supergravity models with M-theory origin. Our setup is similar to the one in \cite{Gentle:2011kv}. By numerical approach, we obtain the asymptotic hairy parameters such as the mass $M$, electric charge $Q$, the chemical potential $\mu$ and the scalar hair $\phi_2$ related to the VEV of the scalar operator.  These parameters are not independent and we can treat $(M,\mu,\phi_2)$ as functions of the electric charge $Q$.  We thus obtain these relations, focusing on the mass-charge relation. Although some boson stars with arbitrarily large charges were constructed in the previous work \cite{Gentle:2011kv,Dias:2011tj}, the focus of those was not on its possible implications to the large charge sector of the dual CFT.

By using the Euclidean action technique, we can determine the free energy of the boson stars and compare it with the extremal RN-AdS black hole.  This allows us to determine the ground state, or the state with the lowest scaling dimensions in the dual CFT for given global $U(1)$ charge.  Moreover, it may also provide some understanding of the Weak Gravity Conjecture in asymptotically AdS spacetime.  When Einstein-Maxwell gravity is coupled to some additional charged matter, the extremal RN-AdS black hole may turn into a metastable state and decay into other solutions representing the true ground state. The Extremal RN-AdS black hole not only suffers from the superradiant instability for small charges, but also suffers from the horizon instability for large charges \cite{Gubser:2008px,Hartnoll:2008vx,Basu:2010uz,Dias:2011tj,Uchikata:2011zz,Wang:2014eha,Dias:2016pma}. It can be argued that charged scalar hairy black holes in general do not have regular extremal limits \cite{Dias:2011tj}, and the boson stars are the preferred candidates instead. Comparing the mass-charge relation of the extremal RN black hole to that of the boson stars can help us to determine the condition on the mass/charge relation of the fundamental charged scalar hair under which the RN-AdS black hole ceases to be solutions of the lowest free energy.

The paper is organized as follows.  In section \ref{sec:setup}, we set up the framework and the techniques of constructing the charged spherically-symmetric and static solutions.  Since we are interested in the lowest energy solutions in the fixed charge ensemble at zero temperature, two candidates emerge: the charged extremal AdS black holes and the AdS boson stars.  We show that extremal scalar hairy black holes do not exist in the theories we consider and hence the only black hole candidate remaining is the extremal RN-AdS black hole.  For the AdS boson stars, we impose the boundary condition corresponding to turning on a VEV in the dual CFT spontaneously breaking the global symmetry. We then use both the Euclidean action and the Iyer-Wald formalism to establish the first law of the boson star dynamics and also determine the Helmholtz free energy of the fixed charge ensemble.  This provides a mechanism to determine the ground state.

In section 3, we present numerical AdS boson star solutions for the phenomenological toy model, namely Einstein-Maxwell gravity coupled to a negative cosmological constant and a massive charged scalar without self-interaction. We restrict the scalar to be conformally massless, ($m^2\ell^2=-2$,) the same as the two supergravity models, but the charge parameter to be free. (AdS$_4$ boson stars in the toy model with $m\ell=0$ and $m\ell=10$ were studied in \cite{Dias:2016pma,Arias:2016aig} and \cite{Hu:2012dx} respectively.) We construct and classify the boson stars, carrying out a quantitatively more detailed study than \cite{Gentle:2011kv}. We analytically approximate the mass-charge relations, which allows us to compare to that of the extremal RN-AdS black holes, and therefore single out the ground state.

In section 4, we present numerical AdS boson stars in two gauged supergravity models, corresponding to the $SU(3)$ \cite{Nicolai:1985hs} and  $U(1)^4$ \cite{Chong:2004ce} truncation of $D=4$ $SO(8)$ gauged maximal supergravity \cite{deWit:1982bul}.  These models are dual to different sectors of the ABJM Chern-Simons matter SCFT \cite{Aharony:2008ug}\footnote{In supersymmetric conformal field theories with nontrivial moduli space, one may think that mass charge relation is linear. However, it needs not to be the case if the charge is magnetic \cite{Aharony:2012nh,Aharony:2015pla,Dyer:2015zha}. This is indeed the case since one of the $U(1)$ gauge fields in these truncations is dual to a magnetic global $U(1)$ symmetry in the CFT \cite{Borghese:2014gfa}.}. New feature arises owing to the nonlinear scalar potentials. We conclude the paper in section 5.  In appendix A, we present the detailed truncation of the $SU(3)$ supergravity model.  In appendix B, we give explicit second-order nonlinear differential equations for the charged spherically-symmetric and static ansatz.  In appendix C, we derive of the generalized Smarr formula without utilizing the standard scaling argument and apply it to understand the mass-charge relation in the large charge limit.  In appendix D, we present some explicit verification of the first law of the boson star dynamics at certain subtle parameter space.  In order not to interrupt the discussion in the main text, we reserve many illustrative figures in appendix E.

\section{The set up}
\label{sec:setup}

\subsection{The theory and equations}

A general class of Einstein-Maxwell-scalar theory, where the complex scalar $\Phi$ is charged under the Maxwell potential $A$, can be written as
\be
{\cal L}=\sqrt{-g} \Big(R - \ft14 F^2 - X(|\Phi|) (D_\mu \Phi) (D^\mu \Phi)^* - Y(|\Phi|)\Big)\,,\qquad
F=dA\,,\quad D_\mu \Phi = \partial_\mu \Phi - {\rm i} e \Phi\,,
\ee
where $e$ is the electric charge coupling and $(X,Y)$ are certain generic coupling functions of the modulus of $\Phi$. Expressing the complex scalar $\Phi=f(\phi) e^{{\rm i} \chi}$, the phase factor field $\chi$ is ``eaten'' by the Maxwell potential, which becomes massive.  Choosing the function $f$ appropriately, one can rewrite the Lagrangian as
\be
{\cal L}=\sqrt{-g}\Big(R - (\partial\phi)^2 - U(\phi) A^2 - \ft14 F^2 - V(\phi)\Big).\label{genlag}
\ee
This is an Einstein-Proca-scalar theory and it reduces to Einstein-Maxwell theory at some common stationary point of $(U,V)$ where $U$ vanishes. The covariant equations of motion are
\bea
\delta \phi:\quad 2\Box\phi=A^2\fft{\partial U}{\partial\phi} + \fft{\partial V}{\partial\phi}\,;\qquad
\delta A_\mu:\quad \nabla_\mu F^{\mu\nu} = 2 U(\phi) A^\nu\,;\qquad
\delta g_{\mu\nu}:\quad E^{\mu\nu} =0\,,
\eea
where
\bea
E_{\mu\nu} &\equiv& R_{\mu\nu} -\ft12 g_{\mu\nu} R - (\partial_\mu\phi\partial_\nu\phi - \ft12 g_{\mu\nu} (\partial \phi)^2) - (A_\mu A_\nu - \ft12 g_{\mu\nu} A^2)\,U\nn\\
&&-\ft12 (F_{\rho\mu} F^{\rho}{}_\nu - \ft14 g_{\mu\nu} F^2) + \ft12 g_{\mu\nu}\, V.\label{Emunu}
\eea
In this paper, we study three explicit models similar to \cite{Gentle:2011kv}. One is the phenomenological toy model and the other two are gauged supergravity models associated with the $SU(3)$ and $U(1)^4$ truncations of $D=4$ $SO(8)$ gauged maximal supergravity.  The $U$ and $V$ functions of these models are
\bea
\hbox{Toy model}:&&\qquad U=\fft{q^2}{\ell^2} \phi^2\,,\qquad\qquad\,\,\qquad V=-\fft{6}{\ell^2} + \fft{m^2}{\ell^2} \phi^2\,;\nn\\
\hbox{$SU(3)$ model}:&&\qquad U= \fft1{2\ell^2} \sinh^2(\sqrt2\phi)\,,\qquad V=-\fft{1}{\ell^2}
\cosh^2(\ft{1}{\sqrt2}\phi) (7-\cosh(\sqrt2\phi))\,;\nn\\
\hbox{$U(1)^4$ model}:&&\qquad U=\fft1{2\ell^2} \sinh^2(\ft1{\sqrt2} \phi)\,,\qquad
V=-\fft{2}{\ell^2} (\cosh(\sqrt2\phi) + 2)\,.\label{threetheories}
\eea
(See appendix \ref{app:su(3)} for more details on the supergravity models.) For the phenomenological toy model, $(m,q)$ are parameters of the fundamental scalar's mass and the electric coupling charge $e$. Performing small $\phi$ Taylor expansion of the $(U,V)$ functions of the supergravity models about $\phi=0$, one can easily see that $m^2\ell^2=-2$ for both theories, but with $q^2=1$ and $1/4$ for the $SU(3)$ and $U(1)^4$ theories respectively.  In the AdS spacetime, the squared mass of $\phi$ can be negative and the system still remains linearly stable.  In particular the scalar with $m^2\ell^2=-2$ is conformally massless and its leading asymptotic falloff is $1/r$.

In this paper, we consider electrically-charged spherically-symmetric and static solutions in four dimensions, with the ansatz
\bea
ds^2 &=& - h(r) dt^2 + \fft{dr^2}{f(r)} + r^2 (d\theta^2 + \sin^2\theta\, d\varphi^2)\,,\nn\\
A &=& a(r) dt\,,\qquad \phi=\phi(r)\,.\label{genans}
\eea
The $E_r{}^r=0$ equation can be solved as
\be
f=\frac{2 a^2 r^2 U-2 h \left(r^2 V-2\right)}{r \left(r a'^2+4 h'\right)+h \left(4-2 r^2 \phi'^2\right)}\,.\label{fsol}
\ee
The remaining functions $(h,a,\phi)$ satisfy three nonlinear second-order different equations.  (See appendix \ref{app:eoms} for the explicit expressions.)

\subsection{The boundary condition}

The theory admits the AdS spacetime of radius $\ell$ and the metric in the global coordinates is
\be
ds_{\rm AdS}^2 = -(r^2\ell^{-2} + 1) dt^2 + \fft{dr^2}{r^2 \ell^{-2} + 1} + r^2 (d\theta^2 + \sin^2\theta\, d\varphi^2)\,.\label{adsvac}
\ee
From now on, we shall set $\ell=1$ for computational convenience.  We also consider only the conformal massless scalar with $m^2 = - 2$ for the toy model, so that the near boundary behavior for scalars in all three models is
\be
\phi=\fft{\phi_1}r + \fft{\phi_2}{r^2} + \cdots\,.\label{scalarfalloff}
\ee
In the AdS/CFT dictionary, the integration constants $\phi_1$ and $\phi_2$ represent the source and VEV of the dual boundary scalar operator respectively. In the standard boundary condition, $\phi_1$ is the source and $\phi_2$ denotes the VEV. It is recalled that in AdS$_4$, $m^2=-2$ lies in the window where alternative boundary condition is allowed under which the role of $(\phi_1, \phi_2)$ is interchanged \cite{Klebanov:1999tb}. This is also the mixed boundary condition $\phi_2\propto\phi_1^2$ corresponding to the triple trace deformation of the CFT. In this paper, to illustrate the main idea, we focus on the boundary condition $\phi_1=0$  while leaving other choices of boundary condition for future study. For supergravity models dual to sectors of ABJM model, this choice of boundary condition implies turning on the VEV for a bilinear fermion operator charged under the global $U(1)$ symmetry.

Therefore, throughout this paper, the solutions take the form near the AdS$_4$ boundary
\be
h\sim f = r^2 + 1 - \fft{2M}{r} + \fft{4Q^2}{r^2} + \cdots\,,\qquad
a=\mu - \fft{4Q}{r} + \cdots \,,\qquad \phi \sim \fft{\phi_2}{r^2} + \cdots\,,\label{falloffs}
\ee
where $\mu$ is the electric or chemical potential and $Q$ is the electric charge, defined by
\be
Q=\fft{1}{16\pi} \int {*F}\,.
\ee
Performing the power series expansion of $1/r$, we can solve the equations of motion order by order.  It can be easily established that there are in general five nontrivial hairy parameters, $(M,Q, \mu, \phi_1,\phi_2)$. After setting $\phi_1=0$, the remaining parameters $(M, Q,\mu, \phi_2)$ specify the characteristics of the solutions.  Some choice of the parameters can lead to an event horizon, giving rise to a charged black hole. Some choice will cause the collapsing of 2-sphere with no curvature singularity, leading to an asymptotic AdS soliton that is geodesically complete.  The general parameters will presumably lead to solutions with naked singularity.

\subsection{Scalar hairy black holes}

The horizon $r=r_+$ is characterized by $h(r_+)=0=f(r_+)$ with finite $a(r_+)$ and $\phi(r_+)$.  The near-horizon structure can be analysed by the Taylor expansion
\bea
&&h=h_1 (r-r_+) + h_2 (r-r_+)^2 + \cdots\,,\qquad f= f_1 (r-r_+) + f_2 (r-r_+)^2 + \cdots\,,\nn\\
&&a=a_0 + a_1 (r-r_+) + \cdots\,,\qquad \phi=\phi_0 + \tilde \phi_1 (r-r_+) + \cdots\,.
\eea
The equations of motion implies that all the coefficients in the Taylor expansion can be expressed in terms of four parameters $(a_0,a_1, \phi_0, \tilde \phi_1)$, with leading coefficients
\be
f_1 = \fft{m^2 \phi_0}{\tilde \phi_1}\,,\qquad h_1 = \fft{m^2 a_1^2 \phi_0 r_+^2}{4\tilde \phi_1 -
4m^2 r_+ \phi_0 - 2 (m^2 \phi_0^2 -6) \tilde \phi_1 r_+^2}\,.
\ee
Note that the coefficient $a_1$ can be viewed as trivial since it can be scaled so that we require that $h\sim r^2$ at the asymptotic infinity. Thus the horizons are specified by three nontrivial parameters $(r_+, \phi_0, \tilde \phi_1)$.  This implies that the asymptotic four hairy parameters $(M,Q, \mu,\phi_2)$ must satisfy one algebraic constraint.

In this paper, we are interested in extremal black holes with $h_1=0=f_1$. We find that the imposing of the equations of motion on the horizon ansatz implies that the extremal scalar hairy solution is not possible for the conformally massless scalar ($m^2=-2$). The horizon structure requires fine turned mass parameter
\be
m^2 = \frac{2 \left(6 \left(q^2+6\right) r_+^4+2 \left(q^2+6\right) r_+^2+1\right)}{6 r_+^4+r_+^2}>0\,.
\ee
This phenomenon is not uncommon.  For example, in the Einstein-Maxwell-dilaton theory
\be
{\cal L} = \sqrt{-g} (R -\ft12 (\partial\phi)^2 - \ft14 e^{a\phi} F^2)\,,
\ee
it was long understood that the inner (Cauchy) horizon of the dyonic black hole will be destroyed unless the dilaton coupling is discrete and satisfies \cite{Poletti:1995yq}
\be
a^2=\ft12 k(k+1)\,,\qquad k=1,2,\cdots\,.
\ee
Indeed, it was established that only for these dilaton couplings can one construct scalar hairy extremal dyonic black hole \cite{Geng:2018jck}.  Analogous phenomenon was recently observed for black holes carrying massive scalar hair, which abhors the Cauchy horizon for the general mass parameter \cite{Hartnoll:2020rwq}.

The scalar can be consistently set to zero for all three models we consider in this paper. For the vanishing scalar, we have the usual RN-AdS black hole:
\be
h=f=r^2 + 1 - \fft{2M}{r} + \fft{4Q^2}{r^2}\,,\qquad a=4Q\Big(\fft{1}{r_+}- \fft{1}{r}\Big)\,,\qquad\phi=0\,.
\ee
Below we briefly review the well studied thermodynamics of extremal AdS RN black holes for readers' convenience since later we will study the competition between extremal AdS RN black holes and Boson stars with the same charge, in the phase diagram at zero temperature. For sufficiently large mass $M$, there are two real roots $r_\pm$ for which $f(r_\pm)=0$.  The event horizon is located at the larger root $r_+$ and the solution satisfies the first law of black hole thermodynamics
\be
dM=T dS + \mu dQ\,,
\ee
where the temperature $T$, entropy $S$ and the chemical potential $\mu$ are
\be
T=\fft{f'(r_+)}{4\pi}\,,\qquad S=\pi r_+^2\,,\qquad \mu = \fft{4Q}{r_+}\,.
\ee
There exists an extremal limit where $r_\pm$ coalesce
\be
r_\pm=\frac{\sqrt{\sqrt{48 Q^2+1}-1}}{\sqrt{6}}\,,
\ee
for which temperature vanishes and
\be
M = \frac{\sqrt{\sqrt{48 Q^2+1}-1} \left(\sqrt{48 Q^2+1}+2\right)}{3 \sqrt{6}}\,,\qquad
\mu = \sqrt{2 \sqrt{48 Q^2+1}+2}\,.\label{RNMQ}
\ee
The factorization in the mass-charge relation was observed in \cite{Loukas:2018zjh}, reminiscent of the ground state energy for vector models on $R\times S^2$ with large global charge.
The first law of black hole thermodynamics at zero temperature reduces to
\be
dM=\mu dQ\,.\label{0Tfirstlaw}
\ee
As we shall see later that the dynamic first law of the boson star takes the same form.

In the large and small $Q$ limits, the mass-charge relation becomes respectively
\bea
Q\gg 1:&& M=2.48161 Q^{3/2}+0.537285 \sqrt{Q}-0.0193876 Q^{-\fft12}+ {\cal O}\big(Q^{-\fft32}\big)\,,\nn\\
Q\ll 1:&& M=2 Q +4Q^3 - 36 Q^5+ {\cal O}\big(Q^7\big)\,.
\eea
Correspondingly, the chemical potential-charge relations are
\bea
Q\gg 1:&& \mu=3.72242  Q^{\fft12} + 0.268642  Q^{-\fft12} + 0.0096938 Q^{-\fft32} + {\cal O}\big(Q^{-\fft52}\big)\,,\nn\\
Q\ll 1:&& \mu= 2 + 12Q^2 -180 Q^4 +  {\cal O}\big(Q^6\big)\,.
\eea
When $Q\ll 1$, the effect of the cosmological constant becomes unimportant and hence the mass-charge relation reduces to $M=2Q$, which is the same as that of extremal RN black holes that are asymptotic to the flat spacetime. We see that the Smarr relation at large and small $Q$ becomes
\be
\lim_{Q\rightarrow 0} \fft{\mu Q}{M}=1\,,\qquad \lim_{Q\rightarrow \infty} \fft{\mu Q}{M}=\fft32\,.
\label{smarrlimit}
\ee
The large $Q$ limit of the RN-AdS black hole becomes equivalent to the AdS planar black hole, where additional scaling symmetry emerge that is responsible for the second Smarr relation \cite{Gentle:2011kv,Fan:2015tua,Liu:2015tqa}. See appendix \ref{app:smarr} for further discussions.

The thermodynamics for the RN-AdS black hole was well understood. In the extremal limit, the Helmholtz free energy $F_H=M- TS$ and Gibbs free energy $F_G=M-TS - \mu Q$ are
\be
\hbox{Helmholtz:}\quad F_H=M\,;\qquad\qquad
\hbox{Gibbs:}\quad F_G = M- \mu Q\,.
\ee
For a system of fixed temperature $T=0$ and given charge $Q$, the Helmholtz free energy or simply the mass is the measure of how high the excitation of a state.  The state with the lowest Helmholtz free energy is the ground state.

It is important to point out that the large $Q$ expansion of the mass for the extremal RN-AdS black hole has vanishing $c_0$, the constant piece that appears in (\ref{mq1}). This is not in general shared by the AdS boson stars that we shall discuss next.

\subsection{AdS boson stars}

The topology of the AdS boson stars is the same as the AdS vacuum in the global coordinates (\ref{adsvac}) with the radial $r$ runs from 0 to infinity. The ``boundary'' condition at $r=0$ is
\be
f=1\,,\qquad h=h_0\,,\qquad a=a_0\,,\qquad \phi=\phi_0\,,\label{boundarycond1}
\ee
with
\be
f'=h'=a'=\phi'=0\,.\label{boundarycond2}
\ee
Note that regularity of the spacetime requires that $f'=0=h'$, and then the equations of motion imply that $a'=0=\phi'$.  In other words, the AdS boson stars are special class of smooth solitons. It can be easily established that for our (even) scalar coupling potential functions $U$ and $V$, the Taylor expansions at $r=0$ all involve only even $r$ powers, with the coefficients determined by $(h_0,a_0,\phi_0)$.  The ansatz is time-scaling invariant and hence $h_0$ is a trivial parameter, but with the understanding that it should be chosen so that $h= r^2 +1 + \cdots $ at the asymptotic infinity.

  The boundary condition at $r=0$ implies that general smooth solitons are specified by two parameters $(a_0, \phi_0)$.
However, a generic choice of $(a_0,\phi_0)$ will turn on both $(\phi_1,\phi_2)$ at the asymptotic infinity and therefore we need to finetune the $(a_0,\phi_0)$ so that $\phi_1$ vanishes.  In this paper, for the purpose of clear terminology, we refer to the smooth solutions with general $(\phi_1,\phi_2)$ as the AdS solitons, while call those with $\phi_1=0$ as AdS boson stars.  Thus in this paper, the boson stars involve only one nontrivial parameter, implying that these solutions trace a line in the four-dimensional parameter space of $(M,Q,\mu, \phi_2)$.  We choose the parameter to be $Q$, with the mass $M$, $\phi_2$ and chemical potential $\mu$ all depending on $Q$.  This is the analogous choice as the extremal RN-AdS black hole discussed earlier.

The focus of this paper is to construct such AdS boson stars and obtain their mass-charge relation $M(Q)$.
We also obtain $\mu(Q)$ and $\phi_2(Q)$. While it is straightforward to obtain numerical data and draw pictures of these function, it is more satisfying to obtain approximate analytical expressions that can fit all these data, especially for the physically more meaningful relation $M(Q)$.  There are three types of boson stars: (1) gapless solutions whose mass and charge can approach zero smoothly, which we call the type $A$ series of boson stars; (2) the type $B$ series of boson stars that has mass gasp; (3) type $C$ series whose mass and charge are bounded both above and below.

We find that the boson stars that are bounded above are typically small with $Q\ll 1$, in which case, a linear expression of $M\propto Q$ can sufficiently capture the essence of the relation.  For the unbounded boson stars, as we discussed in the introduction, for large $Q$, the Smarr relation implies that $M\sim Q^{\fft32}$.  We do not find a close-form expression of $M(Q)$ that can match the data for both large and small $Q$ with less than $1\%$ error for the unbounded boson stars.  However, inspired by the RN-AdS example \eqref{RNMQ}, we find that such an expression can be achieved for the $\mu(Q)$, i.e., the chemical potential-charge relation, which takes the form
\be
\mu(Q)=\sqrt{\beta_1 Q + \beta_2 + \beta_3 \sqrt{Q^2 + \beta_4}}\,,\label{genmuQ}
\ee
where the coefficients $\beta_i$ can be determined using data fitting by the numerical method and they are all positive.  Compared to the $\mu(Q)$ for the extremal RN-AdS black hole (\ref{RNMQ}), we just add to the $\mu^2(Q)$ expression an additional term linear in $Q$, and hence it is a small step generalization, but as we shall see later, it fits the data in striking accuracy with errors typically of the order $\delta\mu/\mu \sim 10^{-3}\sim 10^{-7}$. Numerical values for various $\beta_i$ can be seen later in several typical examples of boson star solution (for instance, Eqs.~\eqref{fit1}, \eqref{fit2} and \eqref{q1B1muq}). We can then use the first law (\ref{0Tfirstlaw}) and obtain a quadrature expression for the mass $M$
\be
M(Q)=\int_0^Q \mu(Q') dQ'\,.\label{massQquadrature}
\ee
We can then easily establish that for small $Q\le 1$, the mass-charge relation is given by
\be
M=\sqrt{\beta _2+\beta _3 \sqrt{\beta _4}} Q+\frac{\beta _1}{4 \sqrt{\beta _2+\beta _3 \sqrt{\beta _4}}} Q^2+\frac{\left(2 \beta _2 \beta _3-\left(\beta _1^2-2 \beta _3^2\right) \sqrt{\beta _4}\right)}{24 \left(\beta _2+\beta _3 \sqrt{\beta _4}\right)^{3/2} \sqrt{\beta _4}} Q^3+{\cal O}\left(Q^4\right)\,.
\ee
For the more interesting large $Q$ limit, we have
\be
M=c_0+ \ft{2}{3} \sqrt{\beta _1+\beta _3}\, Q^{\fft32}+\frac{\beta _2}{\sqrt{\beta _1+\beta _3}}\, Q^{\fft12}-\frac{\left(2 \beta _3 \left(\beta _1+\beta _3\right) \beta _4-\beta _2^2\right)}{4 \left(\beta _1+\beta _3\right)^{3/2}}\, Q^{-\fft12} +{\cal O}\big(Q^{-\fft32}\big)\,.
\ee
We see that the Smarr relations (\ref{smarrlimit}) for small and large $Q$ are satisfied for our data-fitting ansatz.  The constant $c_0$ is determined by
\be
c_0 = \int_0^\infty \Big(\mu(Q) -
\sqrt{(\beta_1 + \beta_3)Q} - \fft{\beta_2}{2\sqrt{(\beta_1 + \beta_3) Q}} \Big) dQ\,,\label{c0formula}
\ee
for $\mu(Q)$ given in (\ref{genmuQ}).  For the extremal RN-AdS black holes, the coefficient $\beta_1$ vanishes and further more, we have $\beta_2=\beta_3\sqrt{\beta_4}$, in which case one can establish that $c_0$ vanishes also.  For general $\beta_i$ parameters; however, the coefficient $c_0$ are not vanishing and we shall determine it for various models by the numerical method.  Since the determination of $c_0$ from (\ref{c0formula}) is achieved by a subtraction of divergent terms to obtain a finite number, this makes it very difficult to obtain $c_0$ in reasonable accuracy using the numerical approach; nevertheless, our analysis together with using the numerical approach can at least demonstrate that $c_0$ does not vanish in general. The non-vanishing $c_0$ term in the tree level boson star mass does not necessarily imply a contradiction with the CFT prediction \eqref{mq1} where the constant term is due to an 1-loop effect. The reason is that the gravity solutions are usually dual to states of strongly coupled gauge theories; however, the formula \eqref{mq1} was derived for CFT described by vector or matrix models. Similar constant term was also encoded in the recent result of \cite{delaFuente:2020yua}, in which the Fig.(2b) shows that the boson star mass subtracting the leading $Q^{3/2}$ and subleading $Q^{1/2}$ term  approaches a negative constant instead of 0 in the large charge limit.  Indeed, in the large $Q$ expansion, we can also perform data fitting, using the function
\be
M(Q) = c_{\fft32} Q^{\fft32} + c_{\fft12} Q^{\fft21} + c_0 + \cdots
\ee
to determine $c_0$ and we find that it is in general nonzero and consistent with \eqref{c0formula}, but with much less accuracy, typically no more than one significant figure. It should be interesting to understand the counterpart of the $c_0$ term in a strongly coupled conformal gauge theory, which is beyond the scope of this paper.

\subsection{Numerical approaches}
\label{sec:num}

There are few known examples of exact solutions that describe the AdS boson stars.  Numerical approaches have typically been adopted in literature to construct these solutions.  A priori, there are two ways of integrating the differential equations.  One is to use the boundary condition at the asymptotical infinity and integrate from certain large value of $r$ to $r=0$.  The advantage is that we can freely choose the asymptotic hairy parameters that specifying the solutions. However, as we discussed earlier, the boson stars at the asymptotic infinity involves four hair parameters, $(M,Q, \mu, \phi_2)$, but they are specified by only one independent parameter.  The boson stars form lines in the four dimensional parameter space.  It is thus an inconceivable task to search such solutions in good accuracy, since it amounts to locate a one-dimensional line in four-dimensional space. For general $(M,Q, \mu, \phi_2)$ parameters, the solution in the spatial middle can have naked singularity or the event horizon, as well as the soliton structure.

In this paper, we shall use the $r=0$ as the initial boundary and integrate the equations from $r=0$ out to some sufficiently large $r$ and then read off the hair parameters by data fitting using the asymptotic falloff structures. The boundary condition at $r=0$ is specified by (\ref{boundarycond1}) and (\ref{boundarycond2}) with two free parameters $(\phi_0,a_0)$.  Although the solutions are regular at $r=0$, the equations break down at $r=0$ since $r$ appears in the denominator of the expressions of $(h'', a'', \phi'')$ in appendix \ref{app:eoms} and the numerical program has to deal with the subtlety of $0/0$. The issue can be resolved by performing the Taylor expansion at $r=0$ to some higher orders and use the Taylor series as the correct solution in the vicinity of $r=0$.  We can therefore integrate the equations not literally from $r=0$, but slightly away from $r=0$, e.g.~$10^{-3}$ where the equations are well defined.  We perform the Taylor expansions to the tenth order of $r$, but present figuratively only a few terms here for illustration
\bea
&& h=h_0(1 + h_2 r^2 + h_4 r^4 + \cdots)\,,\qquad
f=1 + f_2 r^2 + f_4 r^4 + \cdots\,,\nn\\
&& a=a_0 + \tilde a_2 r^2 + \tilde a_4 r^4 + \cdots\,,\qquad
\phi=\phi_0 + \tilde \phi_2 r^2 + \tilde \phi_4^4 + \cdots\,.
\eea
We can set $h_0=1$ without loss of generality, with the understanding that we can use the time scaling to restore the solution to be $h\sim r^2$ asymptotically.  Substituting the above ansatz into the equations of motion and we can solve them order-by-order in terms of $r$.  We find that all the coefficients can be solved as polynomial ratios of $(\phi_0,a_0)$.

For some chosen $(a_0, \phi_0)$, the integration out to infinity may break down, which implies that this spatial middle structure cannot connect to the asymptotic AdS, but it encounters singularity instead.  When the middle structure for $(a_0,\phi_0)$ can be integrated to the asymptotic infinity, we then obtain a smooth AdS soliton and the falloff of the scalar must take the form (\ref{scalarfalloff}). We can use data fitting  at large $r$ to read off both constants $(\phi_1,\phi_2)$. Since we are only interested in solutions with vanishing $\phi_1$, we can finetune the parameters $(a_0,\phi_0)$ so that $\phi_1$ vanishes, or more practically, we require that $\phi_1<10^{-15}$. The function $h$ at large $r$ then must take the form
\be
h=\lambda \Big(r^2 + 1 - \fft{2M}{r} + \fft{4Q^2}{r^2} + \cdots\Big)\,.
\ee
For sufficiently large $r$, the ellipses can be ignored and we can use the remaining function to perform data fitting and read off $(\lambda, M, Q^2)$.  The reason we inevitably have $\lambda\ne 1$ is that we have set $h_0=1$ as our initial boundary condition at $r=0$.  We need to rescale the functions
\be
h\rightarrow \fft{h}{\lambda} \,,\qquad a\rightarrow \fft{a}{\sqrt{\lambda}}\,,
\ee
to obtain the correct numerical solutions for the $(h,a)$ functions.  We can then read off the correct $(\mu, Q)$ from the falloff of the electric potential in (\ref{falloffs}).  This implies that $a_0$ is not the true chemical potential at the origin; the correct answer is
\be
\mu_0 = \fft{a_0}{\sqrt{\lambda}}\,.
\ee
Thus we can read off six important parameters of a boson star, namely
\be
\{\phi_0, \mu_0, M, Q, \mu, \phi_2\}\,.
\ee
The first two specify the spacetime structure in the middle, whilst the last four specify the asymptotic structure. Note that we can have a consistent check on the electric charge, by first reading off $Q$ from $a(r)$ in (\ref{falloffs}) and then compare it with the coefficient of $1/r^2$ in $h$.  We require that they match with at least five significant figures.

\subsection{Euclidean action and the ground state}

For stationary black holes, there is a quantum static relation (QSR) which states that the on-shell Euclidean action is given by $\beta F$ where $F$ is the free energy and $\beta$ is the period of the Euclidean time \cite{Gibbons:1976ue}. We can take the view that the free energy is the action growth rate with respect to the Euclidean time, in which case, the free energy can be directly calculated in the extremal limit.  In this section, we would like to generalize the concept from black holes to the boson stars.

For asymptotically AdS spacetimes, the Euclidean action have three parts, the bulk, the Gibbons-Hawking surface term and the holographic counterterm:
\be
I_{\rm Eucl} = I_{\rm bulk} + I_{\rm surf} + I_{\rm ct}\,.
\ee
It turns out that the bulk action can be written as surfaces term module the Hamiltonian constraint \cite{Donos:2012yu,Cremonini:2014gia}
\be
I_{\rm bulk} = \ft14\tau \int_{r_0}^\infty 2E^0{}_0 dr + \ft14\tau\, r^2 \sqrt{\ft{f}{h}} (h'-a a')\Big|_{r_0}^{\infty}\,,\label{bulkaction}
\ee
where $E_{\mu\nu}$ is given in (\ref{Emunu}). The equation of motion $E^0{}_0=0$ is called the Hamilton constraint.
This identity can generalized to include general higher-derivative terms, see e.g.~\cite{Liu:2017kml,Fan:2018wnv}.
For the black hole solutions, we have $r_0=r_+$; for the AdS solitons, $r_0=0$. The Gibbons-Hawking surface term is
\be
I_{\rm surf}=-\ft14\tau r^2 \sqrt{\ft{f}{h}} \Big( h' + \fft{4h}{r}\Big)\Big |_{r\rightarrow \infty}\,.
\ee
The holographic counter term is
\be
I_{\rm ct} =\tau \fft{g^2 + 2 r^2}{2g} \sqrt{h}\Big|_{r\rightarrow \infty}\,.
\ee
For the RN-AdS black holes, putting together gives the Gibbs free energy
\be
F_G=M- T S -\mu Q\,.
\ee
For the boson stars, we find it is given by
\be
F_G=M - \mu Q\,,
\ee
which takes the precisely same form as the extremal RN-AdS black hole.  In order to obtain the Helmholtz free energy, we need to add a boundary term for the Maxwell field, representing the Legendre transformation,
namely \cite{Gibbons:1976ue}
\be
I_{\rm Legendre}= \fft{1}{16\pi} \int_{\partial M} d\Sigma_\mu F^{\mu\nu} A_\nu\,.
\ee
This term cancels precisely the $aa'$ term in (\ref{bulkaction}). The total Euclidean action then gives rise to the Helmholtz free energy and for the boson stars or extremal black holes, it is precisely the mass
\be
F_H = M\,.
\ee
Thus we see that the zero-temperature ground state is the lowest energy state.

\subsection{The dynamical first law of the AdS boson star}

We now apply the Iyer-Wald formalism \cite{Wald:1993nt} to demonstrate that the dynamical first law of the boson star takes the same form as (\ref{0Tfirstlaw}).  For the theory of the type (\ref{genlag}), the variation of the Hamiltonian in the Wald formalism for the electrically charged spherically-symmetric and static solution is given by \cite{Lu:2013ura,Liu:2013gja,Lu:2014maa,Liu:2014dva}
\be
\delta H_{\infty} = -\ft14 \lim_{r\rightarrow \infty} \Big(-\fft{2}{r} \delta f - 2 \sqrt{h f} \phi' \delta \phi
-\fft{f}{h} a\delta a' -\ft12 a a' \delta(\fft{f}{h})\Big)\,.
\ee
Substituting the asymptotic boundary condition (\ref{falloffs}), we find
\be
\delta H_{\infty} = dM - \mu dQ\,.
\ee
Following the Wald formalism, we must have $\delta H_{\infty}=0$, since the smooth spatial origin $r=0$ is not a boundary.  (For black holes, the Wald-formalism identity is $\delta H_{\infty}=\delta H_{+}$ instead.) We therefore derive the first law of the boson star dynamics, which has the same form (\ref{0Tfirstlaw}) as the extremal RN-AdS black hole.

It is important to note that first law (\ref{0Tfirstlaw}) we derived from the Iyer-Wald formalism is consistent with the Helmholtz free energy derived from the Euclidean action in the previous subsection. Nevertheless, since we do not have an exact solution of the AdS boson stars to verify this first law analytically, we shall result to the numerical approach and obtain the mass and chemical potential as functions of the charge and confirm its validity.  In fact this first law is particularly useful since it allows to derive the mass-charge relation through its integration (\ref{massQquadrature}), as it is easier to find the analytical data-fitting function $\mu(Q)$.

\subsection{The AdS/CFT correspondence}

In this subsection, we present the map between the mass of global AdS solutions and the scaling dimension of heavy operators in the dual CFT. First of all, the dual CFT lives on the conformal boundary of the global AdS solution which is $R_t\times S^2(\ell)$ where $R_t$ is the time direction and $S^2(\ell)$ denotes a 2-sphere with radius $\ell$. Thus the dual CFT is defined on a sphere of the same size as the bulk AdS. The energy of the extremal RN-AdS black hole or AdS boson stars is identified with the energy of the corresponding CFT state. In full units, this is
\be
{\cal E}_{CFT}=\frac{M}{G_N}\,.
\ee
The state-operator correspondence then implies that the spectrum of dilaton operator contains an operator of scaling dimension
\be
\Delta={\cal E\ell}_{CFT}=\frac{\ell^2}{G_N}\frac{M}{\ell}\,.
\ee
Since $\ell^2/G_N$ is related to the CFT central charge as $\pi c_T/32$, the operator is quite heavy. In our computation, we set $\ell=1$. However, one should keep in mind that in the mass-charge relation, the parameters $M$ and $Q$ are in fact weighted by the AdS radius. As an illustration, the conformal dimension of heavy operator dual to the extremal AdS$_4$ RN black hole can be expressed as
\bea
\frac{32\Delta^{\rm RN}}{\pi c_T} &=& \frac{\sqrt{\sqrt{48 Q^2+1}-1} \left(\sqrt{48 Q^2+1}+2\right)}{3 \sqrt{6}} \nn\\
&=&\frac{4 \sqrt{2}}{3^{3/4}} Q^{\fft32}+\frac{1}{\sqrt{2} \sqrt[4]{3}} Q^{\fft12}-\frac{1}{16 \left(\sqrt{2} 3^{3/4}\right)} Q^{-\fft12} +{\cal O}\left(Q^{-\fft32}\right).
\eea
With this understood, we shall only present the bulk $M(Q)$ relation, rather than the CFT $\Delta (Q)$ in the remainder of the paper.  It is important to note that the constant coefficient $c_0$ is absent in the large $Q$ expansion.  As we will see later, this is quite different from the boson stars where the large charge expansion of mass generally involves the $c_0$ term.

\section{The phenomenological toy model}

Having established the motivation and the general set up, we would like to present the results. In this section, we analyse the numerical data of the phenomenological toy model, given in (\ref{threetheories}).  For completeness, we shall not only obtain solutions that have large charge limit, but also solutions whose charges are bounded above. Qualitatively, our results agree with the findings of \cite{Gentle:2011kv}. Even though these small charge solutions do not have immediate applications in the AdS/CFT correspondence, they can  give a more complete picture of the solution space. The more detailed classification given here also reveals how the previously constructed boson stars, {\it e.g.}~those obtained in \cite{Gentle:2011kv,delaFuente:2020yua} emerge in the solution space. The theory is specified by two parameters $(m,q)$ with $m^2=-2$.  Boson stars asymptotic to global AdS$_4$ exist for general $q$, but the properties can be very different depending on $q>q_c$ or $q<q_c$ where $q_c^2=1.261$. (We shall come back to this point in section \ref{sec:genq}.)  In this section, we analyse the boson stars for $q^2=1.4$ and $q^2=1$ in detail and then give some brief discussions on general $q$.

\subsection{$q^2=1.4$}

\begin{figure}[htp]
\begin{center}
\includegraphics[width=250pt]{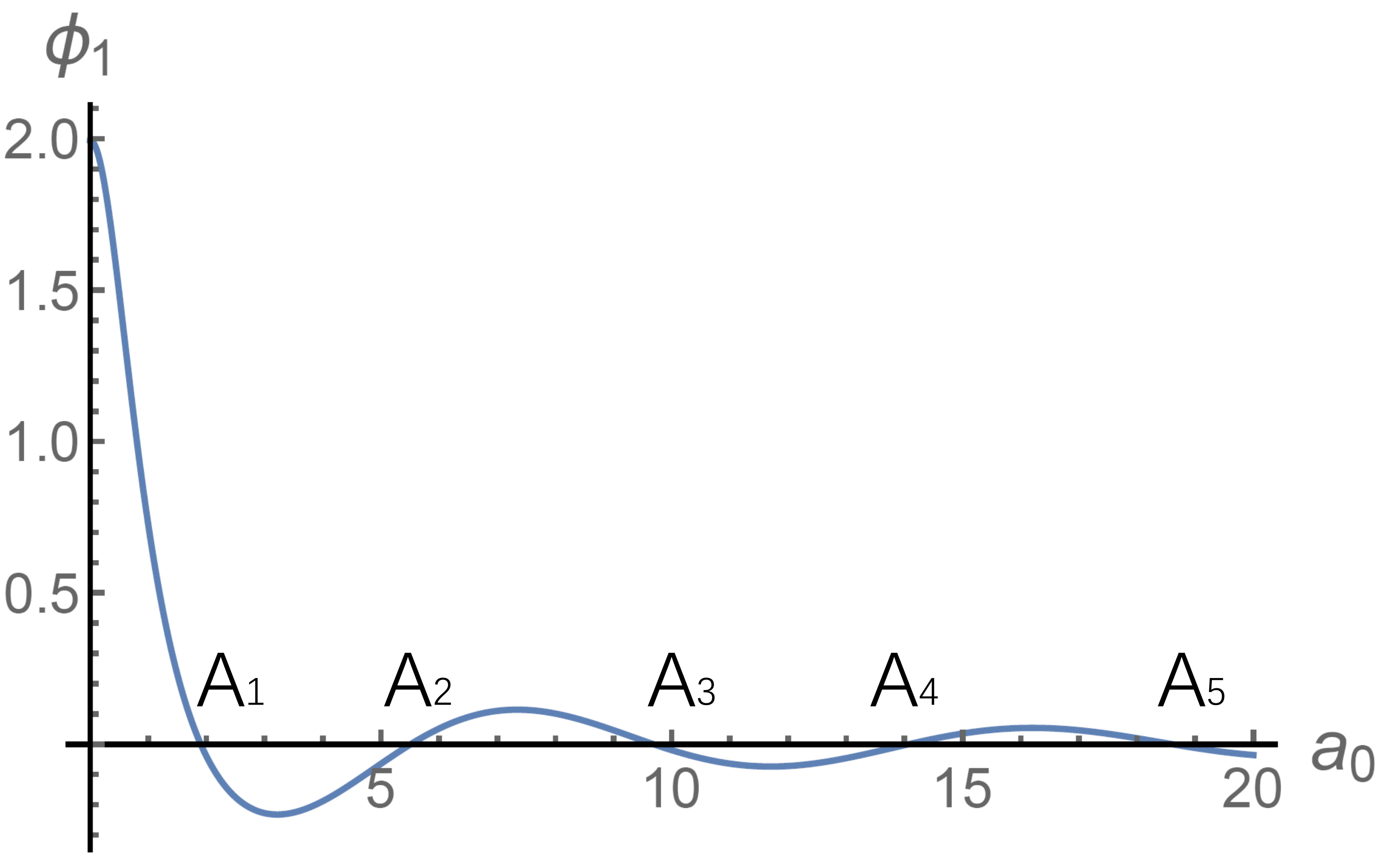}
\end{center}
\caption{\small\it For $q^2=1.4$ and $\phi_0=1$, the asymptotic scalar hair $\phi_1$ is a decaying periodic function of $a_0$.  Each root, labelled as $A_i$, $(i=1,2,3,\ldots,)$ gives a boson star solution. Varying $\phi_0$ then gives rise to the $A_i$ series of solutions.}
\label{q14phi0is1phi1}
\end{figure}

As discussed in section \ref{sec:setup}, the general soliton solutions are characterised by $(\phi_0,a_0)$ defined at the origin $r=0$, which lead to asymptotic scalar hair parameters $(\phi_1,\phi_2)$.  For some specific choices of $(\phi_0,a_0)$, the solitons have vanishing $\phi_1$.  We thus need to scan the parameter plane of $(\phi_0,a_0)$.  For given $\phi_0$, we find that for the $r=0$ boundary condition $a_0=0$, the equations can always be integrated to infinity, implying that $(\phi_0,0)$ always gives a smooth AdS soliton.  We can thus first fix $\phi_0$ and then scan $a_0$ from 0 to certain large number and obtain AdS solitons whose scalar hair $\phi_1$ is a function of $a_0$. The boson star solutions we are looking for are those with $\phi_1(a_0)=0$.  As a concrete example, we consider $\phi_0=1$ and scan $a_0$ from 0 to 20, at an interval of $1/50$.  The function of $\phi_1(a_0)$ is plotted in Fig.~\ref{q14phi0is1phi1}.

We see that for $\phi_0=1$, $\phi_1(a_0)$ is a decaying periodic function with multiple roots (possibly an infinite number of zeros). Each root, labeled as $A_i$, ($i=1,2,3,\ldots$,) gives a boson star solution. If we change $\phi_0$, each root will move continuously with $\phi_0$, giving a continuous series of solutions.  We use $A_i$ to label each series.  We now give some details of the $A_1$, $A_2$ and $A_3$ series at $\phi_0=1$. We have

\smallskip
\begin{center}
\begin{tabular}{|c|c|c|c|c|c|c|c|}
  \hline
   &$a_0$ & $\lambda$ & $\phi_2$ & $\mu$ & $Q$ & $M$ \\ \hline
  $A_1$ &1.902  & 2.890 & 0.9920 & 2.184 & 0.2307 & 0.4479 \\ \hline
  $A_2$ &5.512  & 5.043 & -0.5164 & 2.898 & 0.05024&0.1612 \\ \hline
  $A_3$ &9.686& 6.093 & 0.4417  & 4.194 & 0.02423 & 0.1159\\
  \hline
\end{tabular}
\end{center}
\smallskip

\noindent Note that the sign of the scalar hair $\phi_2$ alternates for the two adjacent series. It should be emphasized again that the chemical potential at the origin is not $a_0$; instead, it is given by
\be
\mu_0= \fft{a_0}{\sqrt{\lambda}}=1.119\,,\qquad 2.455\,,\qquad 3.924\,,
\ee
for the $A_1,A_2,A_3$ solutions respectively. We obtain a matrix of data for the $A_1$, $A_2$ and $A_3$ series of solutions for $\phi_0$ runs from $1/10$ to 3.

\subsubsection{The $A_i$ series: gapless solutions}

As we see later, the pattern of $\phi_1(a_0)$ depicted in Fig.~\ref{q14phi0is1phi1} breaks as $\phi_0$ increases to certain critical value and new pattern will emerge; however, the pattern remains unchanged as $\phi_0$ decreases to 0, corresponding to the AdS vacuum.  Thus the $A$ series of solutions are gapless and their mass and charge can smoothly reach zero.  In this subsection, we discuss the properties of these gapless solutions.  Nevertheless, the $A_1$ series is special, different from the $A_2$ and $A_3$ series and we shall present mostly the $A_2$ and $A_3$ series first, where the electric charge $Q$ is restricted to be small compared to the cosmological constant.

\begin{figure}[htp]
\begin{center}
\includegraphics[width=200pt]{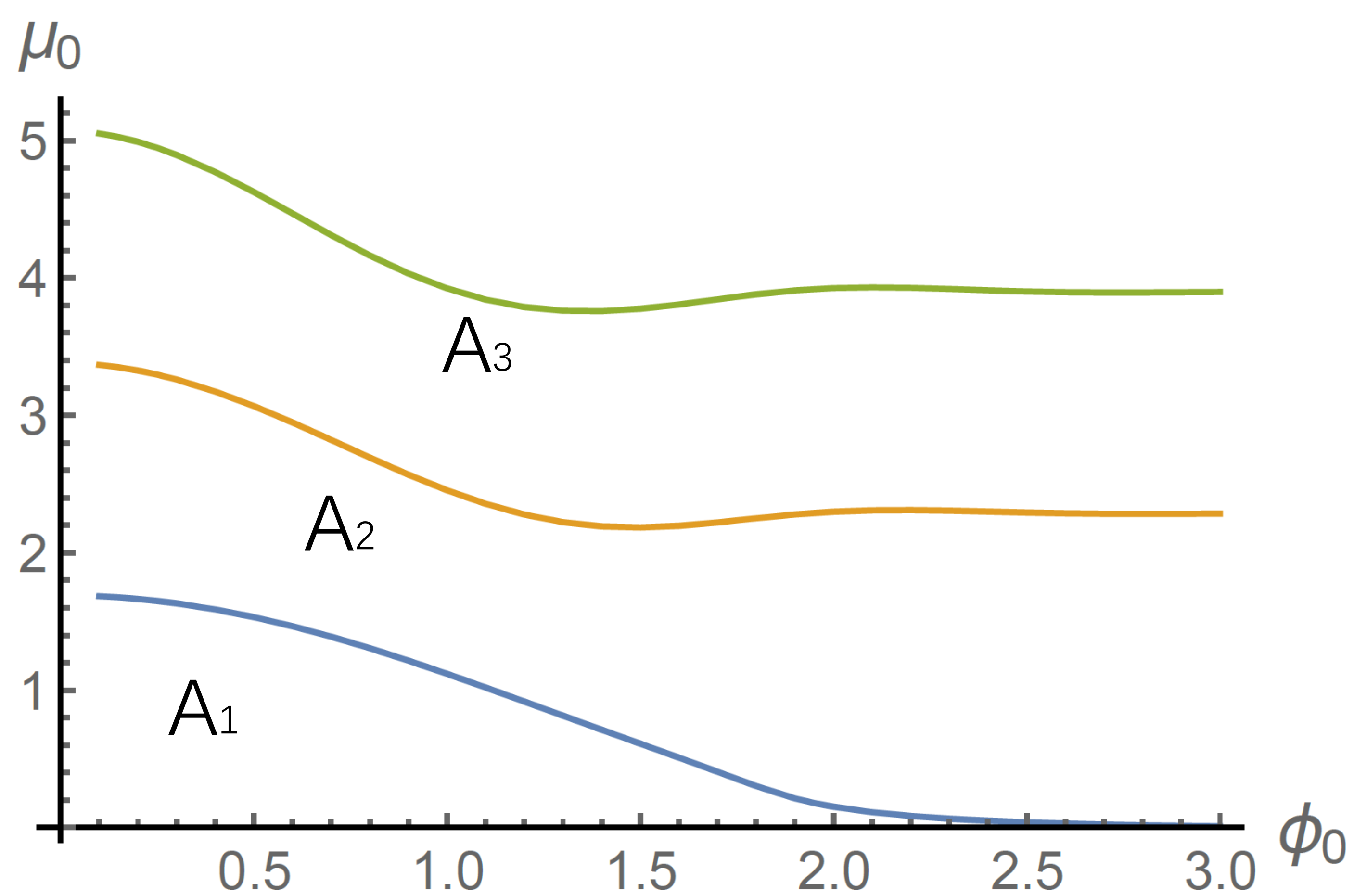}\
\includegraphics[width=200pt]{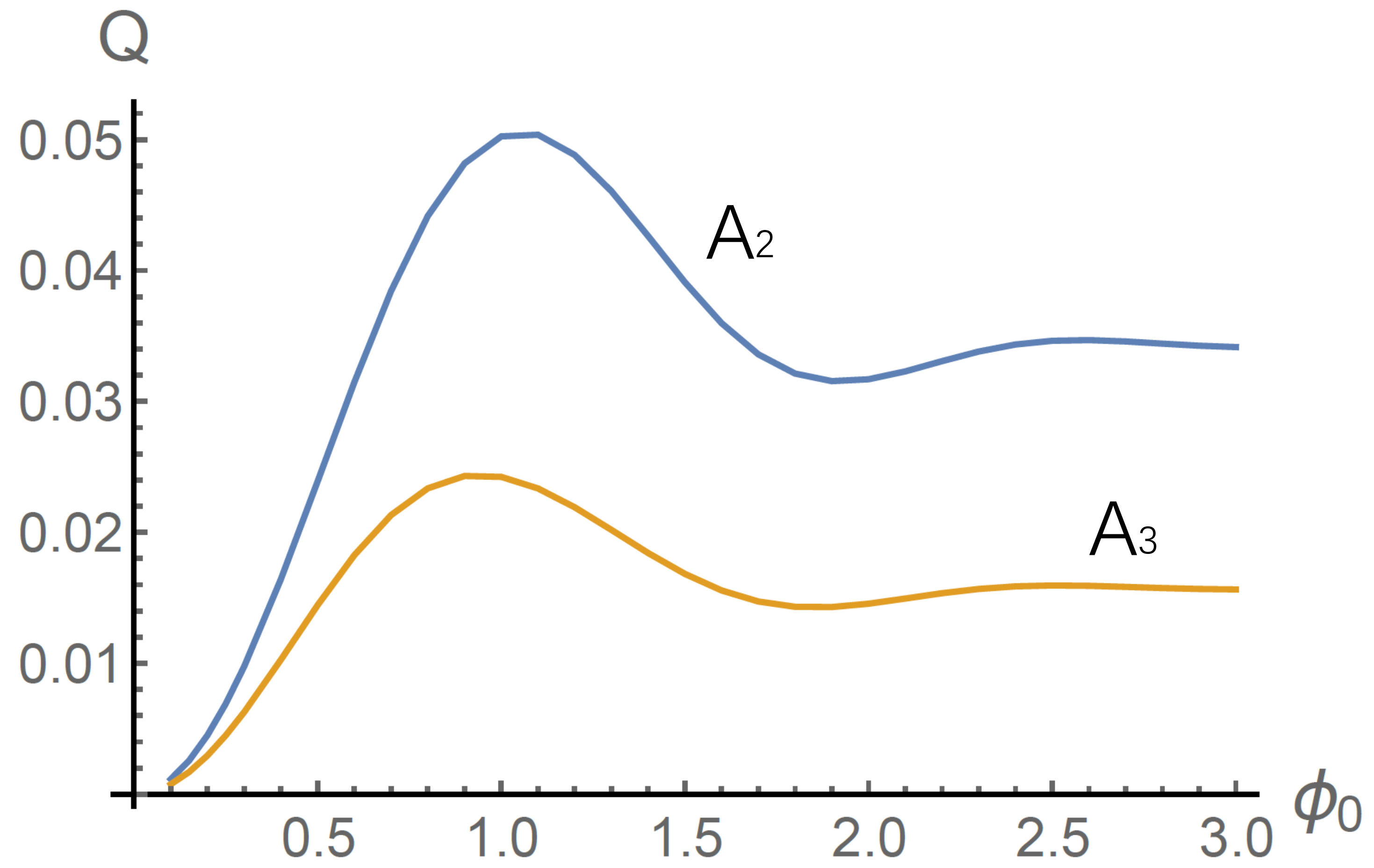}
\end{center}
\caption{\small\it The left plot shows that the $A_{1,2,3}$ series of boson stars trace lines in the $(\phi_0, \mu_0)$ plane. The right plot relates the asymptotic parameter $Q$ to $\phi_0$ defined in the space origin. We see that the charge is restricted to have a maximum that is much less than the AdS radius $\ell=1$. The charge $Q$ for the $A_1$ series is unbounded above and we shall present it later.}
\label{q14A2A3mu0phi0}
\end{figure}

We can see from Fig.~\ref{q14A2A3mu0phi0} that the chemical potential $\mu_0$ at the spatial origin $r=0$ increases as the label of $A_i$ series increases.  For the $A_2$ and $A_3$ series, the electric charge $Q$ is bounded above, with the maximum much less than the AdS radius $\ell=1$.  Furthermore, the charge oscillates with increasingly small magnitude as $\phi_0$ increases.  The electric charge $Q$ is unbounded for the $A_1$ series, which we shall discuss presently.

\begin{figure}[htp]
\begin{center}
\includegraphics[width=200pt]{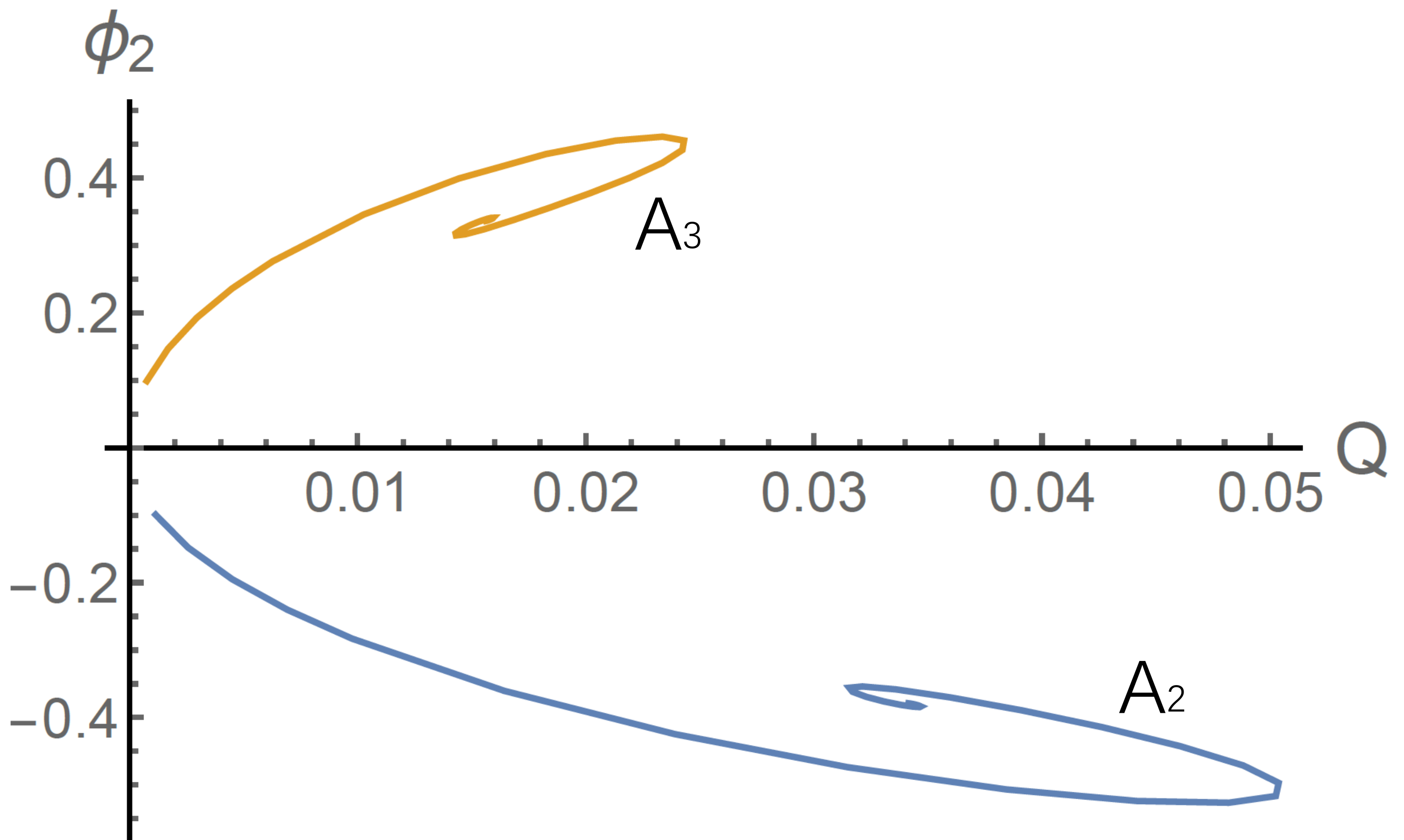}\
\includegraphics[width=200pt]{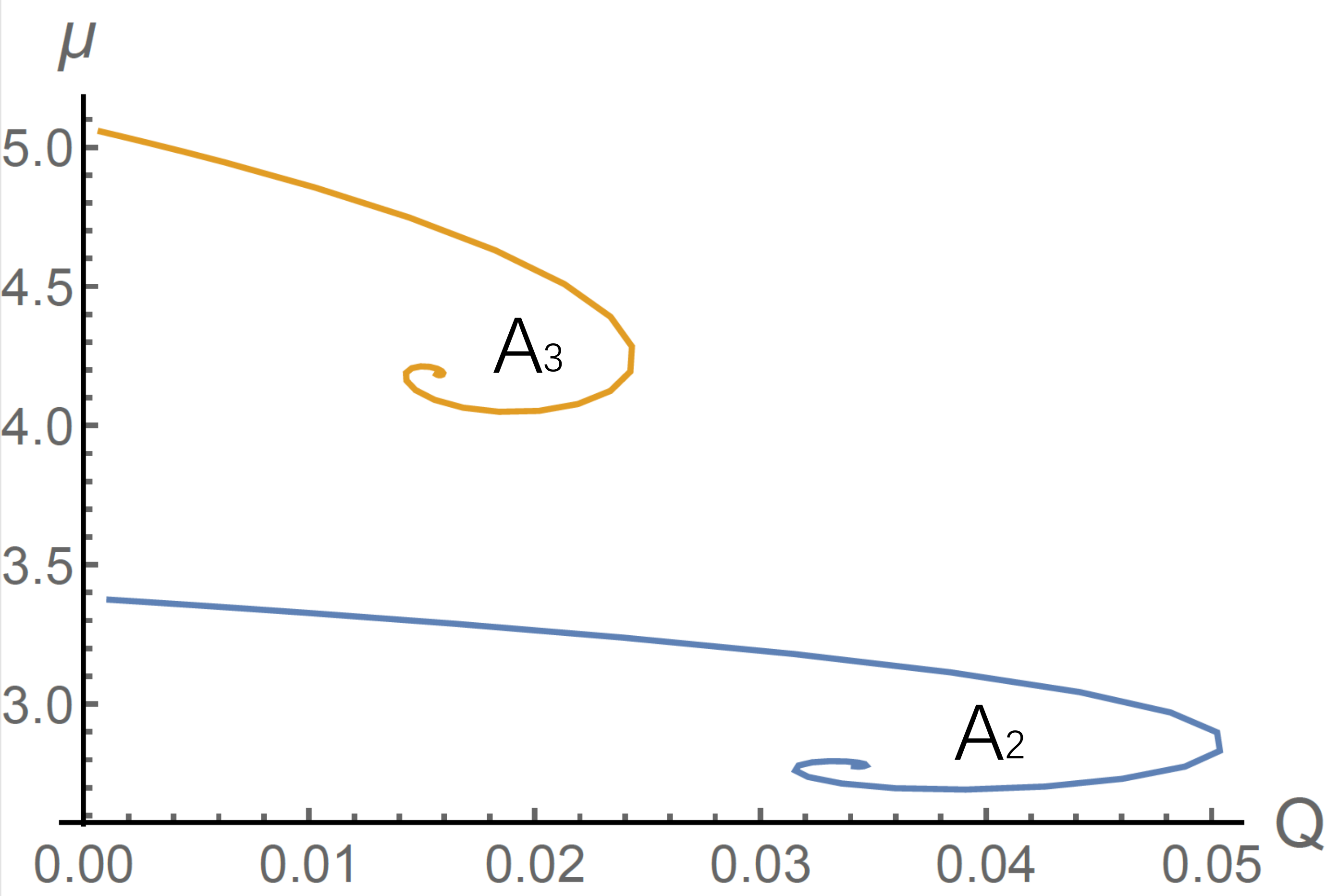}
\end{center}
\caption{\small\it The scalar hair parameter $\phi_2$ and the chemical potential $\mu$ are functions of the electric charge $Q$, for the $A_{2,3}$ series of solutions. The fixed points that the figures spiral into corresponds to taking large $\phi_0$. This is indicative that the charge $Q$ are bounded above. The shape of $\mu_0(Q)$ is similar to $\mu(Q)$. Note further that $\mu\rightarrow \mu_0$ as $Q\rightarrow 0$.}
\label{q14A2A3phi2mu}
\end{figure}

    We now turn to examine the relations of asymptotic hair parameters $(M,Q,\mu,\phi_2)$ of the AdS boson
stars.  We opt to present the one-parameter family of solutions using $Q$ as the variable, since all the boson stars are charged.  As we can see from Fig.~\ref{q14A2A3phi2mu}, for the $A_{2,3}$ series of solutions, the scalar hair $\phi_2$ and the chemical potential spiral into some fixed point, corresponding to large $\phi_0$.  This confirms that the electric charge $Q$ oscillates with $\phi_0$ and is bounded above.  The spiral shapes make it difficult to find an analytical expression for these relations.

\begin{figure}[htp]
\begin{center}
\includegraphics[width=300pt]{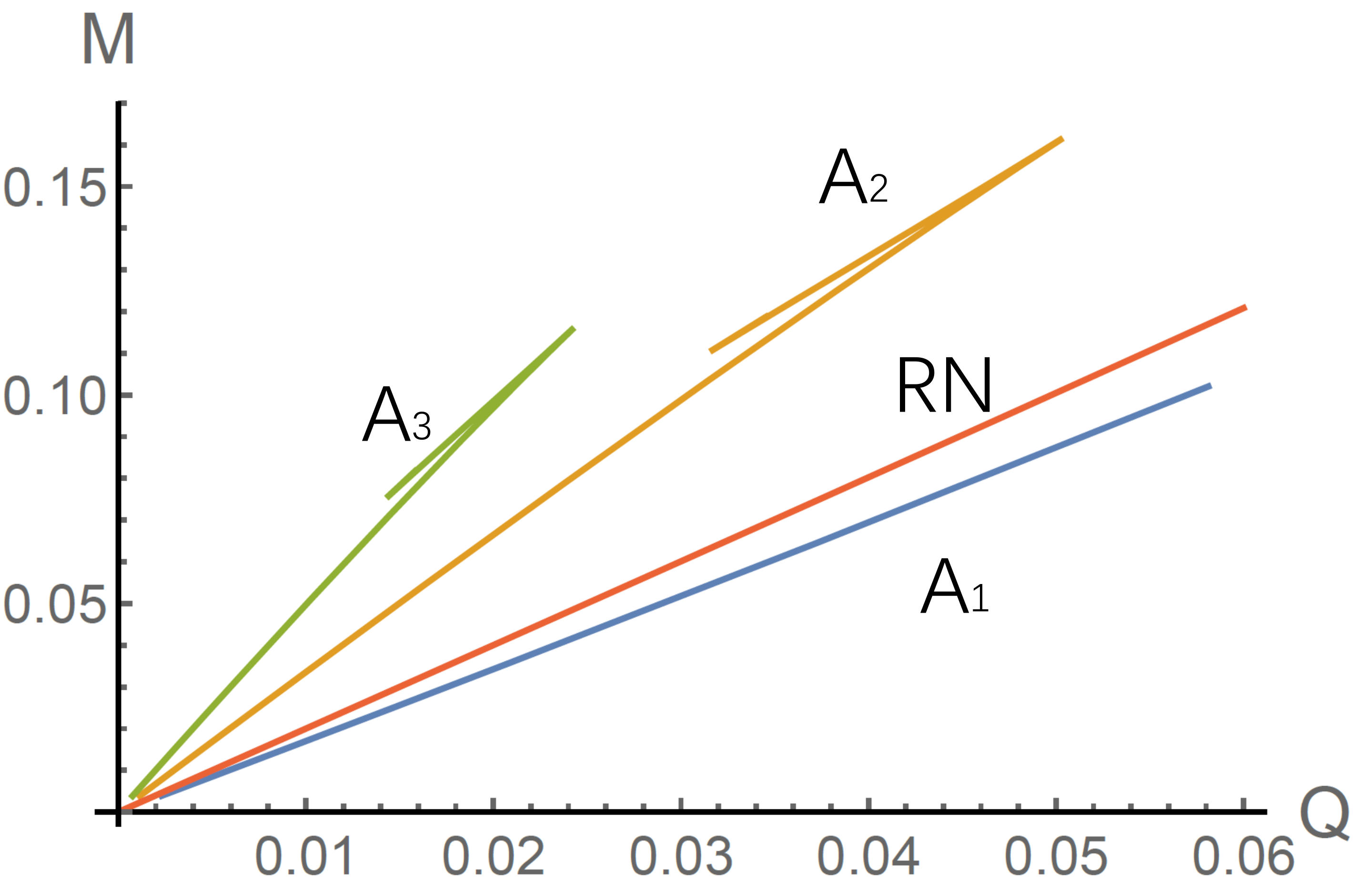}
\end{center}
\caption{\small\it The mass is multi-valued with a spike for the $A_{2,3}$ series.  The $A_1$ series has the lowest mass or energy for given charge $Q$, indicating that it is the ground state for fixed charge.}
\label{q14A1A2A3RNM(Q)}
\end{figure}

The most important physical quantity is the energy or the mass of the boson star for given charge $Q$.  We present the mass-charge relation for the $A_{2,3}$ series of solutions in Fig.~\ref{q14A1A2A3RNM(Q)}. We also present the mass-charge relation of the $A_1$ series and extremal RN black hole in the same charge region.  As we can see from Fig.~\ref{q14A1A2A3RNM(Q)}, there is a cusp in the mass-charge relation for the $A_2$ or $A_3$ series, and mass appears to be linearly dependent on the charge away from the cusp.  This would contradict the $\mu(Q)$ relation depicted in Fig.~\ref{q14A2A3phi2mu}.  A more careful analysis indicates that the $M(Q)$ relation for the $A_{2,3}$ series is not linear and the two lines join tangentially together at the tip so that there is no $\mu(Q)$ discontinuity.  We give a thorough analysis and verify the first law (\ref{0Tfirstlaw}) in appendix \ref{app:cusp}.

We can see from Fig.~\ref{q14A1A2A3RNM(Q)} that for the $A_i$ series of solution, the the upper bound lowers down as $i$ increases, and the mass increases for the same $Q$.  In other words, $M_{A_i}(Q) > M_{A_j} (Q)$ for $i>j$ and consequently the $A_1$ series has the lowest energy. Furthermore the $A_1$ series has less energy than the extremal RN-AdS black hole.  Thus the $A_1$ series represents the ground state, at least in the small charge region we consider here. We therefore would like to study the $A_1$ series in more detail. The $\mu_0$-$\phi_0$ relation was already presented in the left plot of Fig.~\ref{q14A2A3mu0phi0}, which is not significantly different from the those of other series of solutions, except that $\mu_0$ approaches zero for large $\phi_0$ for the $A_1$ series of solutions. However, the distinguishing feature of the $A_1$ series is that the charge $Q$ is unbounded, running from 0 to infinity.

\begin{figure}[htp]
\begin{center}
\includegraphics[width=220pt]{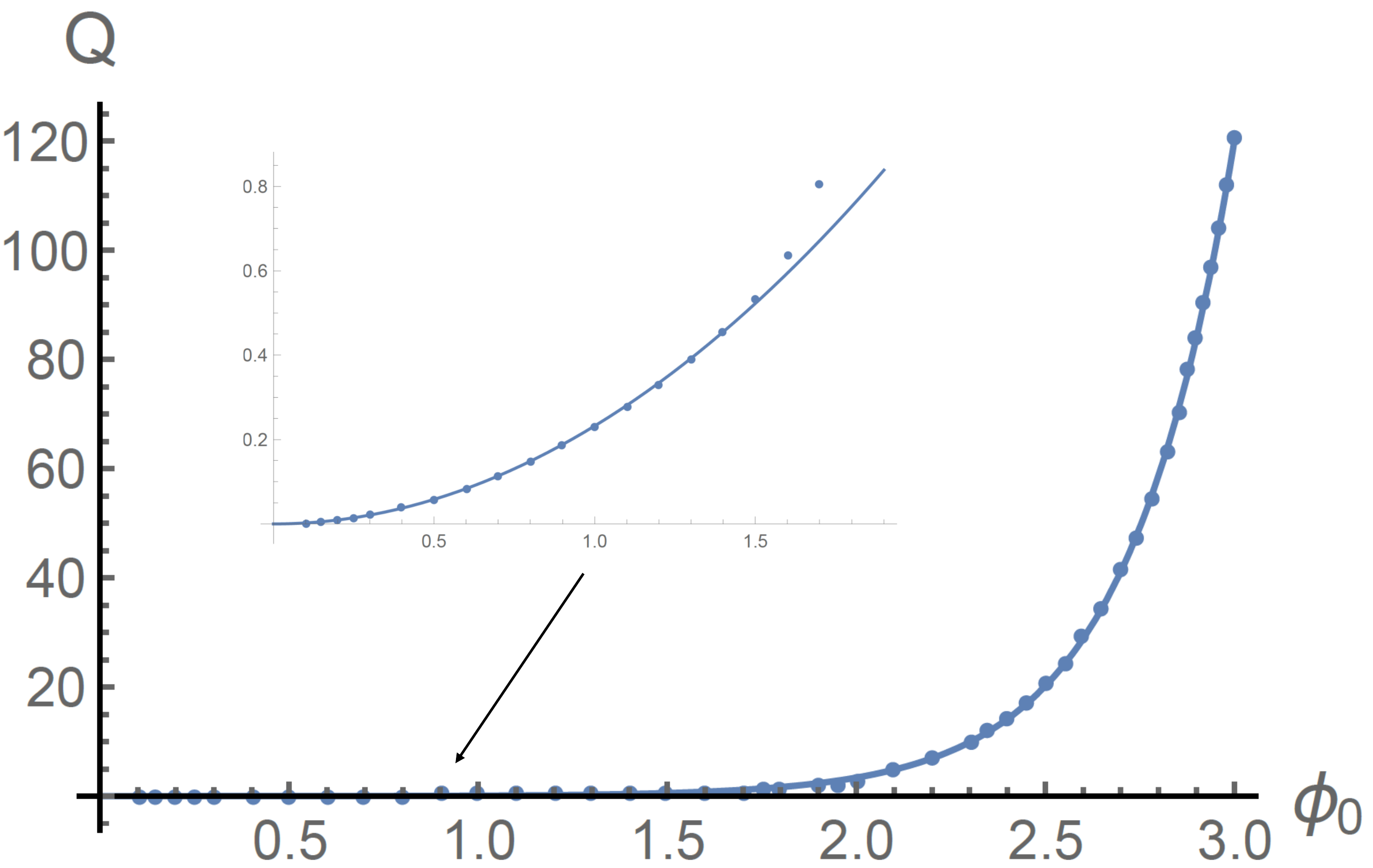}\
\includegraphics[width=220pt]{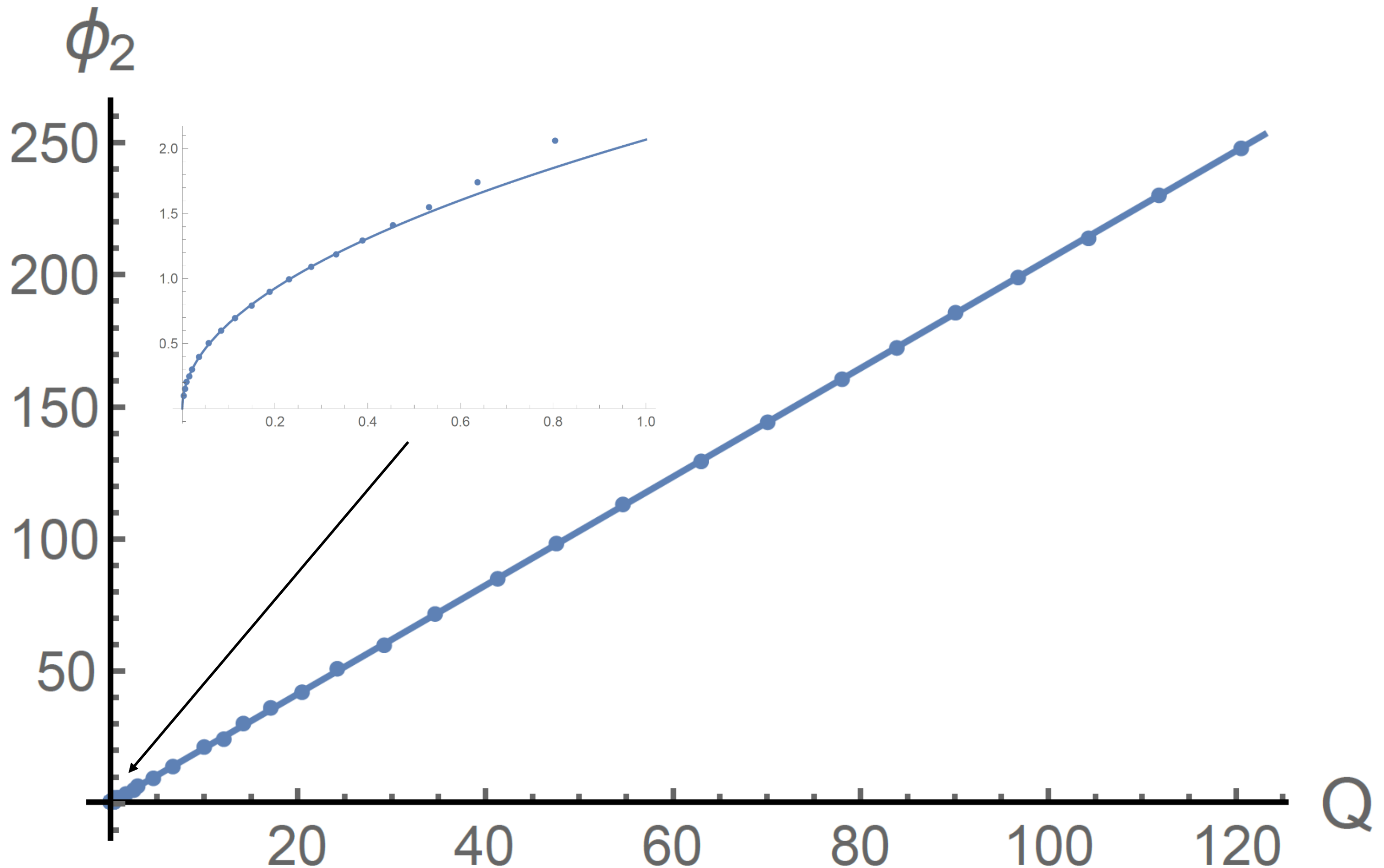}
\end{center}
\caption{\small\it The scalar hair-charge relations for the $A_1$ series of solutions. The left plot shows the charge $Q$ as the function of $\phi_0$.  For large $Q$, it grows exponentially whilst at small $Q$, it behaves as a quadratic function of $Q$. The scalar hair $\phi_2$ behaves linearly at large $Q$ and behaves like $\sqrt{Q}$ as $Q$ approaches 0.  The dots denote the actual numerical data and solid lines are the resulting data fitting functions.}
\label{q14A1Qphi2}
\end{figure}

As we can see from Fig.~\ref{q14A1Qphi2}, for the $A_1$ series of solutions, the charge $Q$ increases exponentially as a function of $\phi_0$ when $Q\gg 1$, but grows quadratically when $Q\ll 1$.  In particular, our data fitting indicates that
\be
Q>10:\qquad Q\sim 0.002717 e^{3.565 \phi_0}\,,\qquad\qquad
Q< 1/2:\qquad Q \sim 0.2323 \phi_0^2\,.
\ee
The scalar hair/charge relation $\phi_2(Q)$, on the other hand, is linear for large $Q$, and becomes $\sqrt{Q}$ as $Q$ approaches zero.  To be specific, we have
\be
Q>1:\qquad \phi_2 \sim 0.3454 + 2.054 Q\,,\qquad\qquad
Q<0.5:\qquad \phi_2 \sim 2.069 Q^2\,.
\ee
Note that the solid lines in Fig.~\ref{q14A1Qphi2} are drawn using the above data fitting functions, and the dots are the actual numerical results.  We have not found explicit data fitting functions $Q(\phi_0)$ and $\phi_2(Q)$ for the whole $Q\ge 0$ that would yield less than $5\%$ error.

\begin{figure}[htp]
\begin{center}
\includegraphics[width=220pt]{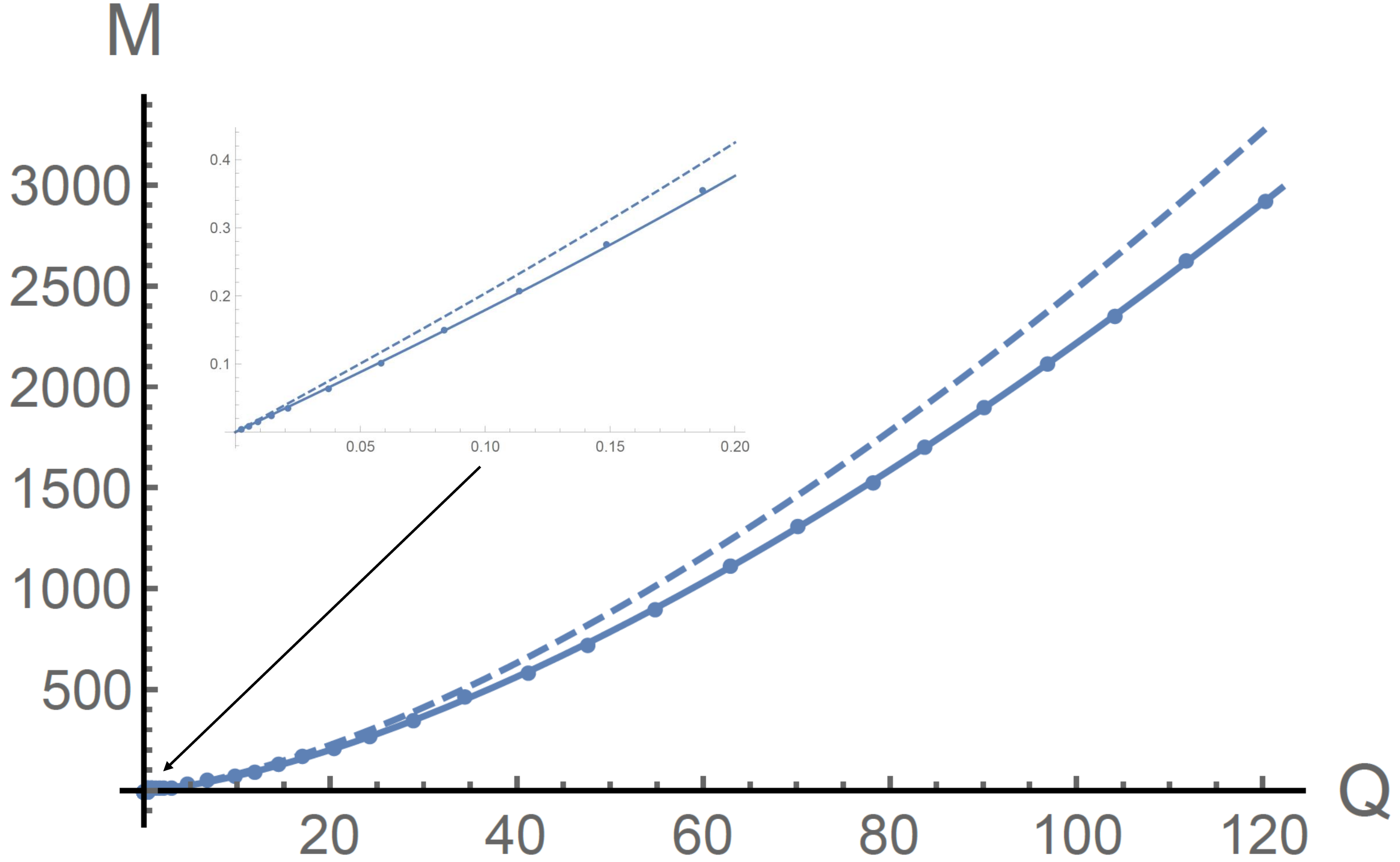}\
\includegraphics[width=220pt]{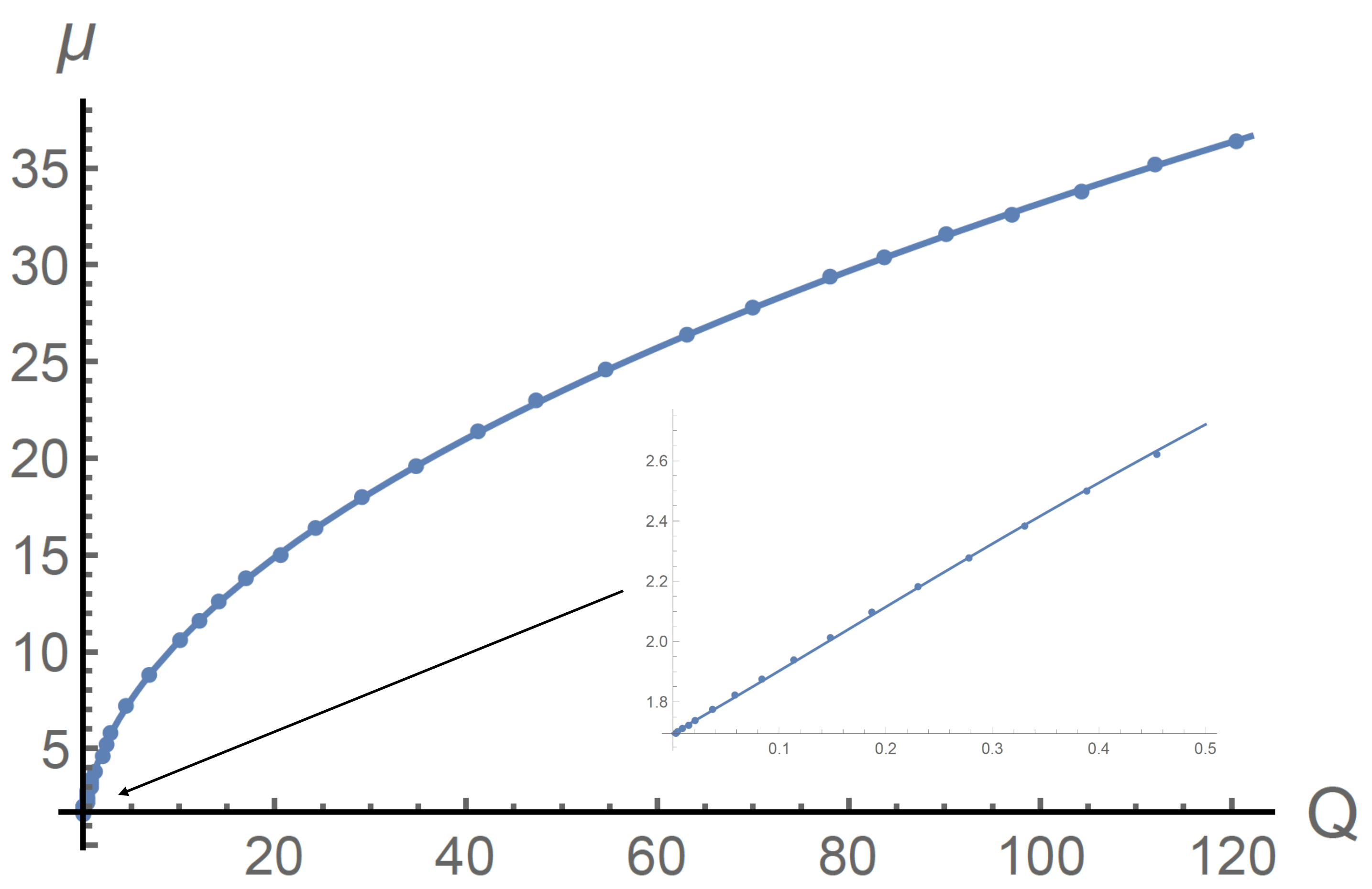}
\end{center}
\caption{\small\it The dots are actual numerical data and the solid lines are the derived data-fitting functions.  The dashed line in the $M(Q)$ graph is the mass-charge relation for the extremal RN-AdS black hole.}
\label{q14A1Mmu}
\end{figure}

The mass $M$ and the chemical potential $\mu$ as functions of charge $Q$ are depicted in Fig.~\ref{q14A1Mmu}.  The dots are actual numerical data and the solid lines are derived data-fitting functions.  The dashed line in the $M(Q)$ graph is the mass-charge relation for the extremal RN-AdS black hole.

\begin{figure}[htp]
\begin{center}
\includegraphics[width=300pt]{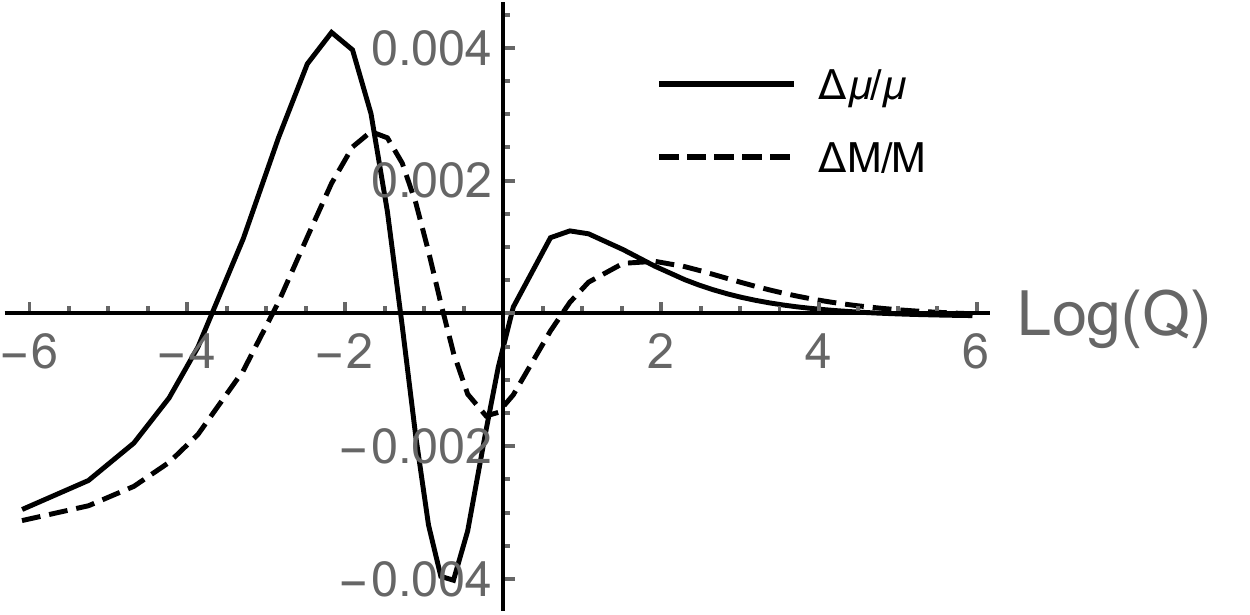}
\end{center}
\caption{\small\it This figure shows that the errors of our data-fitting function $\mu(Q)$ and $M(Q)$ with respect to the numerical data are all less than 0.4\%, in the region of $Q\in [0.0023234, 369.36]$. Furthermore the error reduces significantly as the charge increases. We use the logarithmic charge scale since it spans five orders of magnitude.}
\label{q14muMQerror}
\end{figure}

Inspired by the mass-charge relation for the extremal RN-AdS black hole, we obtain the following data-fitting functions
\be
M(Q) = \int_0^Q \mu(Q') dQ'\,,\qquad \mu(Q)=\sqrt{\sqrt{17.1525 Q^2+1.95582}+6.86161 Q+1.47745}\,.\label{fit1}
\ee
As we can see from Fig.~\ref{q14A1Mmu} that the above data-fitting functions match the data not only for the large $Q$ but also for the small $Q$.  In order to test the error of our data-fitting function for $\mu(Q)$ and $M(Q)$, we obtained the data for charges from 0.0023234 to $Q=370$ and we find that, throughout this charge region of five orders of magnitude, our data-fitting functions $M(Q)$ and $\mu(Q)$ produce less than $0.4\%$ of error, as can be seen by Fig.~\ref{q14muMQerror}.

It can be easily checked that $M(Q)$ for the $A_1$ solutions are less than the mass of the extremal RN-AdS black hole of the same charge, consistent with our earlier claim that the $A_1$ series are the ground states. For large $Q$, we have
\be
M\sim c_0+ 2.2114 Q^{\fft32}+0.445404 Q^{\fft12}-0.0562314 Q^{-\fft12}+{\cal O} \left(Q^{-\fft32}\right)
\ee
where $c_0=0.085282$. Changing to the convention of \cite{delaFuente:2020yua}, ($M=\fft12\tilde M, Q=\fft12 \tilde Q$,) we have
\be
\tilde M \sim 2c_0+ 1.56370 \tilde Q^{\fft32}+0.629896 \tilde Q^{\fft12}-0.159046 \tilde  Q^{-\fft12}+{\cal O} \left(\tilde Q^{-\fft32}\right)\,,
\ee
which agrees very well with \cite{delaFuente:2020yua} for the leading order coefficient.  We also introduce a constant $c_0$ coefficient that was neglected by \cite{delaFuente:2020yua}, although their Fig.~2b clearly indicates the presence of such a constant term. We also find that
\be
\mu = 3.31710 Q^{\fft12}+0.222702 Q^{-\fft12}+0.0281157 Q^{-\fft32}+{\cal O}\left(Q^{\fft52}\right)\,.
\ee
The orders of power expansion are the same as those of the extremal RN-AdS black hole, but with different coefficients.  On the other hand, the small $Q$ expansion of $\mu(Q)$ for the $A_1$ solutions involves a linear dependence of $Q$, which is absent in the extremal RN-AdS black hole.

\subsubsection{The $B_i$ series: solutions with mass-gap}
\label{sec:q14b}

For small $\phi_0$, the pattern of $\phi_1(a_0)$ was depicted in Fig.~\ref{q14phi0is1phi1}. As $\phi_0$ approaches zero, the pattern remains, indicating that the $A$ series of solutions are gapless with respect to the AdS vacuum.  As $\phi_0$ increases, the pattern changes and new type of roots of $\phi_1(a_0)$ emerge.  We present Fig.~\ref{q14BRNphi1(a0)a} in appendix \ref{app:graphs} to illustrate the emergence of the new pattern of the roots.

At $\phi_0=2$, we see a new bump emerges between the $A_1$ and $A_2$ roots, and this new local maximum $B$ rises as $\phi_0$ increases until it touches the $\phi_1=0$ axis, creating a double root $B_1$ of $\phi_1(a_0)$.
The tip of $B$ local maximum continue to rise with the increasing $\phi_0$, so that the double root $B_1$ splits into two single roots.  Then tip of $B$ starts to lower and become a local minimum and its position lowers and touches the $\phi_1(a_0)=0$ axis again, creating the $B_2$ series, {\it etc.}  The emergence of the $B$ series of solutions are very sensitive to the value of $\phi_0$.  If we further increase the $\phi_0$, the left and right groups of the $B$ series breaks up, and there is no soliton solutions at all in the certain middle region of $a_0$.  We illustrate this in Fig.~\ref{q14BRNphi1(a0)b} in appendix \ref{app:graphs}.

Unlike the $A$ roots, whose number remains unchanged, the $B$ roots are continually generated as $\phi_0$ increases.  Thus these solutions all have mass gap and are not smoothly connected to the AdS vacuum.
For $\phi_0=2.5$, we find at least three types of $B$ roots, which we label them $B_1,B_2$ and $B_3$.  For each type, there are $B_i^{-}$ in the left and $B_i^{+}$ in the right. We give the roots $a_0$ and the corresponding boson star charge $Q$ for the $B_i$ solutions at $\phi_0=2.5$:

\smallskip
\begin{center}
\begin{tabular}{|c|c|c|c|c|c|c|}
  \hline
   & $B_1^{-}$ & $B_2^{-}$ & $B_3^{-}$ & $B_3^{\rm +}$ & $B_2^{+}$ & $B_1^{+}$ \\ \hline
  $a_0$ & 0.844 & 0.88955 & 0.8902259 & 2.137304 & 2.1387 & 2.25 \\ \hline
  $Q$ & 3.86 & 3.5556 & 3.550867 & 1.316266 & 1.3159 & 1.27 \\
  \hline
\end{tabular}
\end{center}
\smallskip

We would like call these $B$ solutions as the RN-like boson stars in that their mass-charge relation are very close to the extremal RN-AdS black hole, namely the $M(Q)$ function given in (\ref{RNMQ}).  In the table above, the number of significant figure presented is to signify the number of the significant figures that the mass of the boson star matches that of the extremal RN-AdS black hole of the same mass.  We also find that the scalar hair parameter $\phi_2\rightarrow 0$ for $B_i$ with increasing $i$.  This implies that for these boson stars, the asymptotic behavior becomes increasingly indistinguishable from the extremal RN-AdS black holes.  It should be pointed out that for all the $B_i$ solutions that we examined, although the mass is close to the extremal RN-AdS black hole of the same charge, they are all slightly smaller, indicating that the extremal RN-black hole is the upper bound of all the type-$B$ solutions.  As $\phi_0$ increases, there are more and more type $B$ boson stars that become approximately degenerate state with the extremal RN-AdS black hole.  This may provide a new understanding of the extremal RN-AdS black hole entropy.

\begin{figure}[htp]
\begin{center}
\includegraphics[width=220pt]{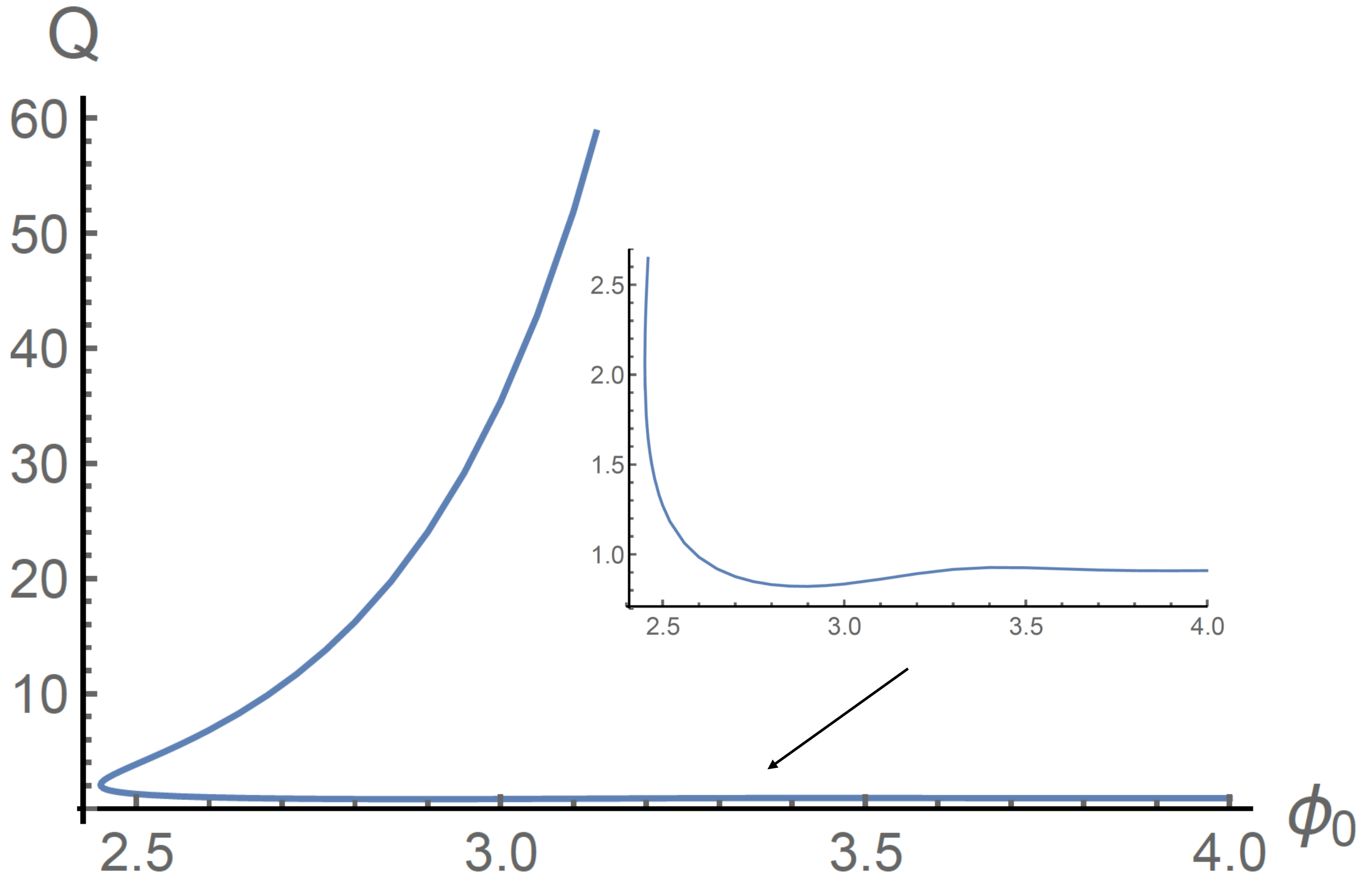}\
\includegraphics[width=220pt]{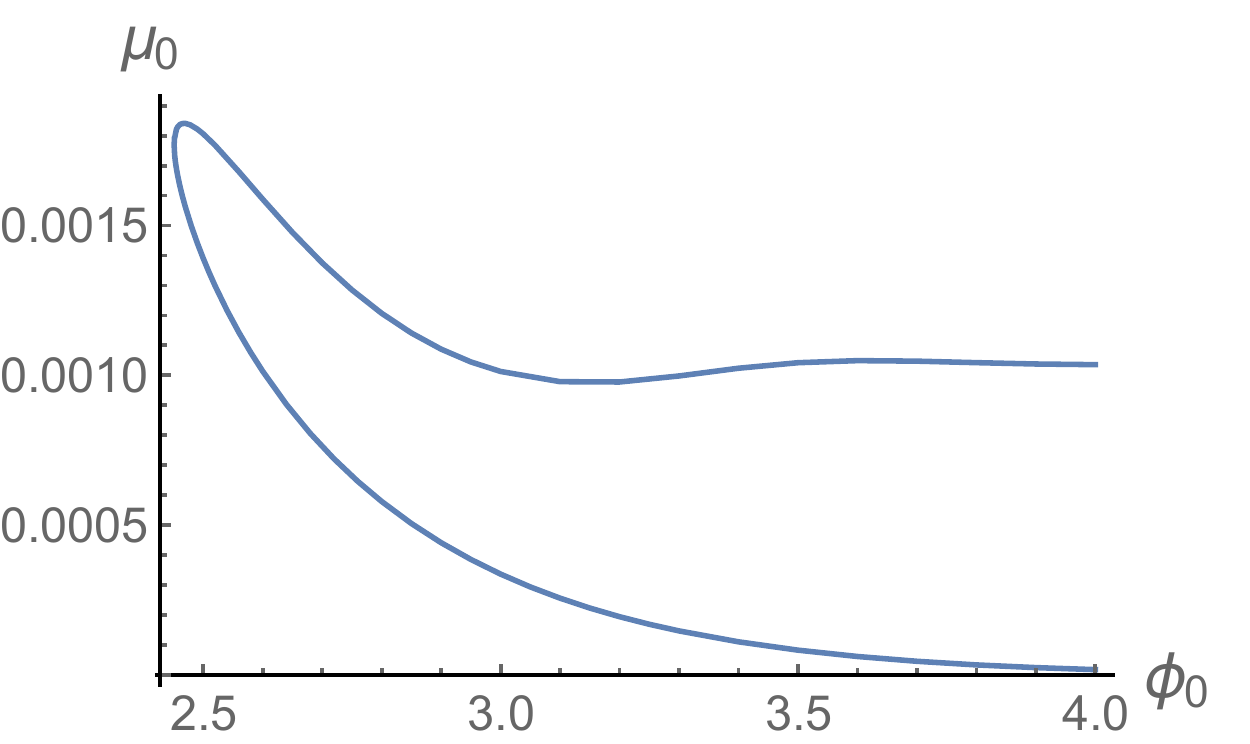}
\end{center}
\caption{\small\it The upper branch of $Q(\phi_0)$ corresponds to the smaller (left) roots of the $B_1$ series, and the lower branch corresponds to the larger (right) roots. The two branches join at where the two roots coalesces at some $\phi_0^{\rm min}$. The charge $Q$ is unbounded in the upper branch, but bounded below in the lower branch with $Q_{\rm min}\sim 0.8222$.  The upper and lower branches of $\mu_0(\phi_0)$ correspond to the larger and smaller roots of the $B_1$ series.}
\label{q14B1Qmu0}
\end{figure}

Since the mass-charge relation of the $B_1$ series of solutions differ the most from the extremal RN-AdS black hole among all the type-$B$ solutions, we shall study this series in greater detail. In general, The $B_1$ series has two roots, and they coalesces at some minimum $\phi^{\rm min}_0=2.451115$. This implies that the $Q(\phi_0)$  and $\mu_0(\phi_0)$ functions are two valued, as shown in Fig.~\ref{q14B1Qmu0}.

The upper branch of $Q(\phi_0)$ are associated with the left (or smaller) roots of the $B_1$ series, and the charge $Q$ is unbounded above and the smaller roots are pushed further to the left in the $\phi_1(a_0)$ graph.  The lower branch are generated by the right (or larger) roots with the charge $Q$ oscillating at a small amplitude.  The minimum value $Q_{\rm min}\sim 0.8222$ does not occur at the double root of $\phi_1(a_0)$ at $\phi_0^{\rm min}=2.451115$, which has $Q=2.06321$, but at the certain right $a_0=6.74301$ root of $\phi_1(a_0)$ with $\phi_0=2.9000$. This makes it rather tedious to locate the minimum charge $Q_{\rm min}$ for the boson stars with a mass gap.

\begin{figure}[htp]
\begin{center}
\includegraphics[width=220pt]{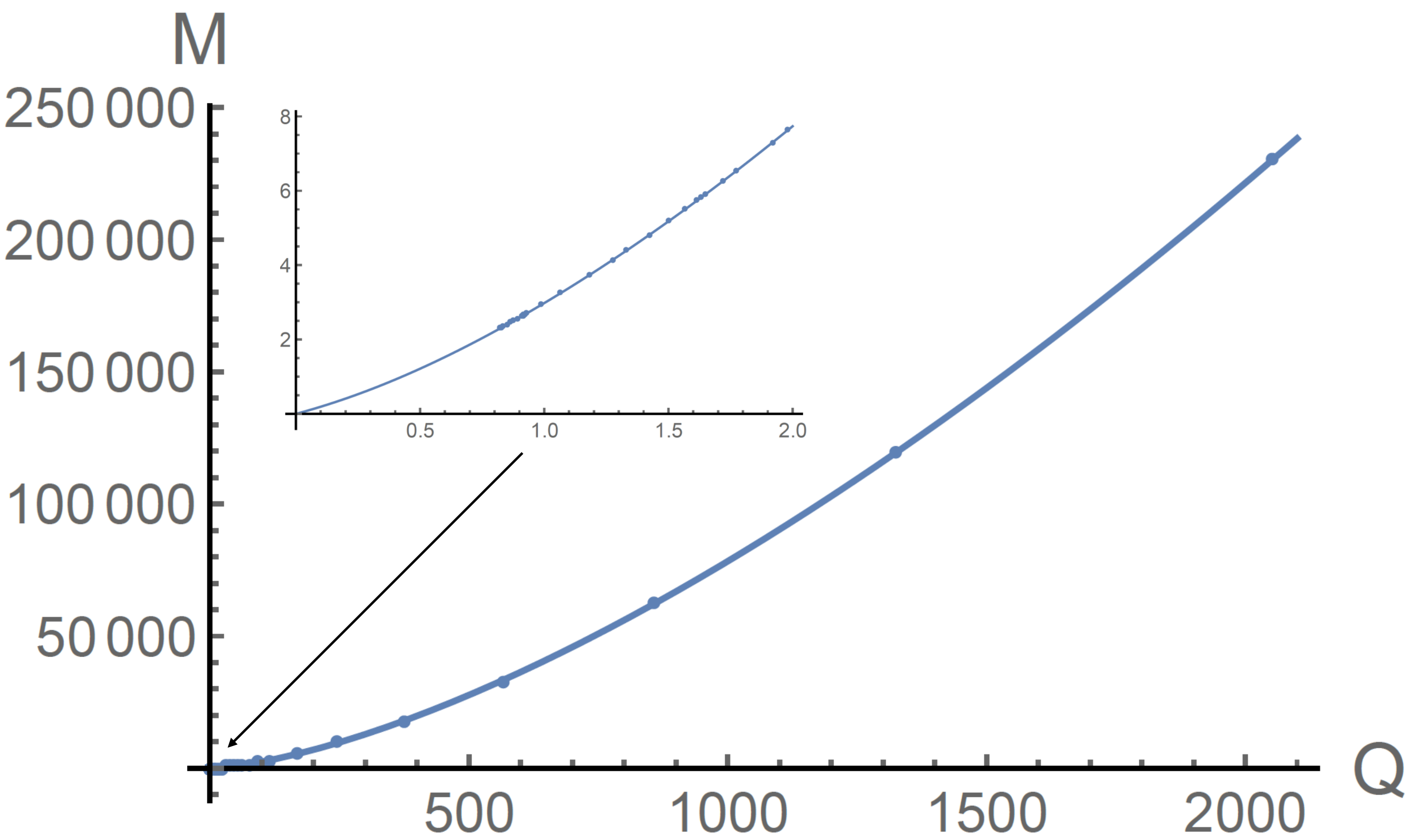}\
\includegraphics[width=220pt]{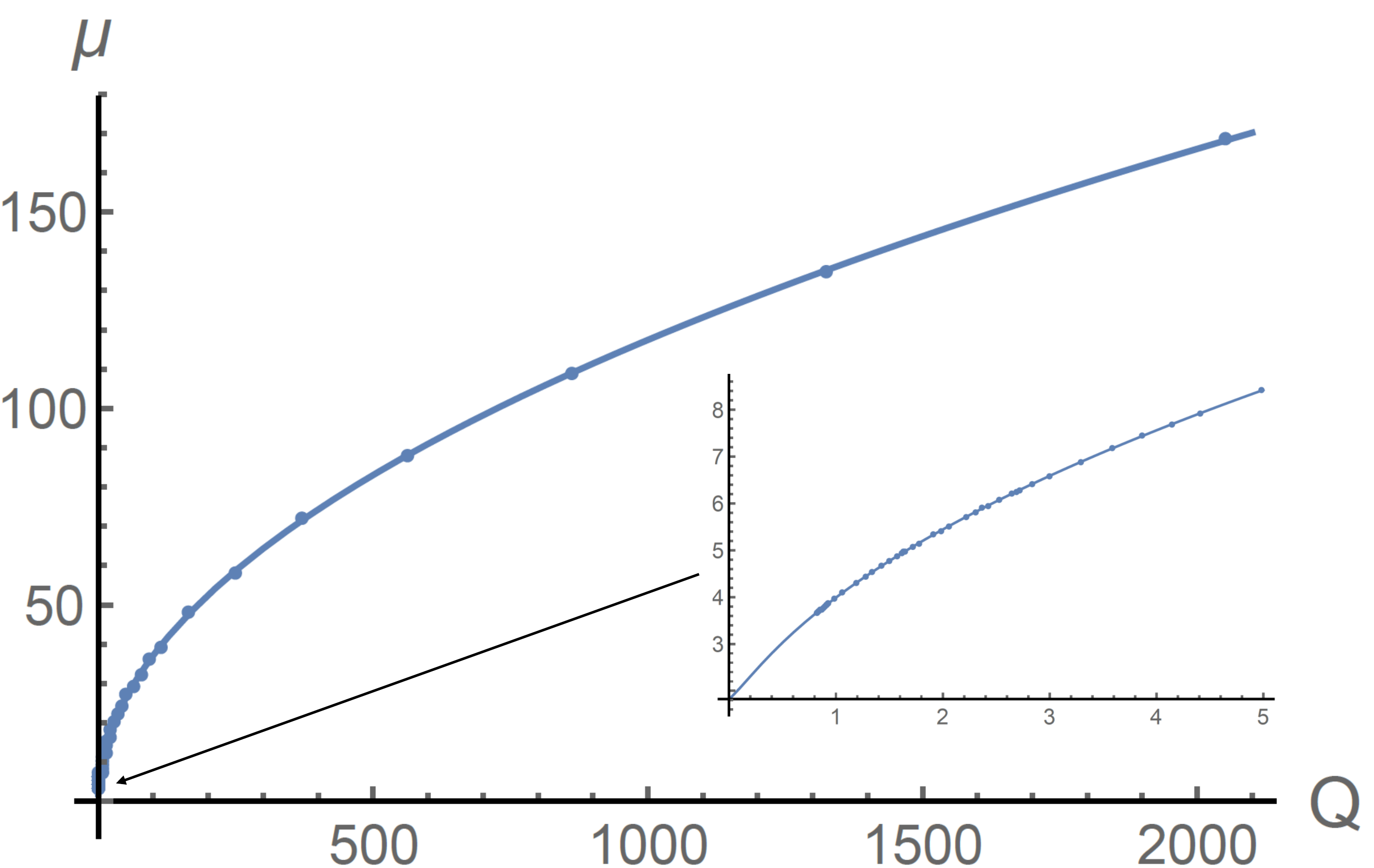}
\end{center}
\caption{\small\it The dots are actual numerical data and the solid lines are the derived data-fitting functions.  The dashed line in the $M(Q)$ graph is the mass-charge relation for the extremal RN-AdS black hole.}
\label{q14B1Mmu}
\end{figure}

Using the numerical method, we obtain a large amount data of the $B_1$ series, with the charge ranging from $Q_{\rm min}\sim 0.8222$ to a very large $Q\sim 2051$, corresponding to $\phi_0$, running from $2.45115$ to 4. We could not find a close-form data-fitting function $M(Q)$ that yields less than $1\%$ of error in this large range of the $Q$ parameter.  On the other hand, we find a close form function $\mu(Q)$, namely
\be
\mu = \sqrt{6.03677 \sqrt{Q^2+0.0481225}+7.75028 Q+1.99386}\,.\label{fit2}
\ee
We find that this function produces a stunning less than $10^{-3}\%$ of error comparing to the actual numerical data. We can then use the first law $dM=\mu dQ$ to define the mass as the quadrature (\ref{massQquadrature}).  The $M(Q)$ and $\mu(Q)$ relations are depicted in Fig.~\ref{q14B1Mmu}.  The solid line are the data-fitting functions and dots are the actual numerical data.  They match perfectly from small to large $Q$. Note that $\phi_2(Q)$ relation can also be approximated by the linear relation $\phi_2=-0.0110485-0.390663 Q$ and we present the matching of the numerical data with this linear relation in Fig.~\ref{q14BRN-phi2-Q} in appendix \ref{app:graphs}.

The large $Q$ behavior can be easily determined, given by
\bea
M &\sim& c_0+ 2.47539 Q^{\fft32}+0.53698 Q^{\fft12}-0.0197047 Q^{-\fft12}+{\cal O}\left(Q^{-\fft32}\right)\,,\nn\\
\mu &\sim & 3.71309 Q^{\fft12}+0.26849 Q^{-\fft12}+0.00985236 Q^{-\fft32}+{\cal O}\left(Q^{-\fft52}\right)\,.
\eea
with $c_0=-0.0135596$. Since the $B_1$ series has minimum $Q_{\rm min}\sim 0.8222$, the small $Q\rightarrow 0$ behavior of the data-fitting function is pointless to analyse.  It should emphasized that we have
\be
M(Q)^{A_1} <M(Q)^{B_1} < M(Q)^{\rm RN}\,,
\ee
for all the charge $Q\ge Q_{\rm min}$, indicating the $B_1$ series is the (slightly) lower energy state than the extremal RN-AdS black hole, but (much) higher than the $A_1$ series.

To conclude, for $q^2=1.4$, there are two types boson stars: type A are gapless and can smoothly connect to the AdS vacuum. The charge and hence the mass of the general $A_i$ series are bounded above, with an exception the $A_1$ series. The $B_i$ series of solutions all have a mass gap with some minimum $Q_{\rm min}$. For given charge, the mass of the type-$B$ solutions are smaller than that of the extremal RN-AdS black hole, but becomes increasingly indistinguishable.  The ground state can be identified as the $A_1$ series of the boson stars.

\subsection{$q^2=1$}
\label{sec:q=1}

We now construct the boson stars with $q=1$.  This case arises in supergravity models when the scalar is linearized.  While some boson stars share the same characteristics as those in $q^2=1.4$, many do not. We shall briefly present the results that are analogous to those of $q^2=1.4$ and analyse the new solutions in greater detail.

\subsubsection{The $A_i$ series: gapless solutions}

For small $\phi_0$, the $\phi_1(a_0)$ function shares the same pattern as in Fig.~\ref{q14phi0is1phi1}, and we can label the roots of $\phi_1(a_0)$ accordingly.  We obtain numerical data for the $A_1$, $A_2$ and $A_3$ series of solutions, and we present the various quantities in Fig.~\ref{q1A1A2A34plots} in appendix \ref{app:graphs}.

An important difference for $q^2=1$ compared to the earlier $q^2=1.4$ case is that the $A_1$ series of solutions are also bounded above, just like the rest $A$ series, as can be seen in Fig.~\ref{q1-A1A2A3RN-M(Q)} in appendix \ref{app:graphs}. Although the $A_1$ series remains the lowest energy state amount the $A_i$ series, it is more excited than the extremal RN-AdS black hole.  Thus for small charges, the extremal RN-AdS black hole is the ground state.
(Recall the $A_1$ series has lower energy than the extremal RN-AdS when $q^2=1.4$.)  In appendix \ref{app:cusp}, we give a detailed demonstration of how the first law $dM=\mu dQ$ is satisfied despite of the existence of a cusp in the mass-charge relation.

\subsubsection{The $B_i$ series: Solutions with mass gap}

The analogous pattern of Fig.~\ref{q14phi0is1phi1} will change when $\phi_0$ increases, but the style of the change is different from the $q^2=1.4$ case.  A new bump emerges between $a_0=0$ and the root of $A_1$, as can be seen in Fig.~\ref{q1-B-phia0} in appendix \ref{app:graphs}. As $\phi_0$ increases, the bump lowers down until its tip touches the $a_0$ axis at $\phi_0=2.621995$, creating the first double root, which we label as $B_1$. As $\phi_0$ continue to increase, the double root split into one smaller and one bigger roots and they widen with the smaller root becoming even smaller and the bigger root bigger.  A new double root $B_2$ is then created at $\phi_0=2.87416052$.  Eventually, the barren land in some middle region of $a_0$ is created where there is no AdS soliton solutions at all, for sufficiently large $\phi_0$ and the smaller roots of the $B_i$ series are pushed further to the left of the barren land while the larger roots are pushed to the right.  As in the $q^2=1.4$ case, the $B$ series are RN-like in that their mass-charge relations are very close to that of the extremal RN-AdS black hole. As concrete examples, the mass and charges for the boson stars associated with the $B_1$ and $B_2$ double roots are
\bea
\hbox{$B_1$ double root:}&& Q=2.97104\,,\qquad M=12.8523\,,\qquad (M^{\rm RN}=13.6235)\,,\nn\\
\hbox{$B_2$ double root:}&& Q=2.88051\,,\qquad M=13.0259\,,\qquad (M^{\rm RN}=13.0327)\,.
\eea
In the above we also list the mass of the extremal RN-AdS black hole of the same charge. We see for the starting of the $B_2$ series, its mass matches that of the corresponding extremal RN-AdS with four significant figures.  The level of degeneracy increases further for the $B_i$ series with higher $i$.  It should be pointed out however, that we find that boson stars of the $B$ type are all less than that of the corresponding extremal RN-AdS black holes and hence correspond to the lower energy states.

Since the mass-charge relation of the $B_1$ series are recognisably different from the extremal RN-AdS black holes, we present their properties in some detail. The $\phi_2(Q)$ function is effective linear and our data fitting indicates that it can be expressed as
\bea
\phi_2=0.170463+1.64755 Q\,.
\eea
The $\mu(Q)$ is more subtle and we find within $10^{-3}\%$ error that
\be
\mu=\sqrt{4.39505 \sqrt{Q^2+0.0560187}+7.83442 Q+1.86301}\,,\label{q1B1muq}
\ee
from which, we can derive the mass-charge function $M(Q)$ using (\ref{massQquadrature}). As we see from Fig.~\ref{q1-B-all} in appendix \ref{app:graphs} that the above functions match the numerical date perfectly.
For large $Q$, we have
\be
\mu=3.49707 \sqrt{Q}+0.266368 Q^{-\fft12Q}+0.00745635 Q^{-\fft32}+{\cal O}\left(Q^{-\fft52}\right)\,,
\ee
and hence we have
\be
M=-0.0314351+ 2.33138 Q^{\fft32}+0.532735 Q^{\fft12}-0.0149127 Q^{-\fft12}+{\cal O}\left(Q^{-\fft32}\right)\,,
\ee
Thus we see that for sufficiently small $Q$, the extremal RN-AdS black hole has the lowest energy. For large $Q$, the $B$ series of boson stars all have smaller mass than the corresponding extremal RN-AdS black hole, with the $B_1$ series being the ground state.

\subsection{General $q^2$}
\label{sec:genq}

\begin{figure}[htp]
\begin{center}
\includegraphics[width=260pt]{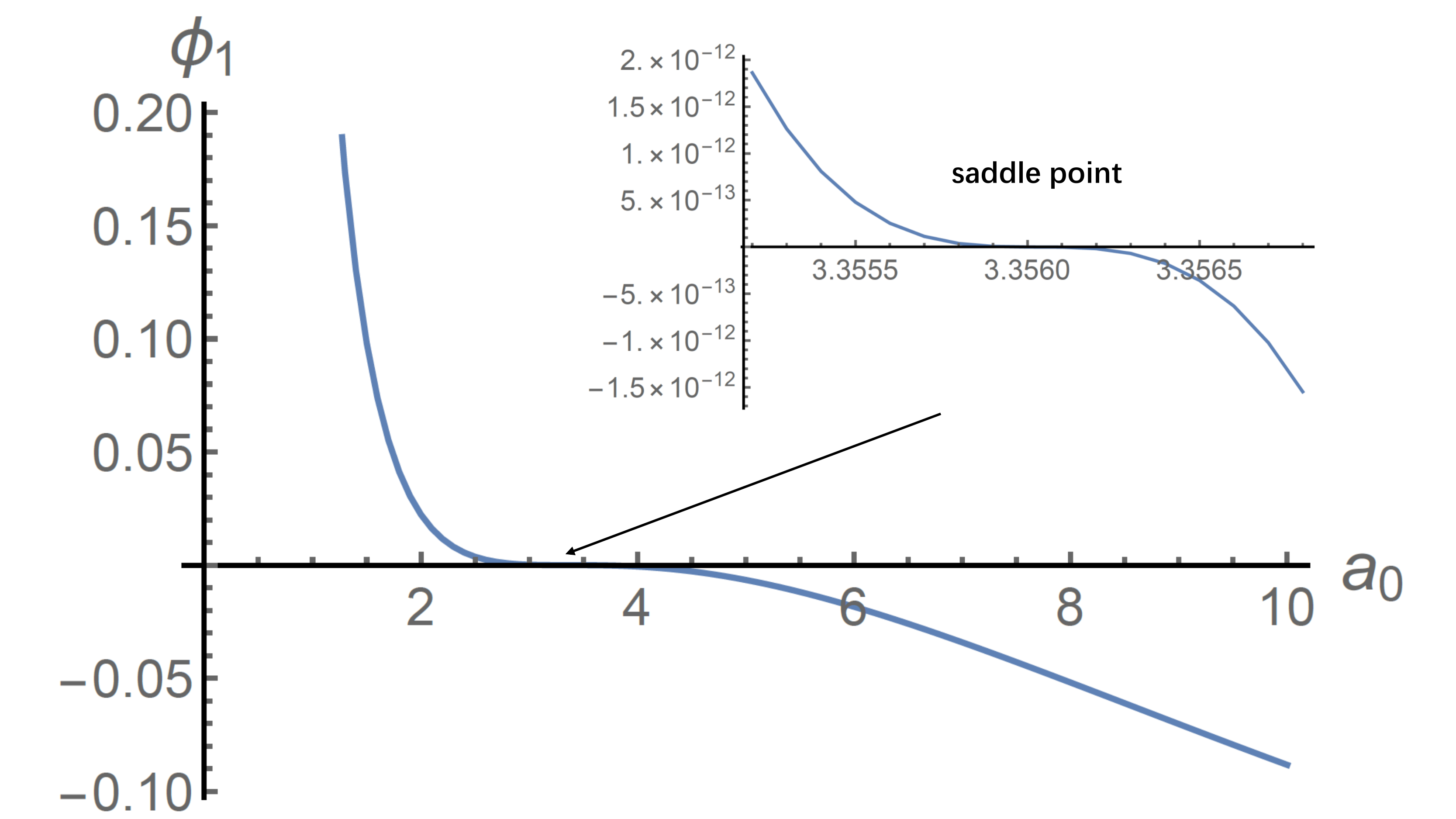}
\end{center}
\caption{\small\it When $q_c^2=1.2609$, vanishing saddle point arises where the $B_1$ double root and $A_1$ root coalesces at $a_0=3.3560$. For $q^2< q_c^2$, the gapless $A_1$ solution is bounded above whilst when $q^2\ge q_c$, its charge is unbounded.}
\label{qc-saddle}
\end{figure}

We have examined the properties of boson stars with $q^2=1.4$ and $q^2=1$ in the previous two subsections, and in both cases, there are two types of solutions.  One is the gapless type $A$ solutions that can connect to the AdS vacuum smoothly and the other is the type $B$ solutions that have mass gap.  The electric charge and hence the mass of gapless solutions are in general bounded above, whilst solutions with mass gap are typically unbounded above.  An important difference is that in the $q^2=1.4$ case, the $A_1$ series of solutions have unbounded charge as well. The cause of the difference can be traced back to how the $B_1$ double root arises.  This was illustrated in Fig.~\ref{q14BRNphi1(a0)a} and Fig.~\ref{q1-B-phia0} for $q^2=1.4$ and $q^2=1.0$ respectively.  Thus it is clear that the critical $q_c$ to separate these two behavior is when the $\phi_1(a_0)$ function has a vanishing saddle point, as illustrated in Fig.~\ref{qc-saddle}.

For $q^2\ge q_c^2$, the $A_1$ series of boson stars, rather than the extremal RN-AdS black hole, are associated with the ground state. Furthermore, the $A_1$ series is the ground state for all the charges, from zero to infinity. The situation is quite different when $q^2< q_c^2$, in which case, the $A_1$ series is bounded above. Furthermore, the extremal RN-AdS black hole becomes the ground state for small charges. These results are consistent with the fact that the superradiant instability occurs for larger values of $q^2$ for the extremal RN-AdS black holes with small charges \cite{Dias:2016pma}.

\begin{figure}[htp]
\begin{center}
\includegraphics[width=250pt]{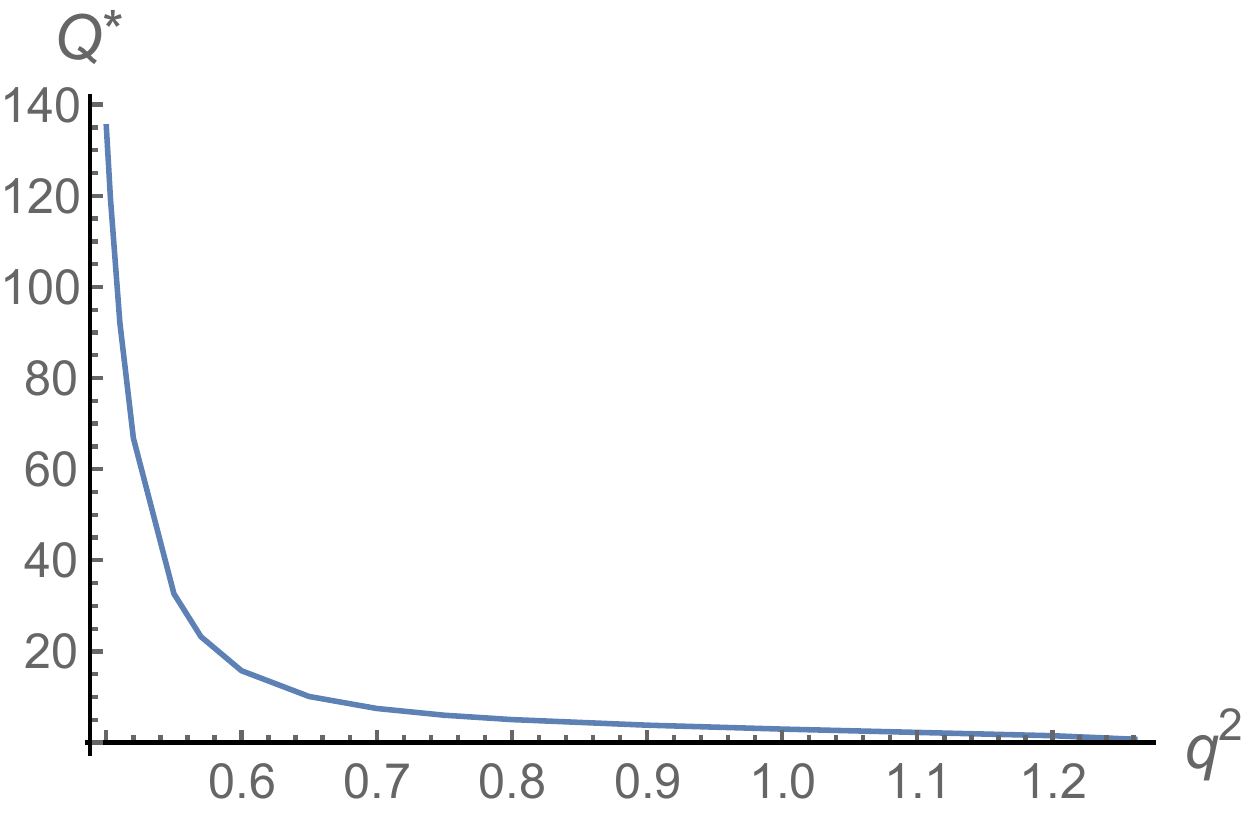}
\end{center}
\caption{\small\it The charge $Q^*$ of the double root boson star $B_1^*$ increases as $q$ decreases from $q_c\sim 1.261$. This shows that the mass gap increases for the $B_1$ series as $q$ decreases.}
\label{gqQq}
\end{figure}

When the charge $Q$ increases and boson stars with mass gap, namely the $B_i$ series of the solutions start to emerge, even for $q^2<q_c^2$.  The $B_1$ series has the lowest energy among all the $B_i$ series of boson stars and it emerges as a double root of $\phi_1(a_0)$ at certain special $\phi_0^*$.  As $\phi_0$ increases further, the double root splits into $B_1^-$ and $B_1^+$ and the smaller root $B_1^-$ moves further left, giving rise to a boson star with ever increasing and unbounded mass and charge.  The larger root $B_1^+$ moves further right and the mass and charges oscillate to some fixed point.  While strictly not quite true, we shall use the charge $Q^*$ of the boson star at the double root $B_1^*$ as the representation of the mass gap of the $B_1$ solutions for given $q$.  We find that $Q^*$ increases monotonically as $q$ decreases from $q_c$, as shown in Fig.~\ref{gqQq}.

\begin{figure}[htp]
\begin{center}
\includegraphics[width=220pt]{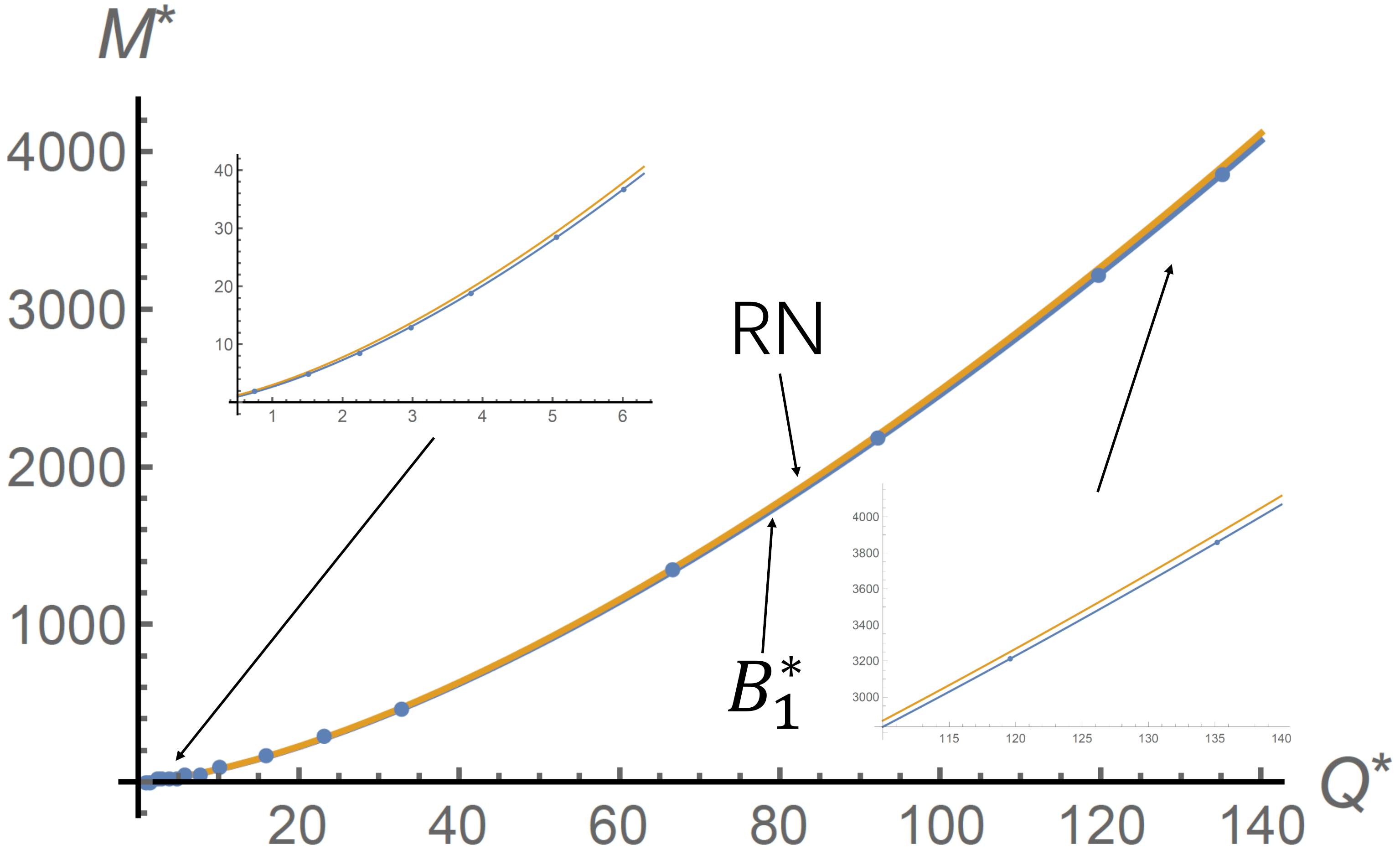}
\includegraphics[width=220pt]{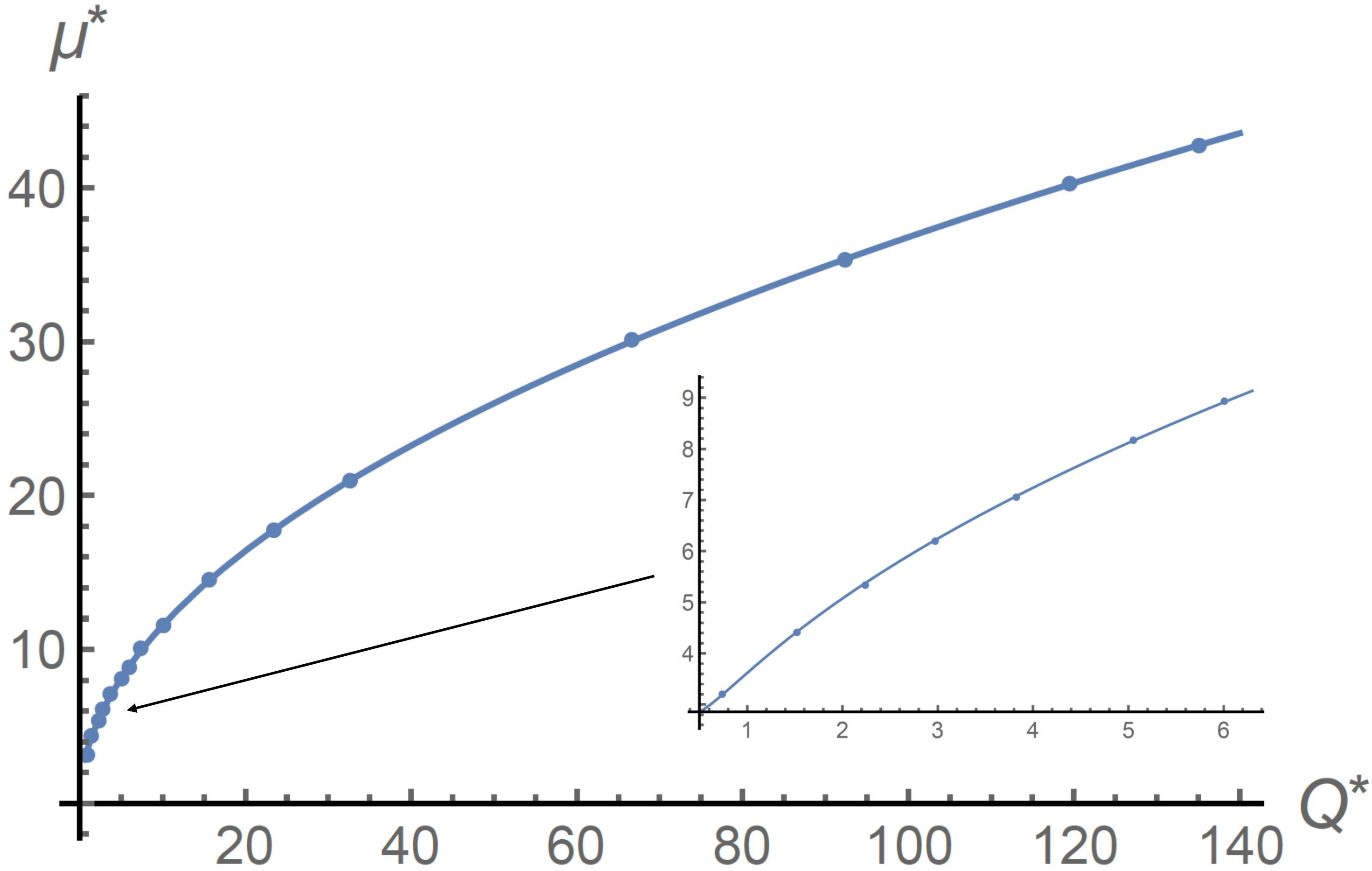}
\end{center}
\caption{\small\it The data-fitting functions (\ref{starrelation}) matches the numerical data within $1\%$ of error. It is important to note that $M^*< M^{\rm RN}$ for all $Q^*$, indicating that the extremal RN-AdS black hole are not the ground state for $Q\ge Q^*$.}
\label{gqMmuQ}
\end{figure}

The increasing of the charge gap becomes exponential as $q$ becomes even smaller.  For example, at $q^2=0.45$, we have $Q^*\sim 10^6$, whilst $Q^*\sim 135.12$ when $q^2=0.5$.  We find that, as shown in Fig.~\ref{gqMmuQ}, there exist very good data-fitting functions within $1\%$ accuracy for the $M^*(Q^*)$ and $\mu^*(Q^*)$ relations:
\bea
M^* &=& +0.275673+2.45511 (Q^*)^{\fft32}+0.144343 (Q^*)^{\fft12}-0.195568 (Q^*)^{-\fft12}\,,\nn\\
\mu^* &=& 3.68348 (Q^*)^{\fft12}- 0.313391 (Q^*)^{-\fft12} + 0.246595 (Q^*)^{-\fft32}\,.\label{starrelation}
\eea
Note that since $Q^*$ has a minimum, so the above is an exact, applicable for all $Q\ge Q^*$.

As we have noticed before, the $B_i$ series of solutions all have less energy than the extremal RN-AdS black hole, and the mass-charge relation becomes indistinguishable from that of the RN solution as the label $i$ increases and the $B_1$ solution has the lowest energy among them.  For $q^2<q_c^2$, the extremal RN-AdS black hole is the ground state for $Q<Q^*$, but concedes this role to the $B_1$ boson star for $Q\ge Q^*$. The exponential increasing of the mass gap for the $B_1$ for smaller $q$ implies that the RN-AdS black hole plays a larger role of the ground state for smaller $q$.

Before ending this section, we would like to observe an intriguing phenomenon.  We do not expect that $dM^*=\mu^* dQ^*$ would hold since different starred values correspond to different $q$, the parameter of the theory.  Nevertheless, we see that $dM^*=\mu^* dQ^*$ holds for the leading order.  This suggests that we may introduce $\tilde \mu$ to extend the first law $dM=\mu dQ + \tilde \mu dq$ such that the $\tilde \mu dq$ term does not contribute to the leading $dM$ term of large $Q$.

\section{The supergravity models}

In this section, we study the AdS boson stars in the two gauged supergravity models, corresponding to the $SU(3)$ and $U(1)^4$ truncations of $D=4$ $SO(8)$ gauged maximal supergravity. (See appendix \ref{app:su(3)} for more detail.)

\subsection{The $U(1)^4$ model}

The relevant Lagrangian is (\ref{genlag}) with $U$ and $V$ functions given by (\ref{threetheories}). At the linear scalar level, it corresponds to the toy model of $q^2=1/4$.  We find that the theory admits only the type-$A$ boson stars whose mass and charge are bounded above.  In fact they are much less than the AdS radius.  For these small boson stars, the nonlinear effect associated with the supergravity scalar potential is negligible.  Indeed, as we can see from the Fig.~\ref{sgrq12A-all} in appendix \ref{app:graphs}, the solutions resemble those of the $A$ series of boson stars in the toy model with $q^2< q_c^2$. It is also important to note that in this case the extremal RN-AdS has the lowest energy for given charge $Q$ compared to the boson star solutions with the designed boundary conditions. Therefore in this model, it should be interesting to consider boson stars with alternative boundary conditions which appears to have lower energy than the extremal RN-AdS \cite{Gentle:2011kv}. We will leave this for future study.

\subsection{The $SU(3)$ model}

The $SU(3)$ model has a rich variety of AdS boson star solutions.  Not only does it have the type $A$ and type $B$ series of solutions, but the type C, a new type with the mass and charge bounded both below and above.

\subsubsection{The $A$ series:  gapless solutions}

We first present the $A$ series of solutions.  For small $\phi_0$, the $\phi_1(a_0)$ function appears analogous to
(\ref{q14phi0is1phi1}) with an exception that the $A_1$ root is always $a_0=2$.  As $\phi_0$ increases and the shape of $\phi_1(a_0)$ alters significantly and most of the $A$ roots increase and move to the right. However, the $A_1$ root remains fixed to be $a_0=2$.  (It should be mentioned here that the chemical potential at the origin is not $a_0$, but $a_0/\sqrt{\lambda}$, as explained in section \ref{sec:num}.)  We obtain numerical data for the $A_1$, $A_2$ and $A_3$ series of solutions and the results are presented in Fig.~\ref{sgrq1A-all} in appendix \ref{app:graphs}.

The $A_1$ series of boson stars deserve further comments.  The $\mu_0(\phi_0)$ relation resembles that of the $q^2=1.4$ case in that $\mu_0$ approaches zero as $\phi_0$ increases.  However, unlike the $A_1$ series of the $q^2=1.4$, here the solutions have bounded mass and charge, with $Q_{\rm max}\sim 0.1702$.  Another intriguing phenomenon is that the chemical potential $\mu$ is fixed to be literally 2, which implies that
\be
M=2Q\,.
\ee
(Our numerical data shows that $\mu\sim 2\pm 10^{-9}$.) This is precisely the mass-charge relation of the extremal RN black hole that is asymptotic to the Minkowski spacetime.  Consequently, the energy for the $A_1$ series of boson stars have less energy than the extremal RN-AdS black hole.  In contrast, as we discussed in section \ref{sec:q=1}, the $q^2=1$ theory, the $A_1$ boson stars have larger energy.  It should be also pointed out, like the rest $A$ series of boson stars,  in the $A_1$ solutions, the charge $Q$ also oscillates around some fixed value as $\phi_0$ increases, but $M=2Q$ relation remains fixed, with no sign of cusps.

\subsubsection{The $B$ series: solutions with mass gap}

As mentioned earlier, the shape of the $\phi_1(a_0)$ graph changes significantly as $\phi_0$ increases to some critical value, $\phi_0^c\sim 1.2465$, for which $a_0=0$ starts to fail to produce a soliton solution.  New roots, which we call the $B$ series, emerge at $0<a_0 \ll 1$.  These roots migrate to the right as $\phi_0$ increases.  The Fig.~\ref{sgrBphi1(a0)} in appendix \ref{app:graphs} shows this migration for the $B_1$ series.

The $B$ series of solutions have mass gap, but unbound above.  However, there is a significant difference compared to the $B$ series of boson stars in the previous toy model.  Here, large $Q$ boson stars occur at small $\phi_0$ with $Q=\infty$ emerging at $\phi=\phi_0^c$.  As $\phi$ increases, $Q$ decreases and then oscillates around some fixed value. Various properties of the solution are plotted in Fig.~\ref{sgrq1B-all} in appendix \ref{app:graphs}. Compared to the $B_1$ series of the toy model, the solutions in the supergravity model have more interesting structures for small $Q$, with $\mu(Q)$ visibly spiraling into a fixed point. This multi-valued function makes it unlikely to find a data-fitting function that can fit all the data within a reasonable error margin. However, for the data before the spiral occurs, we find a good approximate analytical expression, given by
\bea
\mu&=&3.43392 Q^{\fft12} +0.236493Q^{-\fft12} +  0.000744661 Q^{-\fft32} + {\cal O}(Q^{-\fft52})\,,\nn\\
M&=&c_0  +2.28928 Q^\fft32+ 0.472987 Q^{\fft12}  -0.00148932 Q^{-\fft12} + {\cal O}(Q^{-\fft32})\,,
\eea
where $c_0=-0.0133957$ and $dM=\mu dQ$. As we can see from the Fig.~\ref{sgrq1B-all} that $M(Q)$ function (solid line) fits the mass data (the dots) within good error margin even for small $Q$, the $\mu(Q)$ function fits the data very well only before the spiral occurs.

     It should be pointed out that the $B_1$ series of solution has less energy than the extremal RN-black hole with
the same charge.  In fact the general $B_i$ series can be called the RN-like, in that, the mass-charge relation becomes increasingly the same as that of the extremal RN-AdS black hole, as $i$ increases.  For example, we consider $\phi_0=1.5$ and the mass and charge for the $B_1$ and $B_2$ boson stars are
\bea
B_1:&& Q=0.863501\,,\qquad M=2.26344\,,\qquad M^{\rm RN}=2.47028\,,\nn\\
B_2:&& Q=1.44098\,,\qquad M=4.91667\,,\qquad M^{\rm RN}=4.92168\,,
\eea
We see that for the $B_2$ solution, the mass of the boson star matches the corresponding extremal RN-AdS black hole with three significant figures, albeit slightly smaller.  The difference becomes increasing negligible for the higher labelled $B$ series.  This is analogous to the $B$ series of boson stars in the toy models.

\subsubsection{The $C$ series: mass-gapped solutions with an upper bound}

\begin{figure}
\begin{center}
\includegraphics[width=250pt]{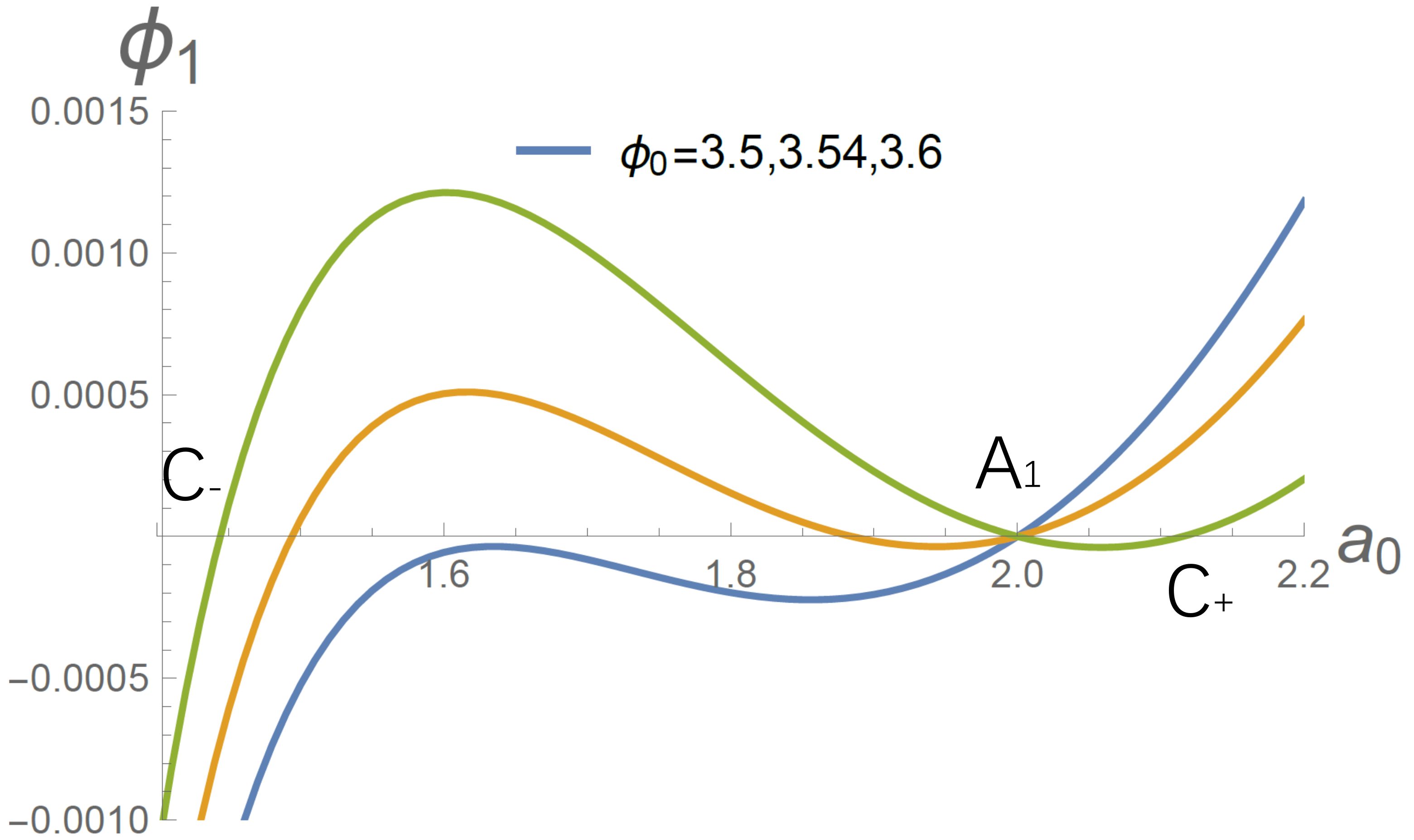}
\end{center}
\caption{\small\it New roots, which we label as $C$, emerge as $\phi_0$ increases further.  They come in pairs $C_\pm$, and coalesces at the staring point $\phi_0=3.5025$.  The $C_-$ roots move to the left whilst the $C_+$ roots move to the right as $\phi_0$ increases.}
\label{sgrCphi1(a0)}
\end{figure}

As we can see from Fig.~\ref{sgrCphi1(a0)}, a new double root start to emerge as $\phi_0$ becomes slightly bigger than 3.5, 3.5025 to be more precise.  As $\phi_0$ increases, the double root splits into two roots, with the smaller $C_-$ root moves to the left, whilst the bigger $C_+$ root moves to the right.  After joining with the $A_1$ roots, the $C_+$ roots moves further to the right.

The properties of the $C$ series of the boson stars are presented in Fig.~\ref{sgrq1C-all} in appendix \ref{app:graphs}. The curves are intriguing and complicated and hard to find an analytical function to fit the data.  The solutions are characterized by that the mass and charges are bounded both above and below.

Compare the energy of the type $A_1$, $B_1$ and $C$ boson stars to the extremal RN-AdS black hole in the commonly allowed charge region, we find
\be
M^{A_1} < M^{C} < M^{B_1} < M^{RN}\,.
\ee
In other words, the $A_1$ series is the ground state for small $Q$, which is very different from the $q^2=1$ toy model.

\section{Conclusion}

In this paper, we revisit the construction of the four-dimensional AdS boson stars in three theories, the phenomenological toy model consisting of a conformally coupled scalar with electric charge $q$ and two supergravity models descending from M-theory. The AdS boson stars we focus on are specified by four asymptotic parameters, namely the mass $M$, charge $Q$, chemical potential $\mu$ and scalar hair $\phi_2$ that is dual to the VEV of a certain scalar operator in the boundary CFT. These boson star solutions are of one-parameter family and we obtain the relations $M(Q)$, $\mu(Q)$ and $\phi_2(Q)$.  We adopted the Euclidean action technique and Iyer-Wald formalism to derive the Helmholtz free energy and the first law of the boson star dynamics.  We show that in the fixed charge ensemble, the free energy is simply the mass of the boson star, the same as the extremal RN-AdS black hole.  This allows us compare the mass-charge relations of the boson stars and also the extremal RN-AdS black hole and identify the ground state for given charge $Q$.

Our analysis shows that the spectrum of boson stars varies significantly in different theories. Nevertheless, we can categorise them in three classes.  The type $A$ boson stars are gapless and can smoothly connect to the vacuum AdS;
the type $B$ boson stars have a mass gap, but are unbounded above; the type $C$ solutions are bounded both above and below.

For the toy model, there appear to exist an infinite $A_i$-series of boson stars (excited boson stars as can be seen from a linear analysis around global AdS vacuum), roughly labelled by $i=1,2,\cdots$. These gapless solutions are typically bounded from above with charge much less than the cosmological constant $\ell$.  The upper bound for each $A_i$ series lowers as $i$ increases, but the mass increases for the same applicable charge, indicating $A_i$ is a higher excited state than $A_j$ if $i>j$. An exception is the $A_1$ series, whose mass and charge are unbounded above when $q\ge q_c$.  Furthermore, for $q\ge q_c$, the $A_1$ series is the ground state for all charge values; its mass is the lowest, smaller than the extremal RN-AdS black hole.

On the other hand, when $q<q_c$, the $A_1$ series is also bounded above, as in the case of the rest $A_i$ series. In these small charge region, the extremal RN-AdS black hole is therefore the ground state, with no boson stars having less energy. This seems to be consistent with the essence of Weak Gravity Conjecture involving a sufficiently charged particle which renders the extremal RN black hole unstable.

The toy model also appears to admit an infinite number of $B_i$-series of boson stars. The origin of the type $B$ boson stars are very different depending on whether $q\ge q_c$ or $q<q_c$, but they share the same characteristics.  They are all bounded below (with mass gap), but unbounded from above. Meanwhile they all have less energy than the corresponding extremal RN-AdS black hole. As the label $i$ increases, the mass-charge relation becomes increasingly indistinguishable from that of the extremal RN-AdS black hole. In fact viewed from the mass-charge relation, the extremal RN-AdS black hole appears to be the upper bound of the $B_i$ series of boson stars.

We find no evidence of the type $C$ boson stars in the toy model. Thus for $q\ge q_c$, the $A_1$ series is ground state. For $q<q_c$, the extremal RN-AdS black hole is the ground state for sufficiently small charge, while for larger $Q$, the ground state switches to the $B_1$ series.  Such switch indicates phase transition occurs at the ground state as $Q$ increases when $q< q_c$; such phase transition is absent when $q\ge q_c$. The mass gap, or the minimum charge of the $B$ series increase (exponentially) as the parameter $q$ decreases. This picture is consistent with the linear instability analysis of \cite{Dias:2011tj}.

The boson star spectrum is quite different in supergravity models due to the scalar self-interactions.  We find that the $U(1)^4$ model, corresponding to $q^2=1/4$, admits only the $A_i$ series of solutions, whose mass and charge are all bounded above.  Consequently, in this theory, the extremal RN-AdS black hole is the ground state.  The spectrum of the $SU(3)$ model, on the other hand, is much richer, containing not only the $A_i$, $B_i$ but also the $C$ series of boson stars.  For small values of $Q$, the $A_1$ series is the ground state, whilst it becomes the $B_1$ series for large charge, analogous to the toy model of $q^2=1$.  In other words, the extremal RN-AdS black hole is never the ground state for any given value of $Q$. (In contrast, in the toy model with $q^2=1$, the extremal RN-AdS black hole is the ground state for sufficiently small charge.)  An intriguing new property in the $SU(3)$ model is that the $A_1$ series has strict $M=2Q$, the same mass-charge relation of the extremal RN black hole in Minkowski spacetime. This suggests a possibility of an exact solution of the $A_1$ series in the $SU(3)$ model.

Since the boson star solutions unbounded from above contain the ground state, namely either the $A_1$ or the $B_1$ series of solutions, we paid particular attention in deriving the mass-charge relation. We found that simple data-fitting ansatz for $\mu(Q)$ (\ref{genmuQ}) can fit the numerical data in striking accuracy for the full range of charge $Q$ and we can then obtain the mass-charge relation by the quadrature (\ref{massQquadrature}).  The large charge expansion leads to the structure of (\ref{mq1}) with some specific leading coefficients and non-vanishing $c_0$.  These provide concrete examples for testing the AdS/CFT correspondence for the $SU(3)$ supergravity model which is believed to have a well-defined CFT dual.

To end this paper, we would like to outline some future directions worth exploring. First of all, in the CFT side, it has been shown that the spontaneously breaking of the global $U(1)$ symmetry yields a relativistic goldstone mode with speed of sound $c_s^2=\ft12$ and multiple gapless non-relativistic goldstone modes. One should be able to see these modes from the gravity side by studying the spectrum of fluctuations around the boson star solutions. Although the presence of the relativistic sound mode in the gravity side is somewhat expected, it is highly non-triival to show the appearance of the gapless non-relativistic modes. Existence of these modes would provide strong evidence for the duality between the AdS boson star solution and the large charge sector of the strongly coupled CFT. Gapless non-relativistic goldstone modes are known to be absent from the spectrum around AdS RN black hole \cite{Edalati:2010hk}. Secondly, the CFT computation \cite{Orlando1} also reveals that the lowest primary state above the large charge vacuum carries spin 2 and scaling dimension $\Delta(Q)+\sqrt{3}$ with $\Delta(Q)$ being the scaling dimension of the vacuum state. This precise relation is very intriguing. It should be interesting to construct its gravity dual. Finally, it is also important to understand the stability of the solutions we constructed. Technically, tackling this problem is more difficult which requires time-evolution of the system for generic choice of the initial data. However, some encouraging results had already appeared in the literature \cite{Buchel:2013uba,Dias:2012tq} suggesting that for the toy model, the global AdS boson star solutions are stable even at fully nonlinear level. We would like to extend their analysis to the supergravity models involving
scalar self-interactions in the future.

\section*{Acknowledgement}

We are grateful to Yue-Zhou Li, Zhan-feng Mai and Yu Nakayama for useful discussions.  H.S.~Liu is supported in part by NSFC (National Natural Science Foundation of China) Grant No.~11675144.,  H.~L\"u is supported in part by NSFC Grants No.~11875200 and No.~11935009.

%\vskip 30pt

\section*{Appendix}

\appendix

\section{The $SU(3)$ model from $SO(8)$ gauged maximal supergravity}
\label{app:su(3)}

Both the $U(1)^4$ and $SU(3)$ gauged supergravity model can be obtained from  $D=4$ $SO(8)$ gauged maximal supergravity
constructed by \cite{deWit:1982bul}.  The $U(1)^4$ truncation was performed in \cite{Chong:2004ce}. Further truncation by setting all the four $U(1)$ gauge potentials equal were given in \cite{Donos:2011ut}.  In this appendix, we present the $SU(3)$ reduction.

\subsection{Detail of the truncation}

The theory originates from an ${\cal N}=2$ truncation of the $SO(8)$-gauged maximal supergravity in $D=4$ \cite{deWit:1982bul}. It is obtained by keeping fields invariant under the $SU(3)$ subgroup of $SO(8)$. We refer to \cite{Bobev:2009ms} for the detail of this truncation. Here we first sketch some necessary ingredients. The bosonic sector consists of an ${\cal N}=2$ supergravity multiplet, a vector multiplet and a hypermultiplet. The Lagrangian is the following sum
\be
{\cal L}={\cal L}_{\rm Ein}+{\cal L}_{\rm kin}-e{\cal P}+{\cal L}_{\rm gauge}\,,
\ee
where ${\cal L}_{\rm Ein}=\ft12R$ is the Einstein-Hilbert term, ${\cal L}_{kin}$ is the scalar kinetic term, ${\cal L}_{\rm gauge}$ is Lagrangian for the gauge fields and ${\cal P}$ is the scalar potential. We use $z$ to denote the pair of scalars parametrizing the special K\"{a}hler coset $SU(1,1)/U(1)$, and $\zeta_1\,, \zeta_2$ to denote the two complex scalars parametrizing the quaternionic coset space $SU(2,1)/({SU}(2)\times {U}(1))$. Their kinetic takes the form
\begin{equation}\label{kinetic}
e^{-1}{\cal L}_{\rm kin.}= -g_{z{\bar z}}\,\partial_\mu z \partial^\mu{\bar z}-g_{\zeta_i{\bar \zeta}_j}\nabla_\mu \zeta_i\nabla^\mu{\bar \zeta}_j\,,
\end{equation}
where the unique $SU(1,1)$ and $SU(2,1)$-invariant K\"ahler metrics are given by
\begin{equation}\label{kahmetrics}
ds^2_{\rm SK}= 3 \frac{dz\,d{\bar z}}{ (1-|z|^2)^2} ,
\end{equation}
and
\begin{equation}\label{quatermetr}
ds^2_{\rm QK}=
 \frac{d\zeta _1 d\overline\zeta {}_1+d\zeta _2d\overline\zeta {}_2}{1-|\zeta _1|^2-|\zeta _2|^2} +
\frac{(\zeta _1d\overline\zeta {}_1+\zeta _2d\overline \zeta {}_2)(\overline \zeta {}_1d\zeta _1+\overline \zeta {}_2d\zeta _2)}{ (1-|\zeta _1|^2-|\zeta _2|^2)^2}\,,
\end{equation}
respectively.
The covariant derivative  of the charged scalar fields is
\begin{equation}
\label{redcovder}
\nabla_\mu\zeta_i= \partial_\mu\zeta_i+ g \sum_{\alpha=0}^1 A_\mu^\alpha K_\alpha^{\zeta_i}\,,\qquad i=1,2\,,
\end{equation}
where $g$ is the coupling constant of the gauged supergravity, and the $K_\alpha$ are the Killing vectors on the quaternionic coset and is defined below
\begin{equation}\label{killvect}
K_0 = i\,\zeta_1 \partial_{\zeta_1}- i \,\zeta_2\partial_{\zeta_2}+{\rm c.c.}  \,,\qquad  K_1=\sqrt 3\,i\,\zeta_1 \partial_{\zeta_1}+ \sqrt 3 \,i \,\zeta_2\partial_{\zeta_2}+{\rm c.c.}
\,.
\end{equation}
There are two $U(1)$'s inside $SO(8)$ which commute with $SU(3)$. They are embedded in the $SO(8)$ gauged vectors as follows
%%%%%%%%%%%%%%%%%%%%%%
\be\label{gaugefield}
(A^{IJ})= {\rm diag}\,
\left( \hbox{$ {1\over \sqrt 3}\left(\begin{matrix}
0 & A^1\\-A^1 & 0
\end{matrix}\right),{1\over \sqrt 3}\left(\begin{matrix}
0 & A^1\\-A^1 & 0
\end{matrix}\right),{1\over \sqrt 3}\left(\begin{matrix}
0 & A^1\\-A^1 & 0
\end{matrix}\right),\left(\begin{matrix}
0 & A^0\\-A^0 & 0
\end{matrix}\right) $}\right)\,.
\ee
%%%%%%%%%%%%%%%%%%%%
Accordingly, their kinetic terms inherit from those of the $SO(8)$ gauged vectors and take the form in the $SU(3)$-invariant truncation
\bea
\label{maxwell}
e^{-1}\,{\cal L}_{\rm gauge} &=& -{1\over 4} \sum_{\alpha,\beta=0}^1\left( \tau_{\alpha\beta}\, F_{\mu\nu}^{+ \alpha} F^{+\beta}{}^{\mu\nu}+\overline \tau_{\alpha\beta}\, F_{\mu\nu}^{- \alpha} F^{-\beta}{}^{\mu\nu}\right)\nn\\
&&= -{1\over 4} \sum_{\alpha,\beta=0}^1 \left({\rm Re}(\tau_{\alpha\beta})F_{\mu\nu}^\alpha F^{\beta\, \mu\nu}+{\rm Im}(\tau_{\alpha\beta})F_{\mu\nu}^\alpha\widetilde F^{\beta \, \mu\nu }\right)\,,
\eea
where
\begin{equation}
\label{}
F_{\mu\nu}^\alpha= \partial_\mu A_\nu^\alpha-\partial_\nu A_\mu^\alpha\,,
\end{equation}
and\footnote{We use $\eta^{\mu\nu\rho\sigma}=e^{-1}\epsilon^{\mu\nu\rho\sigma}$, $\epsilon^{0123}=1\,.$}
\begin{equation}
\label{fielst}
F_{\mu\nu}^{\pm\alpha}= {1\over 2}\left(F_{\mu\nu}^\alpha\pm\,{i\over 2}\eta_{\mu\nu}{}^{\rho\sigma} F_{\rho\sigma}^{ \alpha}
\right)\,,\qquad \alpha= 0,1\,.
\end{equation}
The symmetric tensor, $\tau_{\alpha\beta}$,  is given by:
\begin{align}
 \tau_{00} & =  {\left(1+2{\bar z} +3 z {\bar z} +3z ^2+2z ^3+z ^3{\bar z} \right)\over (1+z )^2\left (1-2z +2{\bar z}-z  {\bar z} \right)}\,,\\[6pt]
  \tau_{11}& = {\left(1-2{\bar z} +3 z {\bar z} +3z ^2-2z ^3+z ^3{\bar z} \right)\over (1+z )^2\left (1-2z +2{\bar z}-z  {\bar z} \right)}\,,\\[6pt]
 \tau_{01}&= \tau_{10}= \,-{2\sqrt{3}\,z (1+z {\bar z} )\over (1+z )^2\left (1-2z +2{\bar z}-z  {\bar z} \right)}\,.
 \label{tauans3}
\end{align}
The scalar potential of the $SU(3)$-invariant sector is given by
\begin{equation}
\label{poten}
{\cal P}= 2\, g^2 \bigg[ {4 \over 3} (1 - |z|^2)^2 \left|{\partial | {\cal W} | \over \partial z}\right|^2  + (1 -
|\zeta_{12}|^2)^2  \left|{\partial  | {\cal W} | \over \partial \zeta_{12}} \right|^2\  -  3\,  |{\cal W}|^2 \bigg] \,,
\end{equation}
where
\be
\label{zetavar}
\zeta_{12} ={|\zeta_1|+ i\,|\zeta_2|\over 1+\sqrt{1-|\zeta_1|^2-|\zeta_2|^2} }\,,
\ee
and the superpotential is defined as
\be
\label{superpot}
{\cal W}= (1-|z|^2)^{-3/2}(1-|\zeta_{12}|^2)^{-2}\left[(1+z^3)\,(1+\zeta_{12}^4)+6\, z \, \zeta_{12}^2(1+z)\right]\,.
\ee
One can of course also choose $\overline{\cal W}$ as superpotential.

\subsection{Further truncation}

It is straightforward to check that one can set $z\,,\bar{z}$ consistently to 0. Consequently, the gauge kinetic terms become
\be
\tau_{00}=\tau_{1,1}=1\,,\quad \tau_{01}=0\,.
\ee
By introducing the coordinates
\be
\zeta_1=\tanh\rho\cos\ft12\theta e^{i(\phi+\chi)/2}\,,\quad \zeta_2=\tanh\rho\sin\ft12\theta e^{i(\chi-\phi)/2}\,,
\ee
the metric on $SU(2,1)/({SU}(2)\times {U}(1))$ can be recast into the form
\be
ds^2_{\rm QK}=d\rho^2+\ft14\sinh^2\rho(d\theta^2+\sin^2\theta d\phi^2) +\ft1{16}\sinh^22\rho(d\chi+\cos\theta d\phi)^2\,.
\ee
Meanwhile, the Killing vectors defined in (1.6) becomes
\be
K_0=2\partial_\phi\,,\quad K_1=2\sqrt{3}\partial_\chi\,.
\ee
Therefore in terms of the new variables, the scalar kinetic term takes the form
\bea
{\cal L}_{\rm kin}&=&(\partial\rho)^2+\ft14\sinh^2\rho\left((\partial\theta)^2+\sin^2\theta (\partial\phi+2g A_0)^2\right)\nn\\
&&+\ft1{16}\sinh^22\rho\left((\partial\chi+2g\sqrt{3}A_1)+\cos\theta (\partial\phi+2gA_0)\right)^2\,.
\eea
The scalar potential is given by
\be
{\cal P}=\frac{g^2}4\left( (7+\cos\theta)\sinh^4\rho-16\sinh^2\rho-24     \right)\,.
\ee
Then it is easy to see that one can further set $\theta=0$. Once this is done, we can redefine the remaining scalars and vectors as
\be
\tilde{\chi}=\chi+\phi\,,\quad \tilde{A}_0=\ft12(A_0+\sqrt{3}A_1)\,,\quad \tilde{A_1}=\ft12(-\sqrt{3}A_0+A_1)\,.
\ee
In terms of the new vector fields, the vector kinetic still takes diagonal form. Since no field is charged under $\tilde{A}_1$, we can set it to 0 consistently. In the end, we arrive at the simplified model with only one $U(1)$ gauge field
\be
e^{-1}{\cal L}=\ft12 R-(\partial\rho)^2-\ft1{16}\sinh^22\rho(\partial\tilde{\chi}+4g\tilde{A}_0)^2
-\ft14\tilde{F}_0^{\mu\nu}\tilde{F}_{0\mu\nu}-{\cal P}|_{\theta=0}\,,
\ee
which is Einstein-Maxwell coupled to a charge scalar with a particular potential.

We now perform a change of convention
\be
{\cal L}\rightarrow 2{\cal L}\,,\qquad \rho=\ft1{\sqrt{2}} \phi\,,\qquad \tilde A=\ft{1}{\sqrt2} A\,,\qquad g\rightarrow \ft1{\sqrt2} g\,,
\ee
and absorb the Stuckelburg field $\tilde{\chi}$ into the gauge field.
The resulting Lagrangian becomes the one used in the bulk of the paper
\be
\label{trunc2}
e^{-1} {\cal L}=R - (\partial \phi)^2 -\ft12 g^2 \sinh^2\sqrt{2}\phi\, A^2 - \ft14 F^2 +
2g^2 \Big(3+ 2 \sinh^2 (\ft{\phi}{\sqrt{2}}) -\sinh^4 (\ft{\phi}{\sqrt{2}} )^2\Big)\,.
\ee
We recall that the model obtained here comes from a truncation of $D=4$ $SO(8)$ gauged maximal supergravity. Thus it can be embedded in 11 dimensional supergravity  compactified on $S^7$. One the other hand, the same model \eqref{trunc2} also arises from a consistent reduction of even dimensional supergravity on a seven dimensional Sasaki-Einstein manifold preserving 8 supercharges \cite{GSW,GKVW}, although the full-fledged ${\cal N}=2$ models arising from the seven dimensional round sphere and Sasaki-Einstein reductions are different \cite{Bobev:2010ib}.

\section{The explicit equations of motion}
\label{app:eoms}

In this paper, we construct electrically-charged spherically-symmetric solutions in Einstein-Maxwell gravity coupled to a charged scalar.  The ansatz (\ref{genans}) involves four functions $(f,h,a,\phi)$.  The function $f$ can be solved algebraically, given in (\ref{fsol}).  The remaining functions satisfy the following second-order nonlinear differential equations
\bea
h''&=& \fft{1}{4r h \left(r^2 \left(h V-a^2 U\right)-2 h\right)}\Big[2 h^2 r V \left(2 r^2 a'^2+r^2 \phi '^2 \left(r h'-2 h\right)+4 h\right)\nn\\
&&-r a^2 U \left(r^2 a'^2 \left(r h'+6 h\right)+4 \left(-2 h^2 \left(r^2 \phi'^2-2\right)+r^2 h'^2+3 h r h'\right)\right)\nn\\
&&-4 h^2 \left(r a'^2+h' \left(r^2 \phi'^2-4\right)\right) \Big],\nn\\
%%%%
%%%%
a''&=&\fft{1}{4r h \left(r^2 \left(h V-a^2 U\right)-2 h\right)}\Big[2 h^2 r^2 V a' \left(r^2 \phi '^2-4\right)-4 h^2 a' \left(r^2 \phi'^2-4\right)\nn\\
&&+a r U \left(4 ra a' \left(h-r h'\right)-4 r^2 h a'^2-r^3 a a'^3+8 h \left(h \left(r^2 \phi'^2-2\right)-2 r h'\right)\right)
\Big],\nn\\
%%%
%%%
\phi'' &=& \fft{1}{4r h \left(r^2 \left(h V-a^2 U\right)-2 h\right)}\Big[r h V_{,\phi} \left(-r^2 a'^2-4 r h'+2 h \left(r^2\phi '^2-2\right)\right)\nn\\
&&+r a^2 U_{,\phi} \left(r^2 a'^2+4 r h'-2 h r^2 \phi '^2+4 h\right)+r^2 a^2 U \phi ' \left(4 h-r^2 a'^2\right)\nn\\
&&+2 r^2 h V \phi ' \left(h \left(r^2 \phi '^2-4\right)-2 r h'\right)+4 h \phi ' \left(2 r h'+h \left(4-r^2 \phi '^2\right)\right)
\Big].
\eea
Note that a bared $r$ appears in the denominator, indicating that the equations become singular at the space origin $r=0$, where we need to resolve the ambiguity of $0/0$ in the numerical approach.  We resolve this by performing the Taylor expansion at $r=0$ and use it to move the initial integration point slightly away from $r=0$.

\section{Generalized Smarr formula}
\label{app:smarr}

In this section, we provide a derivation of the generalized Smarr formula for solutions in an Einstein-Proca-scalar theory. Unlike the discussion in \cite{Fan:2015tua,Liu:2015tqa}, our method does not rely on the scaling symmetry associated with the effective Lagrangian for planar case. We consider the Einstein-Proca-scalar theory of form
\be
{\cal L}=\sqrt{-g} \Big(R-(\partial\phi)^2- U(\phi)\, A_{\mu}A^{\mu}-\ft14 F^2-V(\phi)\Big)\,.
\ee
The corresponding Einstein equation can be put in the form
\be
R_{\mu\nu}-\partial_\mu\phi\partial_{\nu}\phi-\ft12(F^2_{\mu\nu}-\ft14g_{\mu\nu}F^2)
-U(\phi)\,A_{\mu}A^{\mu}-\ft12g_{\mu\nu}V(\phi)=0\,.
\ee
Given a Killing vector $\xi$, we have
\bea
*d*d\xi&=&-2R_{\mu\nu}\xi^{\mu}dx^{\nu}\,,\nn\\
*(i_{\xi}F\wedge*F-i_{\xi}*F\wedge F)&=&-2\xi^{\mu}(F^2_{\mu\nu}-\ft14g_{\mu\nu}F^2)dx^{\nu}\,.
\eea
Using $d*(UA)=0$ on shell, we now define a closed 3-form $H_{(3)}$
\be
UA=*H_{(3)}=*dB_{(2)}\,,
\ee
which satisfies
\be
*(i_{\xi}H\wedge *U^{-1}H+i_{\xi}*U^{-1}H\wedge H)=U(2A_{\mu}A_{\nu}-g_{\mu\nu}A^2)\xi^{\mu}dx^{\nu}\,.
\ee
Using
\be
{{\cal L}}_{\xi}F=0\,,\quad {{\cal L}}_{\xi}H=0\,,\quad d*F=2H\,,\quad d(*U^{-1}H)=F\,,
\ee
one can show that the following expression is closed, namely, locally
\be
(i_{\xi}F\wedge*F-i_{\xi}*F\wedge F)-2(i_{\xi}H\wedge *U^{-1}H+i_{\xi}*U^{-1}H\wedge H)=d{\cal P}\,,
\ee
where
\be
{\cal P}=i_{\xi}*F\wedge A-i_{\xi}A*F\,.
\ee
Since ${\cal L}_{\xi}\phi=0\,,{\cal L}_{\xi}A^2=0$, there exists a 2-form $\omega$ such that
\be
*d*\omega=(UA^2+V)\xi_\nu dx^{\nu}\,.
\ee
Adding all the pieces together, we have the conservation law
\be
-\ft12*d*d\xi+\ft14*d{\cal P}-\ft12*d*\omega=0\,.
\ee

We now consider the ansatz
\be
ds^2 = - h dt^2 + \fft{dr^2}{f} + r^2 d\Omega_2^2\,,\qquad
\phi=\phi(r)\,,\qquad A=a(r) dt\,,
\ee
we get the generalized Smarr relation
\be
r^2(h'\sqrt{f/h}-\ft12aa'\sqrt{f/h}+S)|_{r=r_i}^{r=r_f}=0\,,
\ee
where $S$ is given by
\be
r^2 S=\int^{r}dr'\,r'^2 (UA^2+V)\frac{\sqrt{h}}{\sqrt{f}}+{\rm constant}.
\ee
To make the discussion more general, we switch to the ansatz
\be
ds^2 = - a^2(r) dt^2 + b^2(r)dr^2 + c(r)^2 d\Omega_{2\,,\lambda}^2\,,\qquad
\phi=\phi(r)\,,\qquad A=A_0(r) dt\,,
\ee
in which $d\Omega_{2\,,\lambda}^2$ is a two dimensional maximally symmetric space with $R_{ab}=\lambda g_{ab}$. In the previous discussions, we learnt that applying the $tt$ component of the Einstein equation, the bulk action can be reduced to a surface term
\be
I_{\rm bulk}=-4\Omega_2(\frac{2c^2a'}{b}-\frac{c^2A_0A_0'}{ab})\Big|^{r={\infty}}_{r=r_0}\,.
\ee
On the other hand, if one substitutes the $rr$ component of the Einstein equation to the action, the on-shell action can also be written as
\be
I_{\rm bulk}=-4\Omega_2(\frac{2c^2a'}{b}+\frac{4acc'}{b})\Big|^{r={\infty}}_{r=r_0}+4\Omega_2\int_{r_0}^{\infty}dr(4\lambda ab-2Vabc^2+\frac{2UA_0^2bc^2}{a})\,.
\ee
There is also a third way to reexpress the onshell action by using the components of the Einstein equation in the $d\Omega_{2\,,\lambda}^2$ direction. One gets
\be
I_{\rm bulk}=-4\Omega_2(\frac{2acc'}{b})\Big|^{r={\infty}}_{r=r_0}+4\pi\int_{r_0}^{\infty}dr(2\lambda ab)\,.
\ee
Equating the above 3 different expressions for the same onshell bulk action, one can obtain useful relations between bulk integral and surface terms. In particular, when $\lambda=0$, these include the two Smarr relations. When applied to extremal charged (hairy) branes, one of the Smarr relation takes the form
\be
3M=2\mu Q\,.
\ee
In the region where the solution exibits scaling symmetry, one can assume $\mu=cQ^{\alpha}$, where c is a $Q$-independent parameter. It is easy to see that the Smarr relation above and the first law $dM=\mu dQ$ fix $\alpha=\ft12$ without utilizing the detailed structure of the solution.

\section{Verifying $dM=\mu dQ$ at the tip of the cusp}
\label{app:cusp}

\begin{figure}[htp]
\begin{center}
\includegraphics[width=220pt]{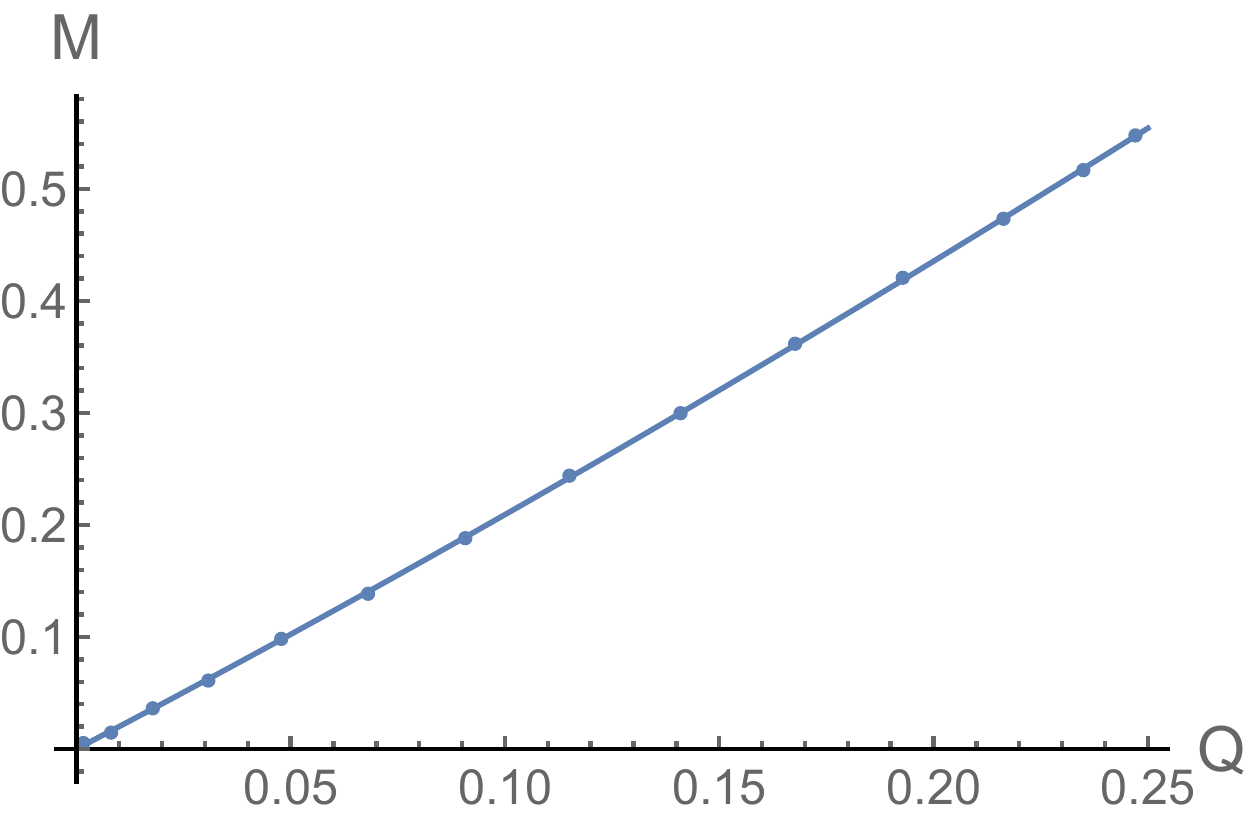}
\includegraphics[width=220pt]{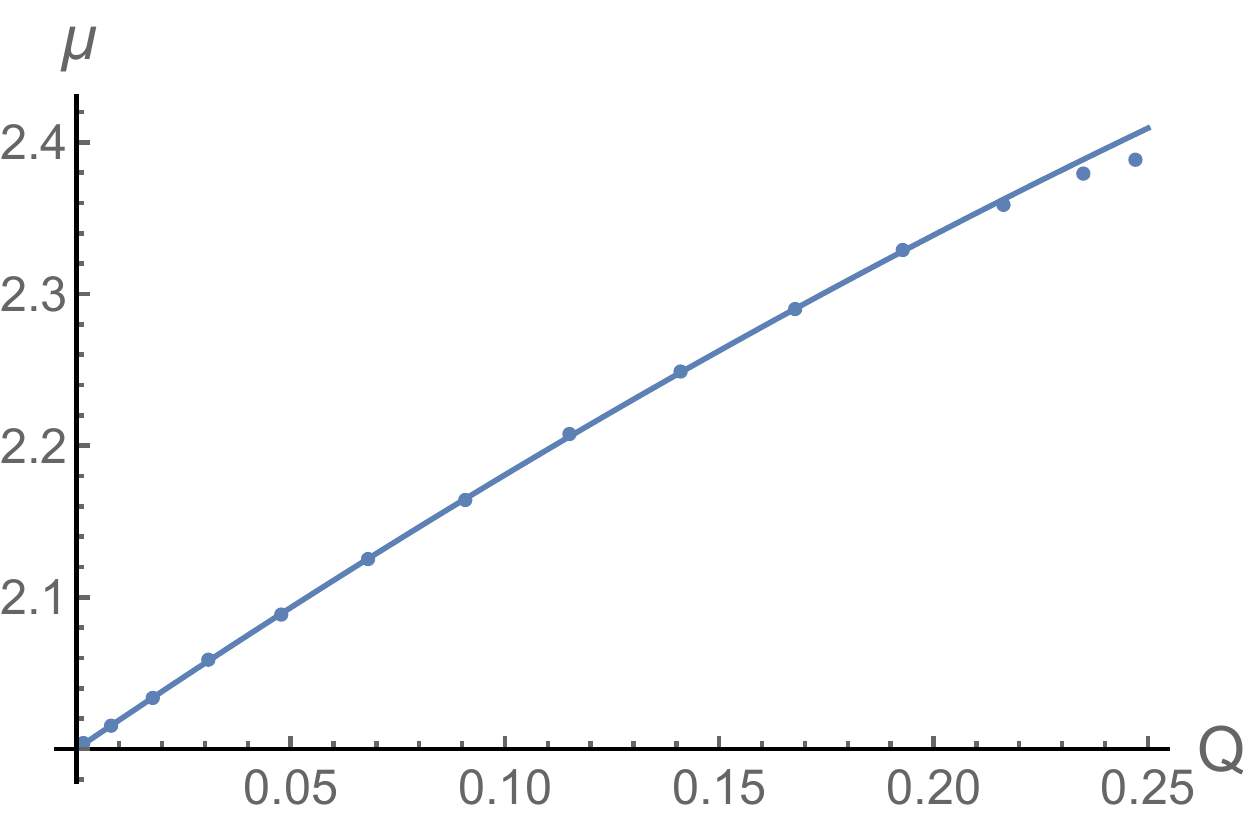}
\end{center}
\caption{\small\it These are the mass and chemical potential dependence on $Q$ for the $A_1$ solutions of $q^2=1$, corresponding to Fig.~\ref{q1-A1A2A3RN-M(Q)}. The dots are numerical data and the solid lines are data-fitting functions satisfying $dM=\mu dQ$.}
\label{q1A1Mmu(Q)smallQ}
\end{figure}

As we see in the main text that for some gapless type-$A$ solutions, the mass and charge are bounded above and are much less than the AdS radius.  The mass-charge relation appears to be two straight lines with a common tip, shown in Figs.~\ref{q14A1A2A3RNM(Q)} and \ref{q1-A1A2A3RN-M(Q)}.  If our first law $dM=\mu dQ$ is correct, this implies that $\mu(Q)$ would be constant and discontinuous, which contradicts the Figs.~\ref{q14A2A3phi2mu} and \ref{q1-A1A2A3RN-M(Q)} respectively.  In this appendix, we use the $A_1$ series in the $q^2=1$ theory as a concrete example to resolve the puzzle.  We first note that the longer line of $A_1$ in Fig.~\ref{q1-A1A2A3RN-M(Q)} that connects to the origin is not straight, but with the mass-charge function
\be
M=2 Q + 0.960472 Q^2 - 0.378403 Q^3 + {\cal O}(Q^4)\,,
\ee
From this, we can obtain the function $\mu(Q)=dM/dQ$.  We see clearly from Fig.~\ref{q1A1Mmu(Q)smallQ} that the data-fitting functions match with the numerical data perfectly for small $Q$, which establishes our first law for small $Q$ and its $Q\rightarrow 0$ limit. Our data-fitting mass-charge relation also implies that the $A_1$ series of solutions have bigger energy than the corresponding extremal RN-AdS black holes in this $q^2=1$ theory.

\begin{figure}[htp]
\begin{center}
\includegraphics[width=250pt]{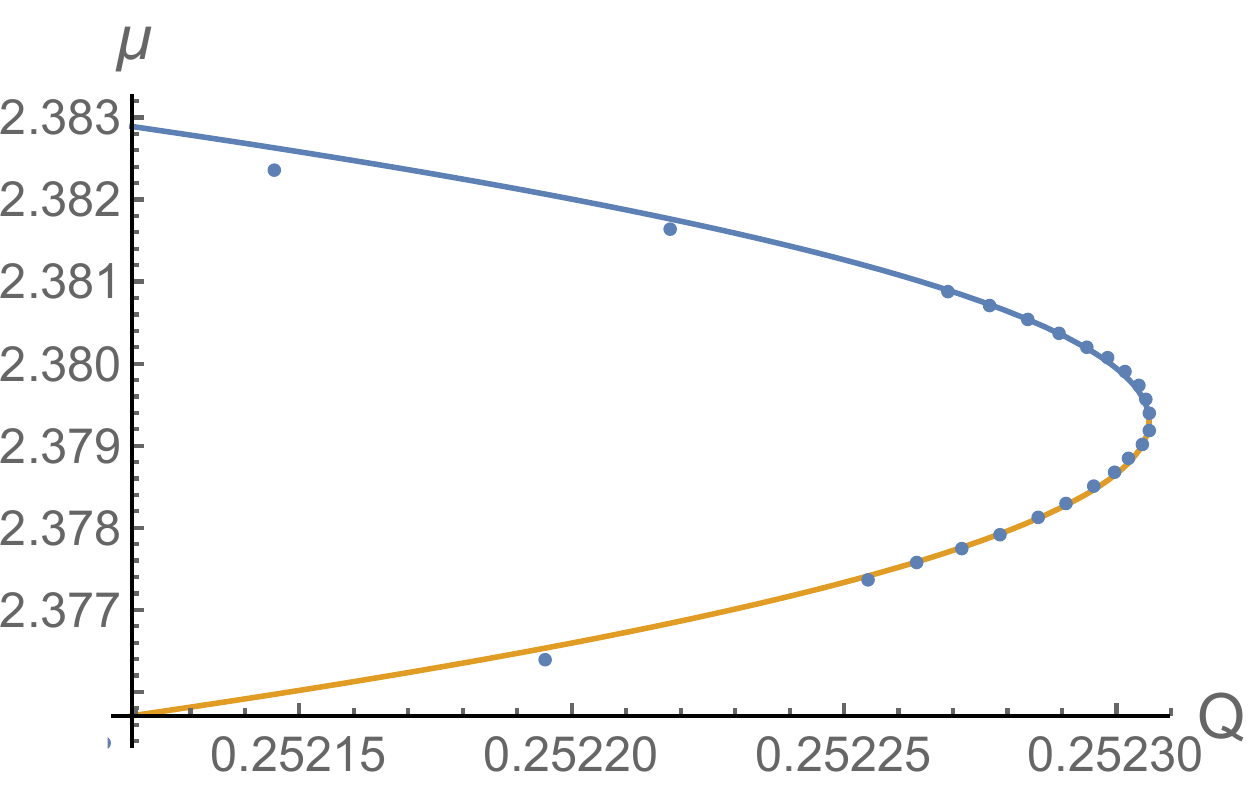}
\end{center}
\caption{\small\it The figure shows that the formula (\ref{mupm}) fits the numerical data perfectly at the tip. The top branch is $\mu_+$ and the lower one is $\mu_-$. The divergence of $d\mu/dQ$ at the tip creates the $A_1$ cusp in Fig.~\ref{q1-A1A2A3RN-M(Q)}, where the two mass lines join in a tangential manner: they are continuous up to and including the first derivative.}
\label{q1A1mu(Q)cusp}
\end{figure}

To establish the validity of the first law $dM=\mu dQ$ at the vicinity of the cusp in the mass-charge relation, shown in Fig.~\ref{q1-A1A2A3RN-M(Q)}, we need to observe that the two lines of $M(Q)$ are tangent to each other at the tip.  In other words, $dM/dQ$ is continuous at the tip, but not the second derivative $d^2M/dQ^2$.  This can be realized by considering
\be
Q_{\rm max} - Q = \alpha (\mu - \mu^*)^2\,,
\ee
at the tip $Q_{\rm max}$.  This implies that
\be
\mu_\pm =\mu^* \pm \sqrt{\fft{Q_{\rm max} - Q}{\alpha}}\,,\qquad M_\pm = M_{\rm max} + \mu^*(Q-Q_{\rm max})
\mp \fft{2(Q_{\rm max}-Q)^{\fft32}}{3\sqrt{\alpha}}\,.
\ee
Thus we see that the two ($\pm$) branches join smoothly for the chemical potential $\mu$, but create a cusp for the mass.  The origin of the cusp is that $d^2M/dQ^2 = d\mu/dQ$ diverges at the cusp. Concretely, for the $A_1$ solution of the $q^2=1$ theory, near the top tip of the mass-charge relation in Fig.~\ref{q1-A1A2A3RN-M(Q)}, we have
\be
\mu_{\pm}=2.3793 \pm 0.262785 \sqrt{0.252306\, -Q}\,,\label{mupm}
\ee
which, can be seen in Fig.~\ref{q1A1mu(Q)cusp}, fits the numerical data near the tip perfectly.

\section{More graphs}
\label{app:graphs}

This is the depository of figures of physical quantities obtained from numerical construction of various AdS boson star solutions.  We collect them here so that we do not interrupt the flow of description in the main text.

\begin{figure}[htp]
\begin{center}
\includegraphics[width=220pt]{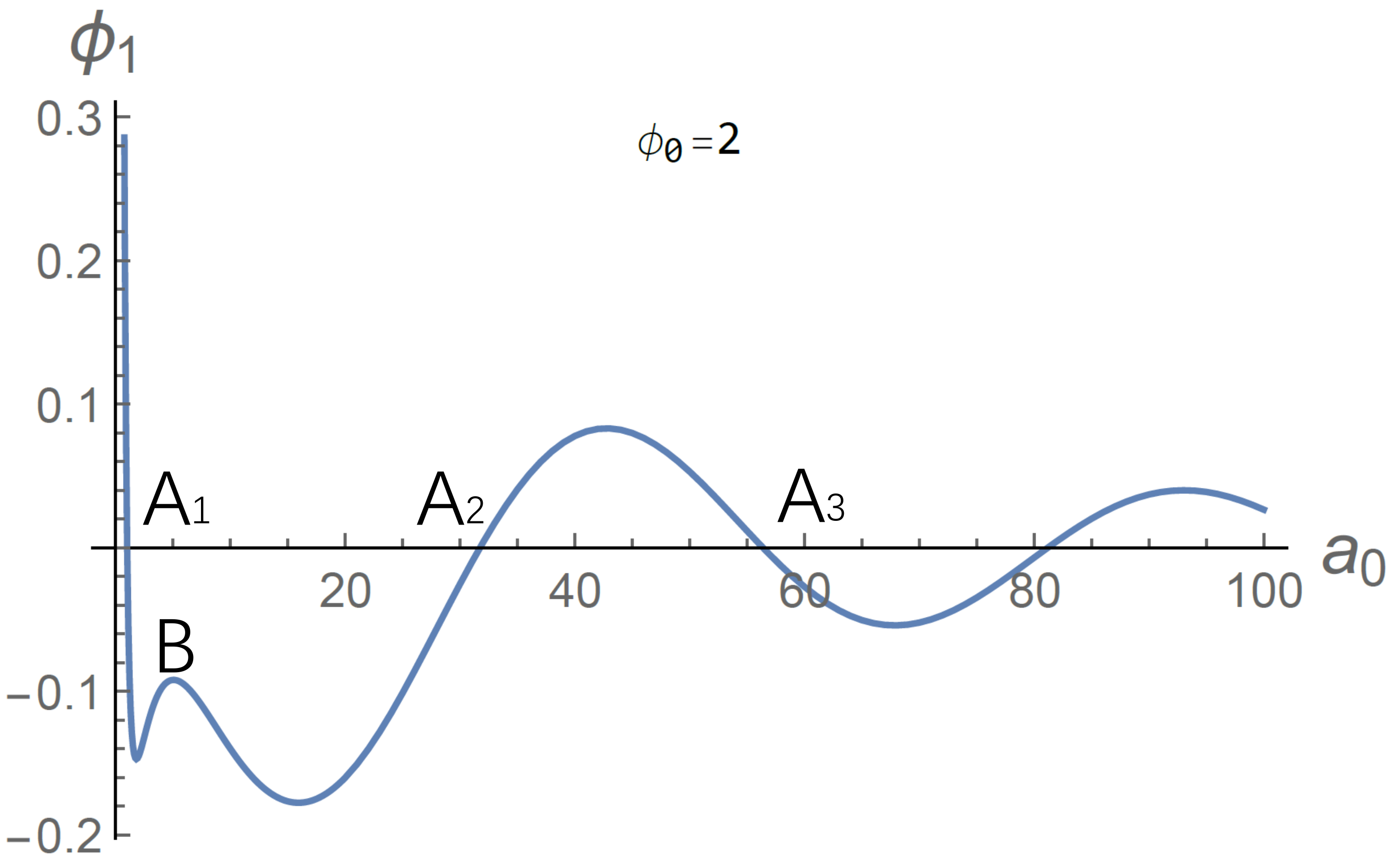}
\includegraphics[width=220pt]{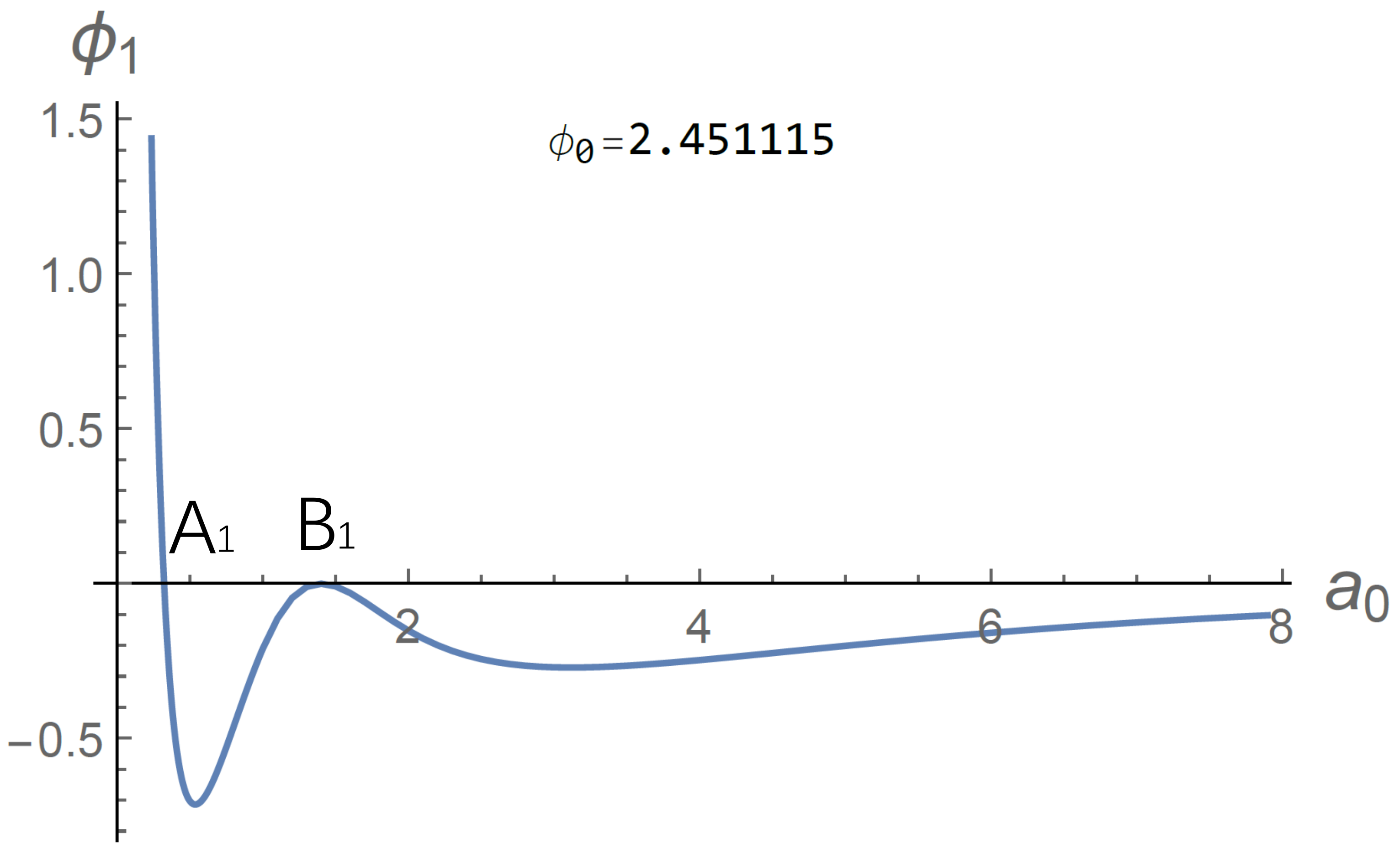}
\includegraphics[width=250pt]{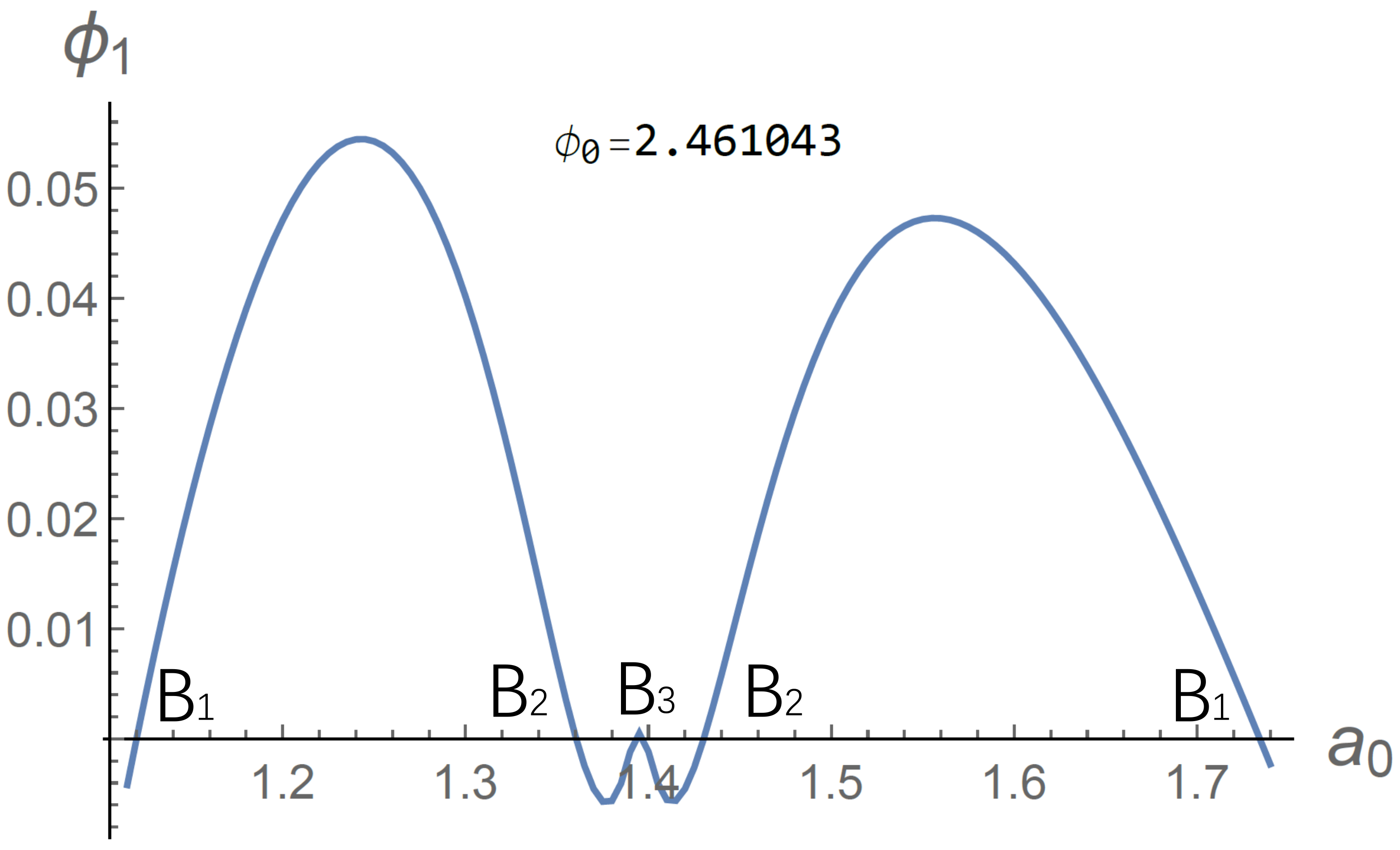}
\end{center}
\caption{\small\it New type of roots of $\phi_1(a_0)$ emerge as $\phi_0$ increases and we label the corresponding boson stars as the $B$ series. The $A_1$ root stays in the left of the $B$ roots while the $A_i$ roots with $i=2,3,\ldots,$ are pushed to the right of all the $B_i$ roots. The theory has $q^2=1.4$.}
\label{q14BRNphi1(a0)a}
\end{figure}

\begin{figure}[htp]
\begin{center}
\includegraphics[width=250pt]{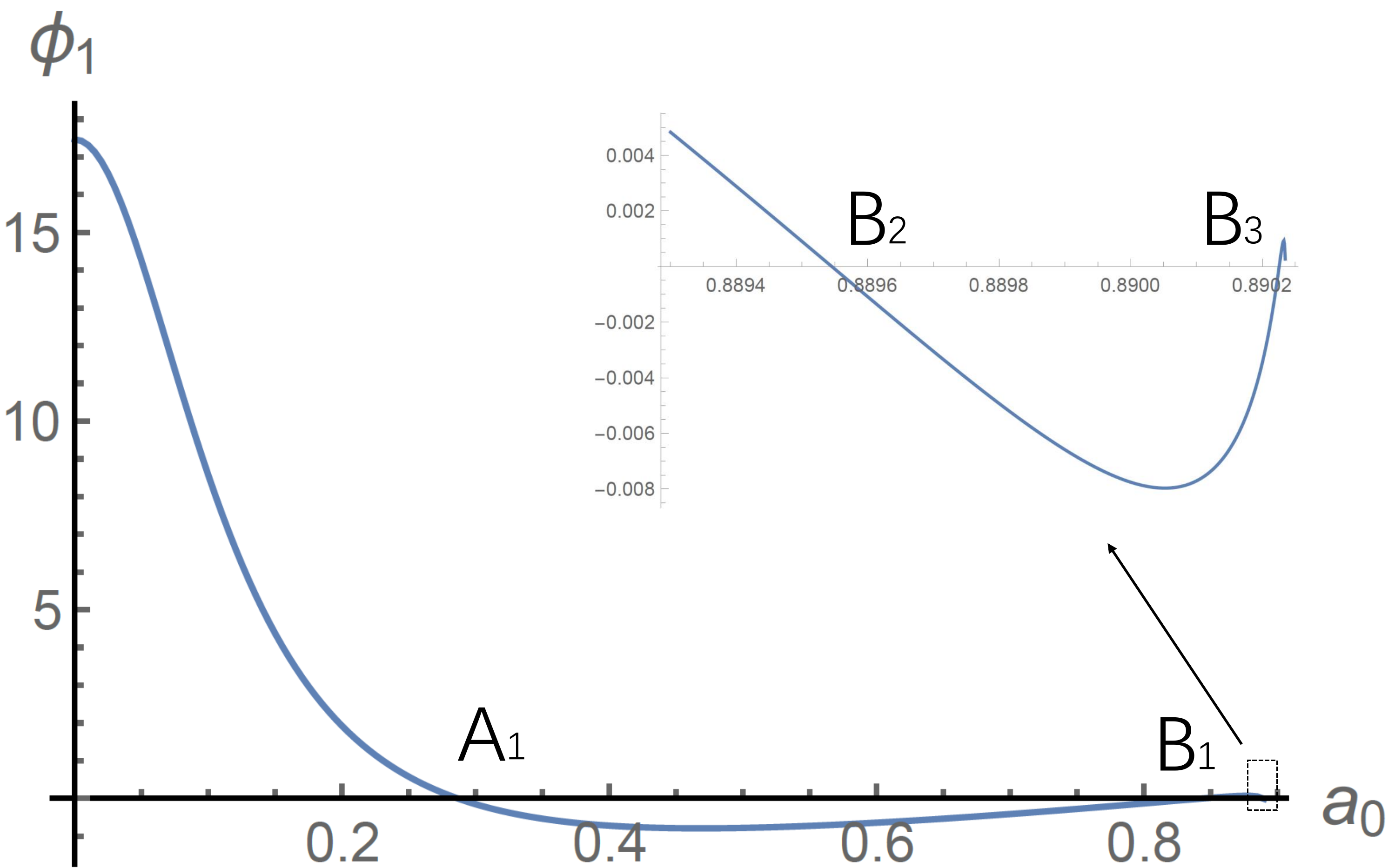}\
\includegraphics[width=250pt]{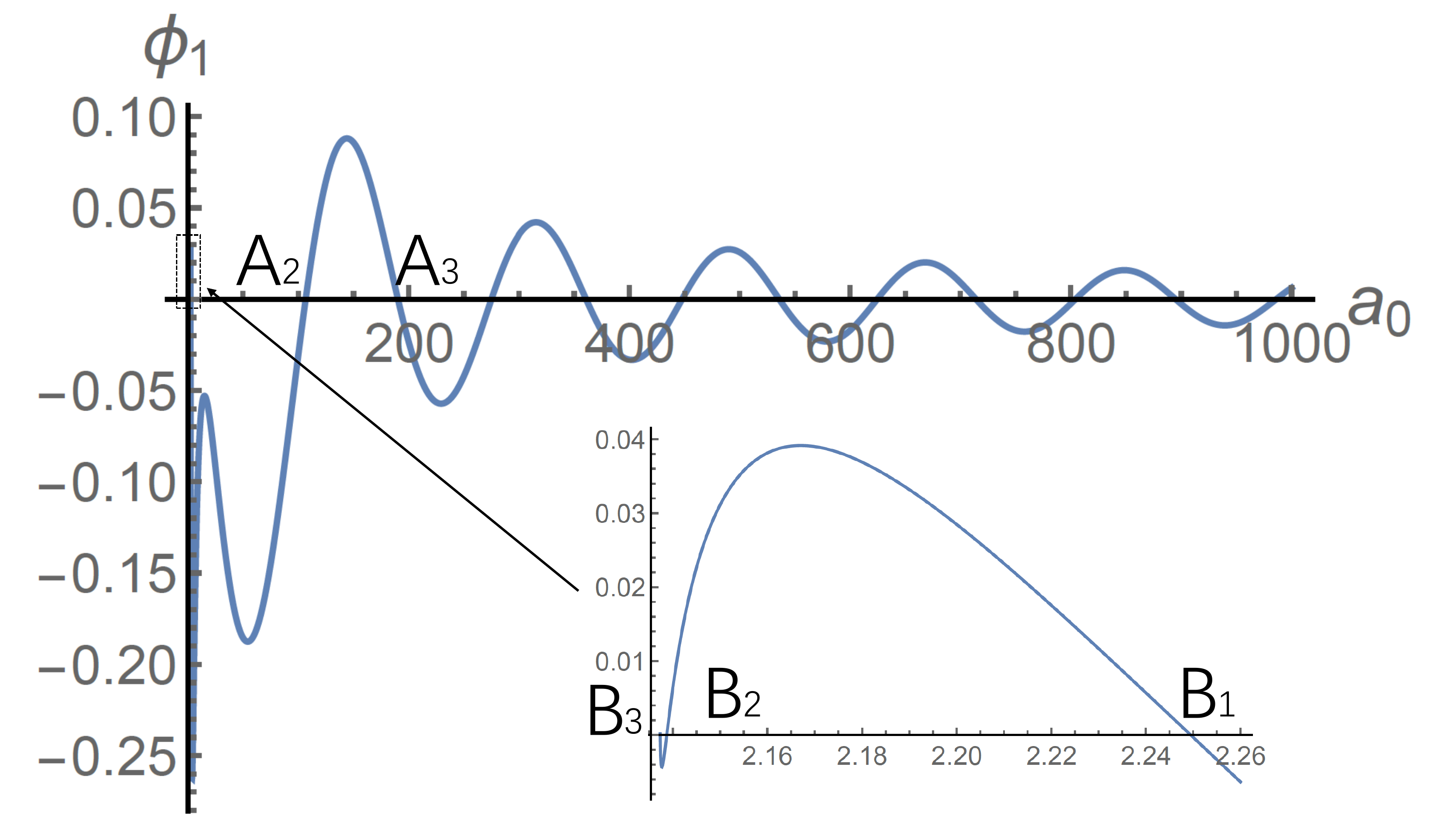}
\end{center}
\caption{\small\it At $\phi_0=2.5$, there is no solution in the region $0.891<a_0<2.135$ and the $B$ series split into the left and right groups.  The $A_1$ roots stay in the left and all the $A_{2,3,\ldots}$ roots are pushed to the far right. Compared to the $A$ roots, the $B$ series occupy only a tiny region in the $a_0$ axis. The theory has $q^2=1.4$}
\label{q14BRNphi1(a0)b}
\end{figure}

\begin{figure}[htp]
\begin{center}
\includegraphics[width=250pt]{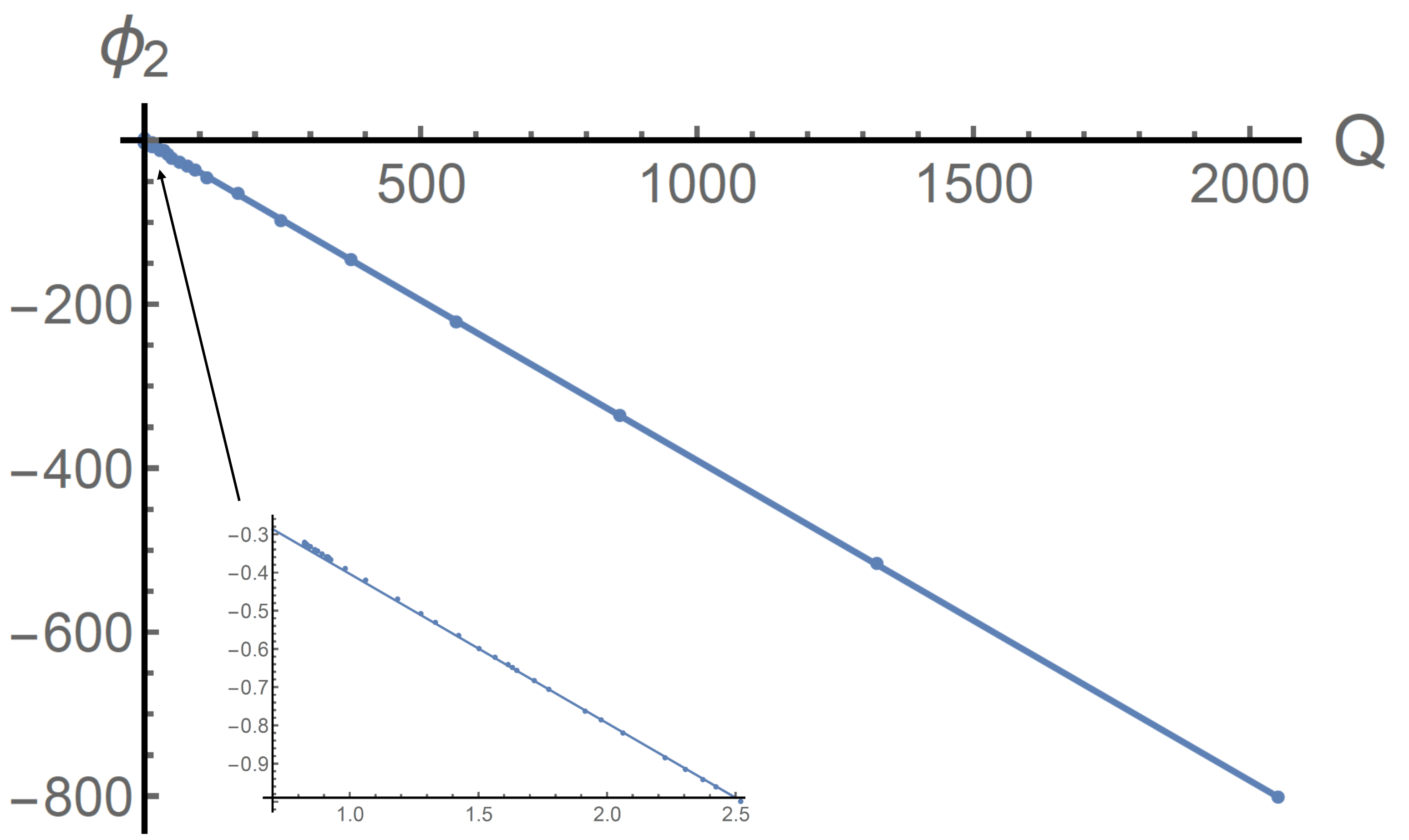}\
\end{center}
\caption{\small\it This shows the matching of the numerical data to the linear approximation $\phi_2=-0.0110485-0.390663 Q$ for the $B_1$ series with $q^2=1.4$, discussed in section \ref{sec:q14b}.}
\label{q14BRN-phi2-Q}
\end{figure}

\begin{figure}[htp]
\begin{center}
\includegraphics[width=220pt]{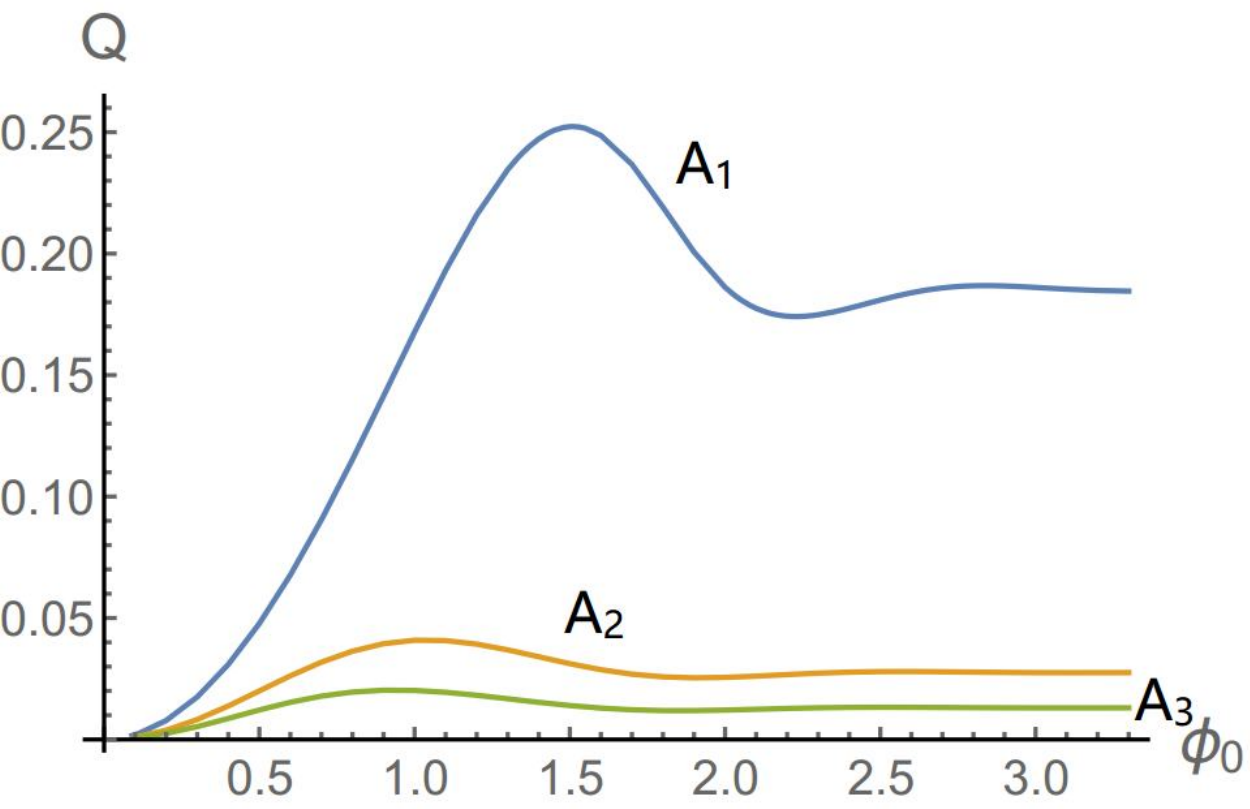}\
\includegraphics[width=220pt]{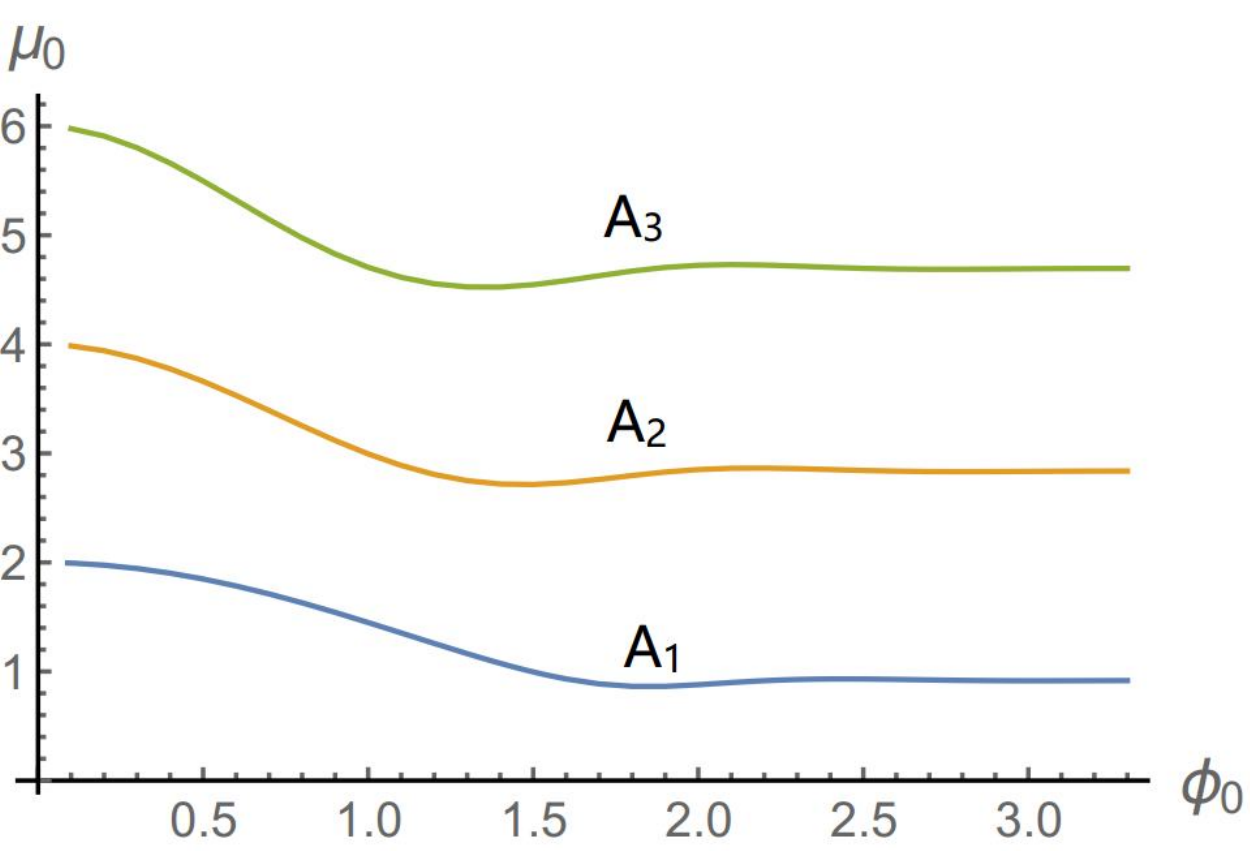}\
\includegraphics[width=220pt]{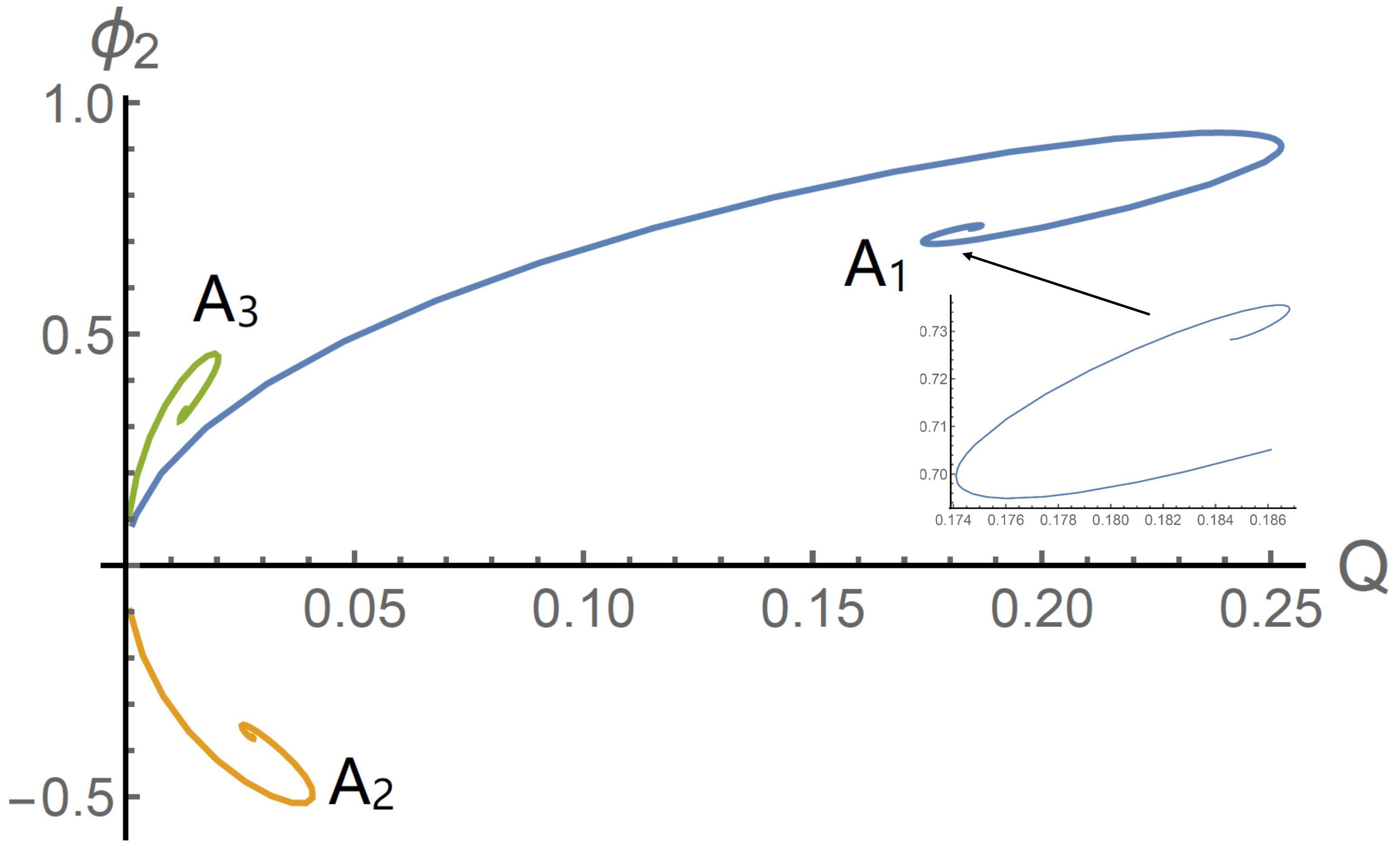}\
\includegraphics[width=220pt]{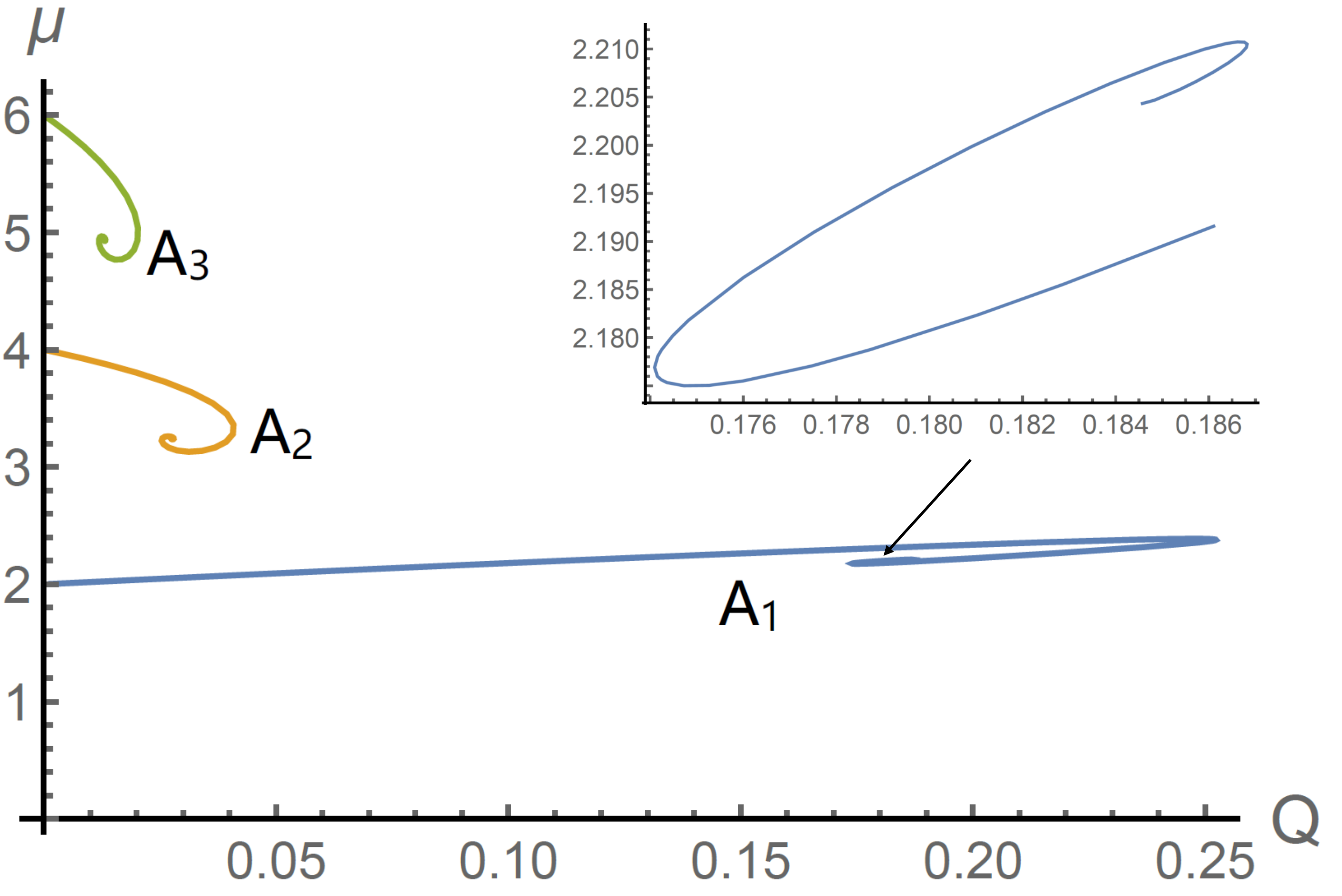}
\end{center}
\caption{\small\it These pictures show the properties of the $A$ series of boson stars when $q^2=1$. Unlike in the case of $q^2=1.4$, the $A_1$ here has the same characteristics as the rest of the $A$ series of solutions. The scalar hair and chemical potential spiral with the charge into some fixed points. The difference $\mu-\mu_0$ vanishes as $Q\rightarrow 0$.}
\label{q1A1A2A34plots}
\end{figure}

\begin{figure}[htp]
\begin{center}
\includegraphics[width=260pt]{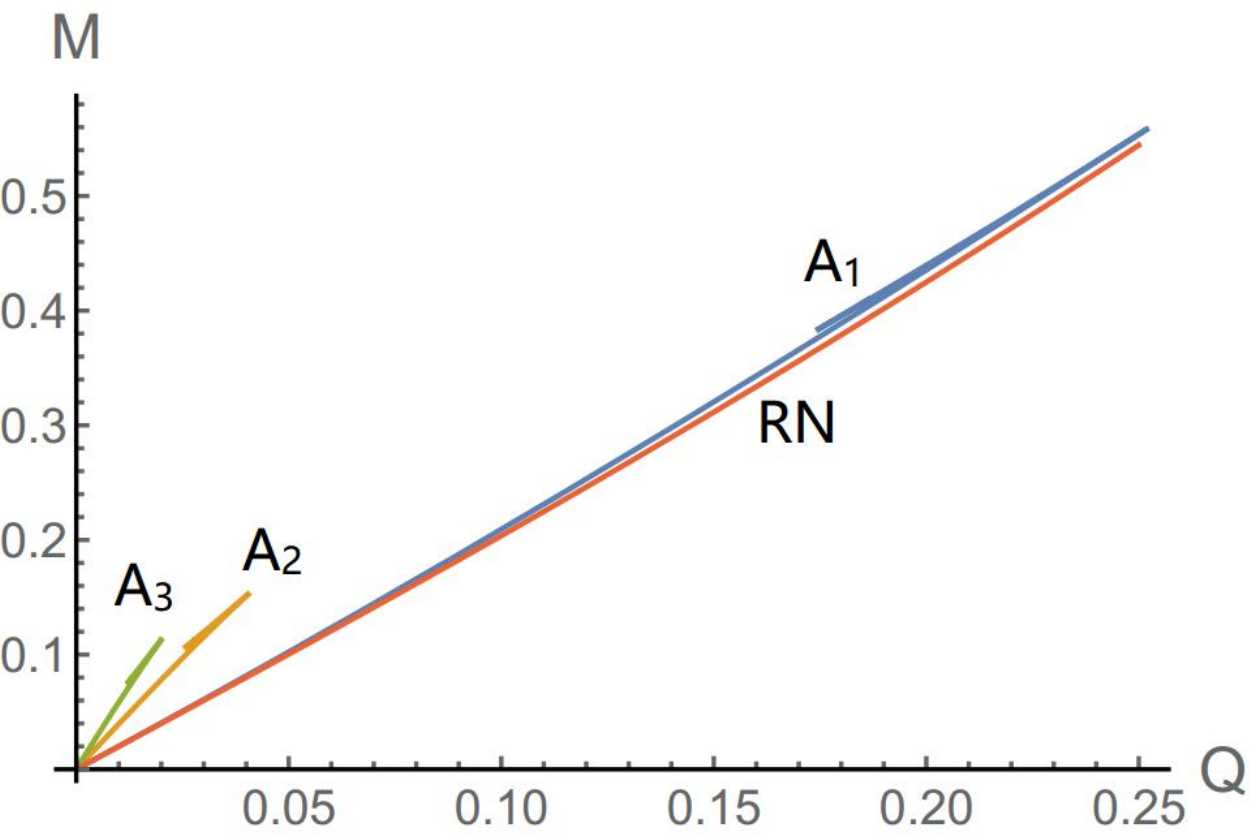}
\end{center}
\caption{\small\it Here are the mass-charge relations of the $A$ series of boson stars of $q^2=1$. All the masses and charges are bounded above and the masses are all larger than that of the extremal RN-AdS black hole.}
\label{q1-A1A2A3RN-M(Q)}
\end{figure}

\begin{figure}[htp]
\begin{center}
\includegraphics[width=220pt]{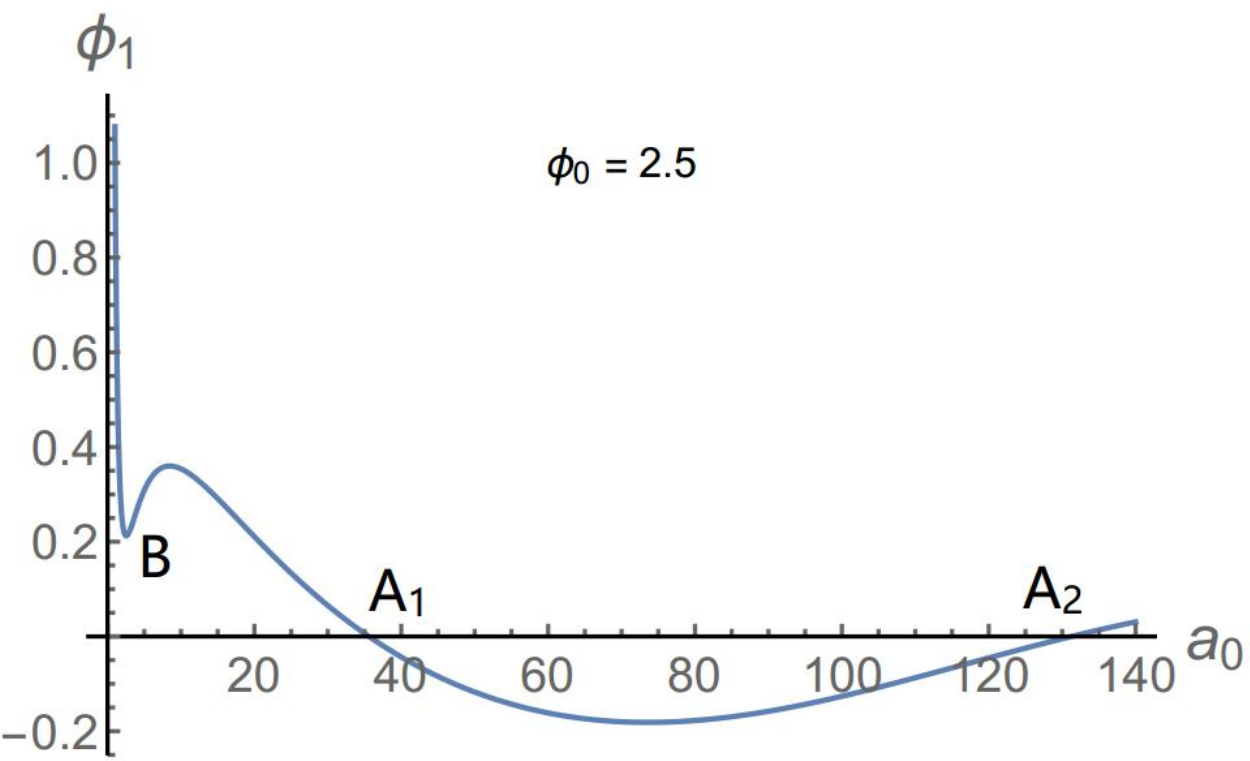}\!
\includegraphics[width=220pt]{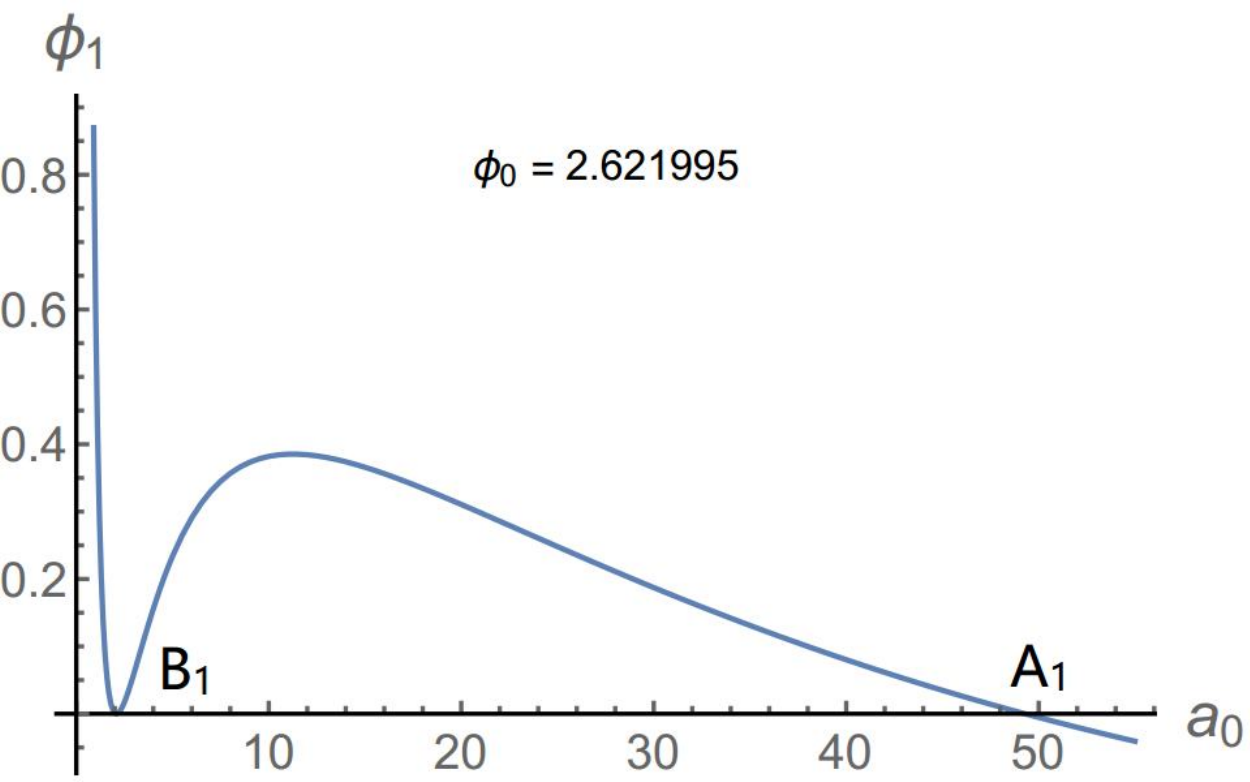}\!
\includegraphics[width=260pt]{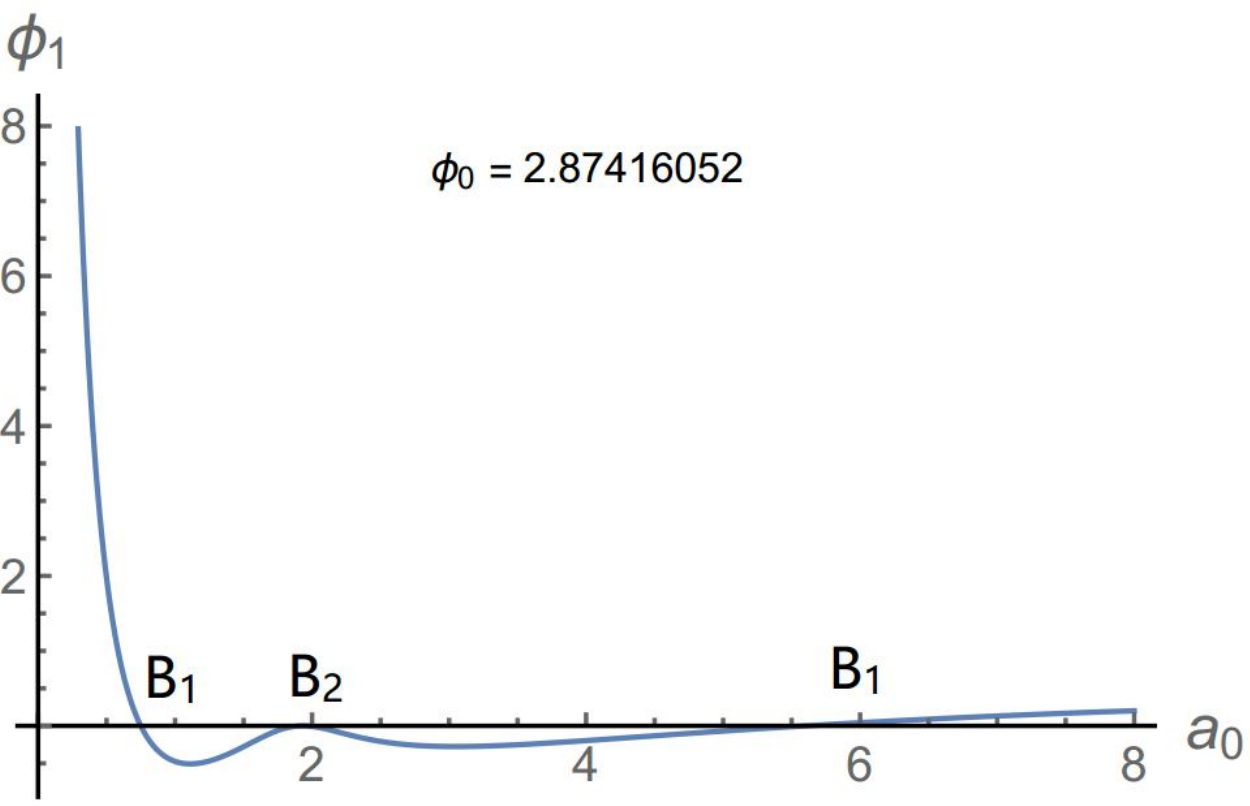}
\end{center}
\caption{\small\it This shows the creation of mass-gap solutions for $q^2=1$, which is very different from that of $q^2=1.4$, depicted in Fig.~\ref{q14BRNphi1(a0)a}. The new bump emerges between $a_0=0$ and the $A_1$ root, and the tip lowers down towards $a_0$-axis and then starts to oscillate as $\phi_0$ increases. All the $A$ roots are pushed to the right to the all the $B$ roots.}
\label{q1-B-phia0}
\end{figure}

\begin{figure}[htp]
\begin{center}
\includegraphics[width=220pt]{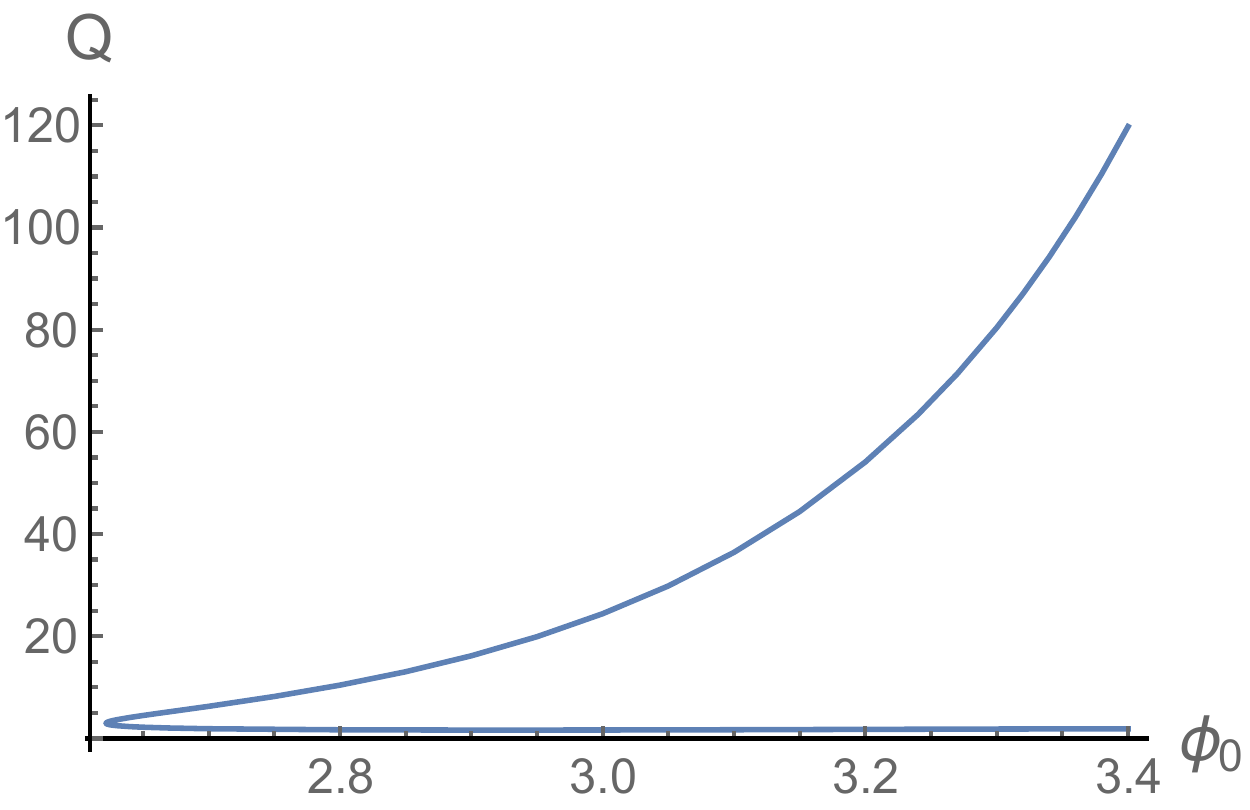}\!
\includegraphics[width=220pt]{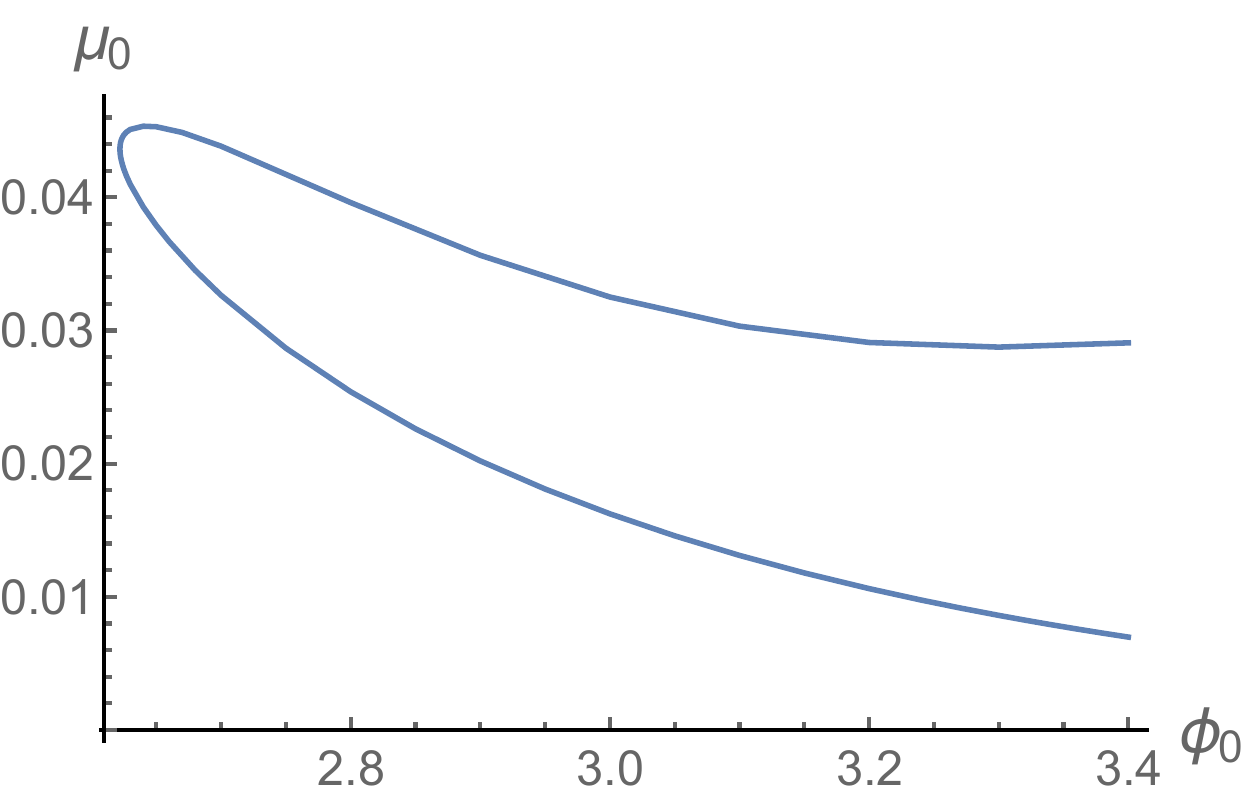}\!
\includegraphics[width=220pt]{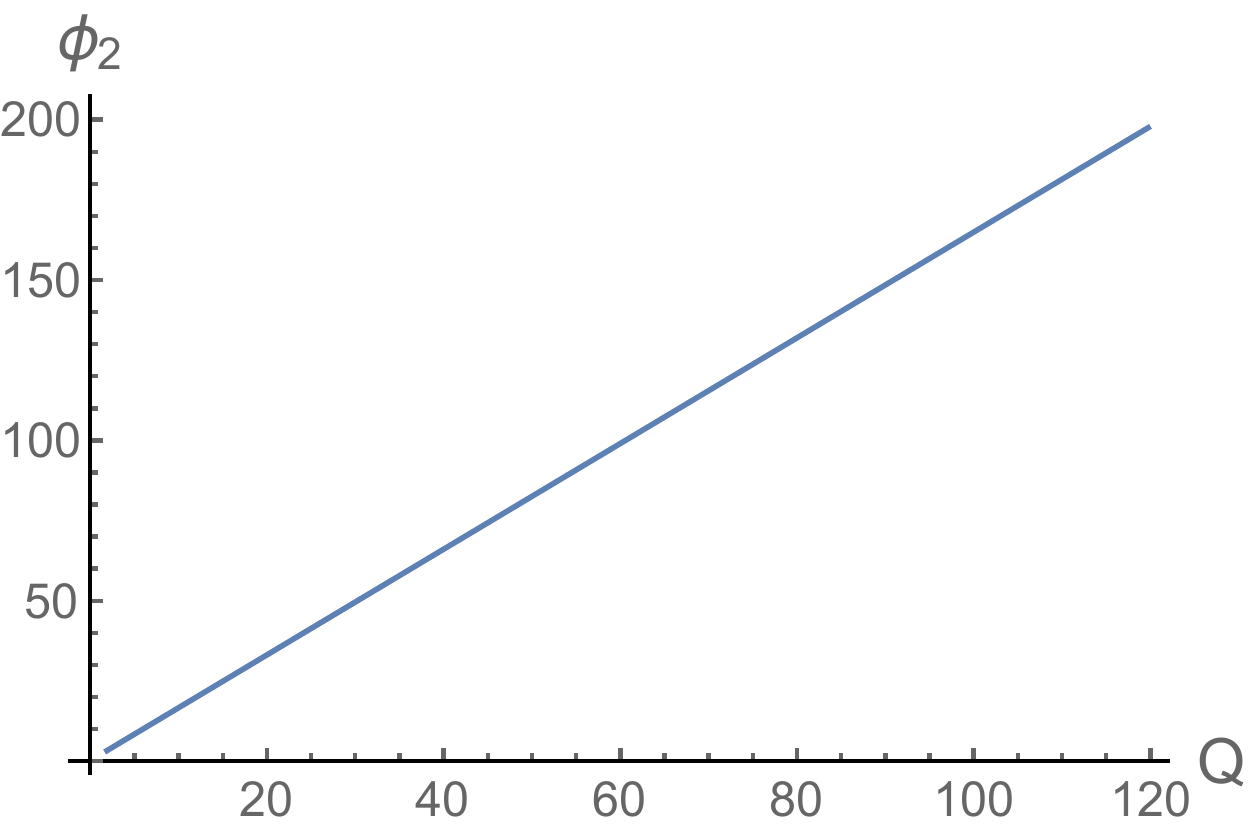}
\includegraphics[width=220pt]{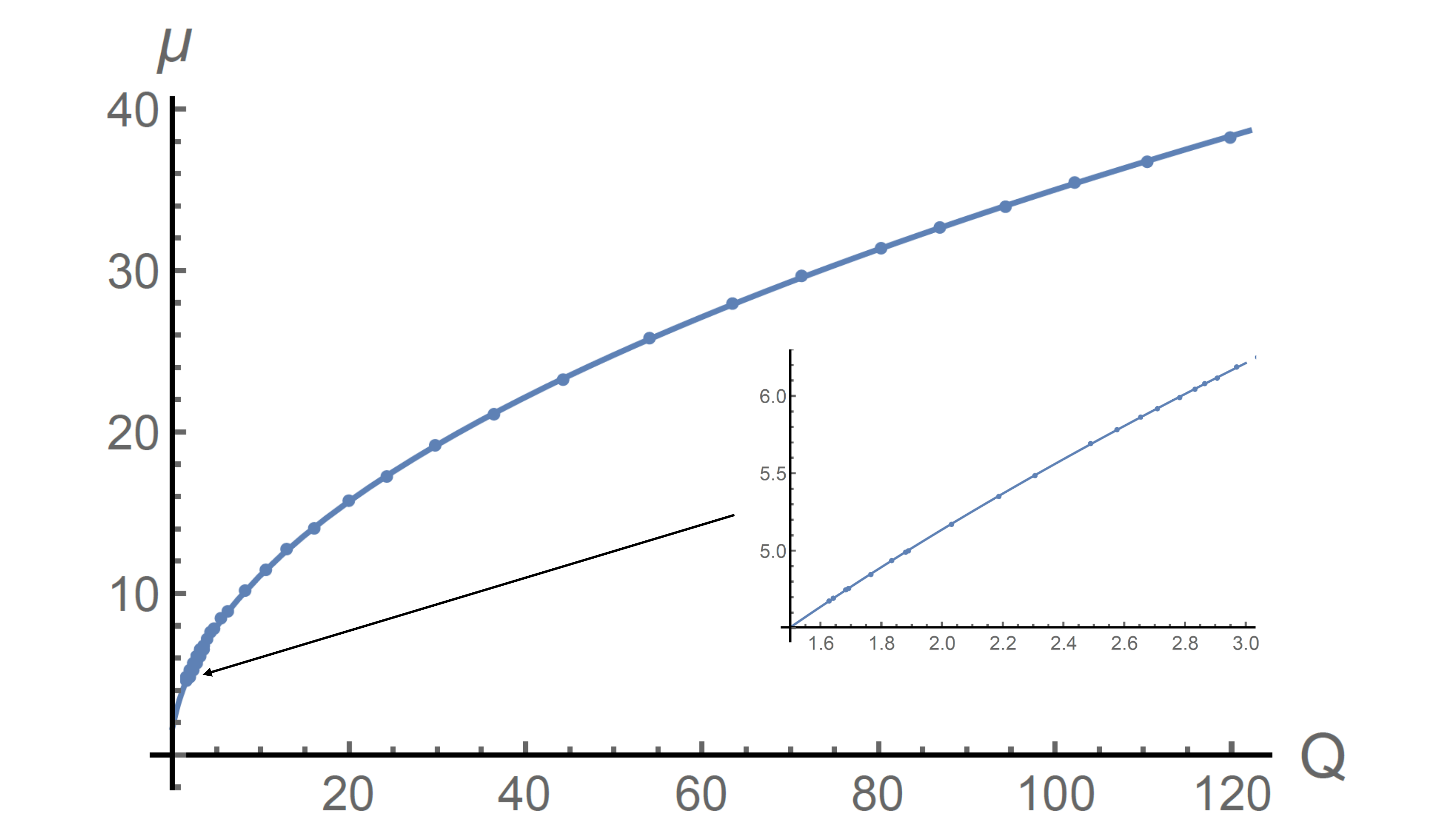}
\includegraphics[width=250pt]{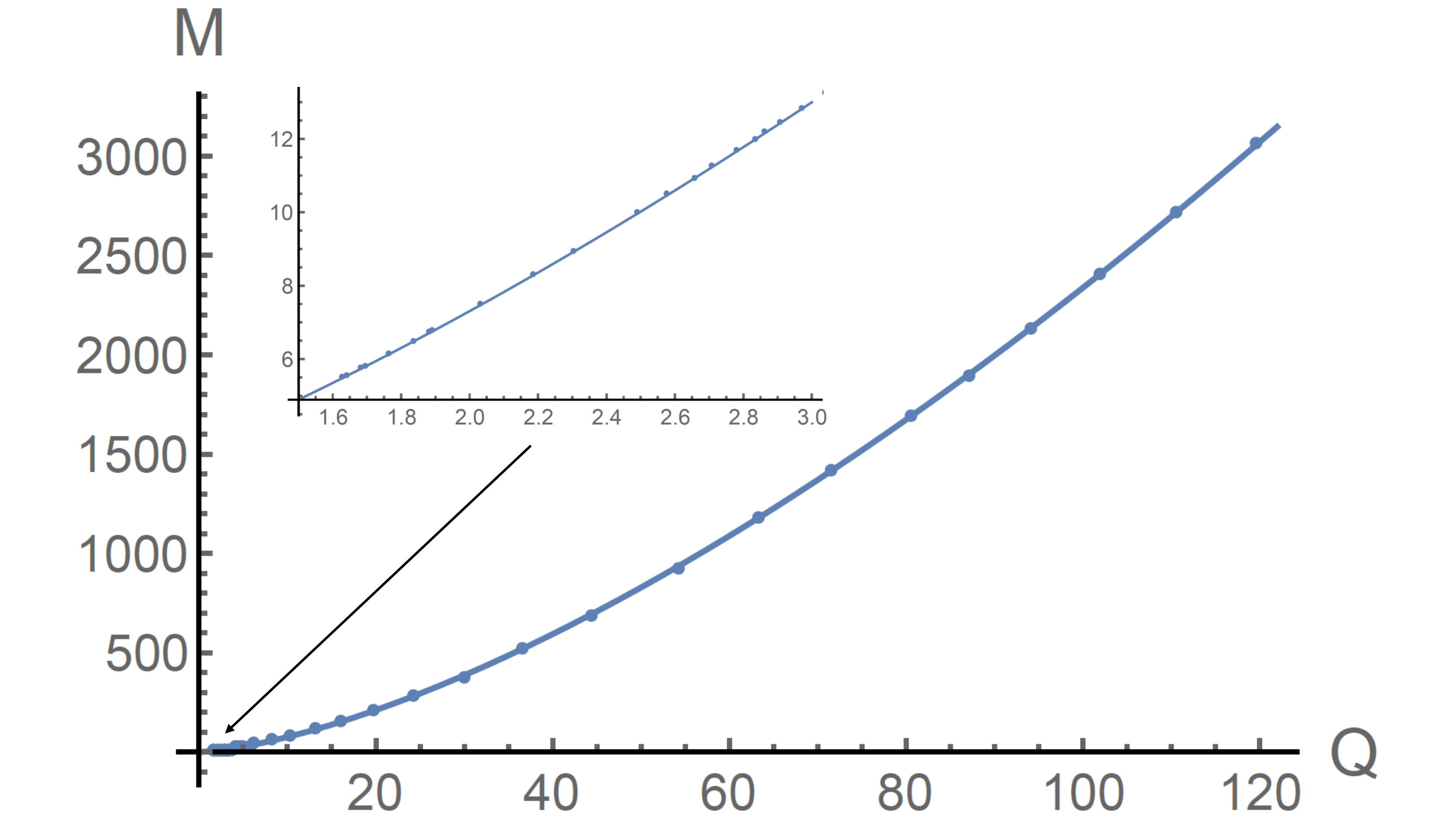}
\end{center}
\caption{\small\it These graphs describe the various properties of the $B_1$ series of boson stars for the $q^2=1$ toy model. Our data-fitting functions, given by (\ref{q1B1muq}) and below, match the numerical data (the dots) of $\mu(Q)$ and $M(Q)$ perfectly.}
\label{q1-B-all}
\end{figure}

\begin{figure}
\begin{center}
\includegraphics[width=220pt]{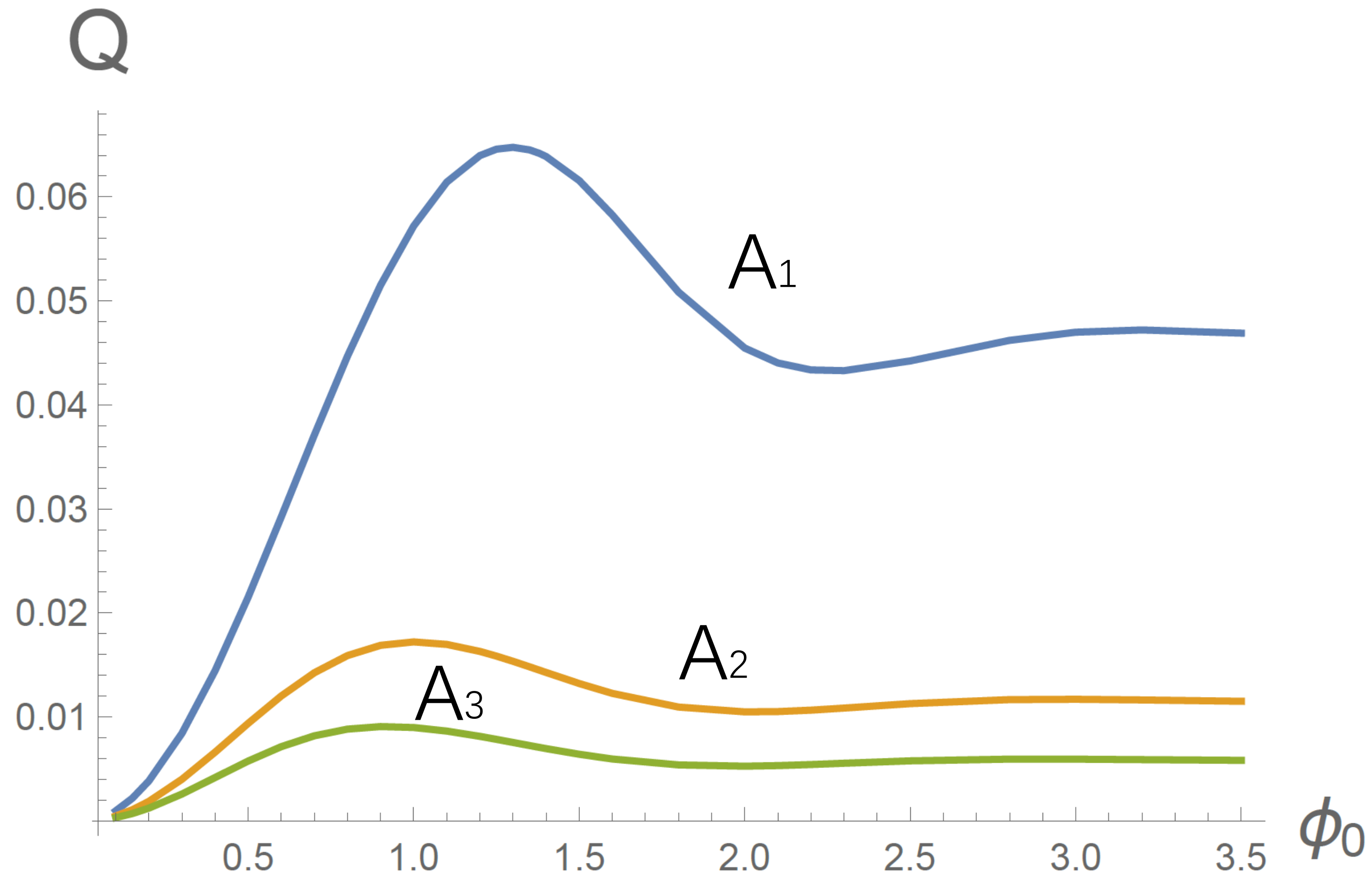}\!
\includegraphics[width=220pt]{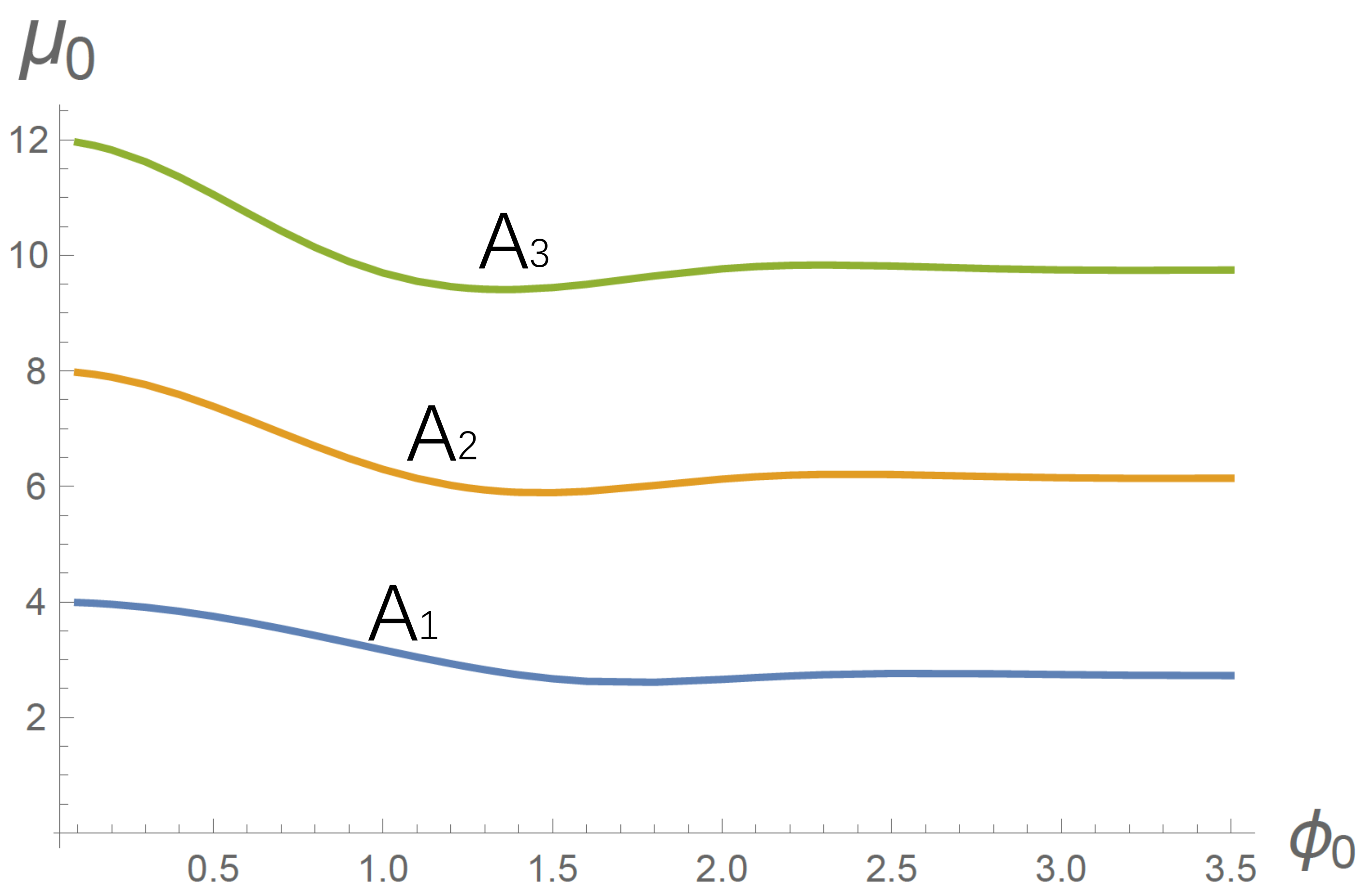}\!
\includegraphics[width=220pt]{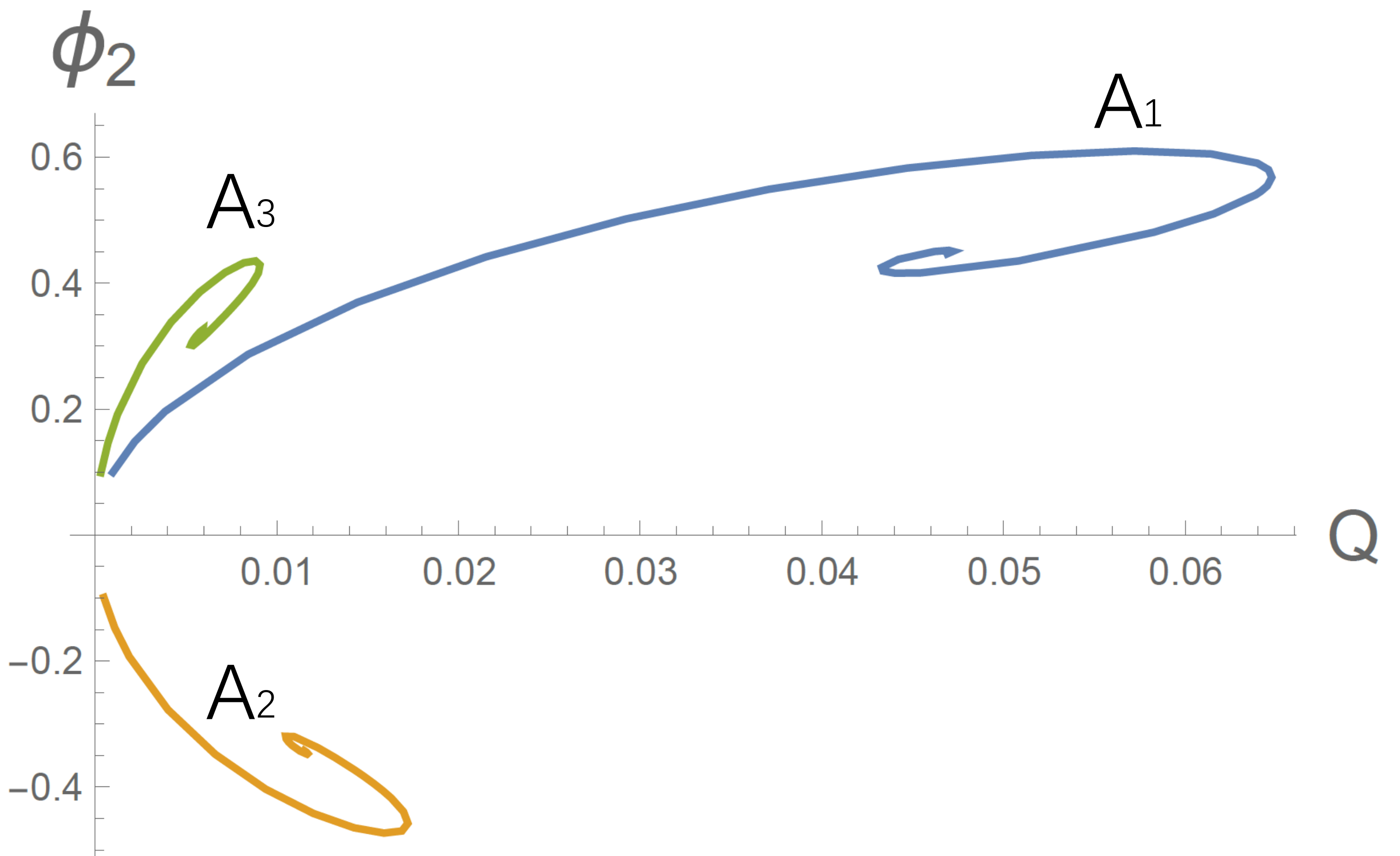}
\includegraphics[width=220pt]{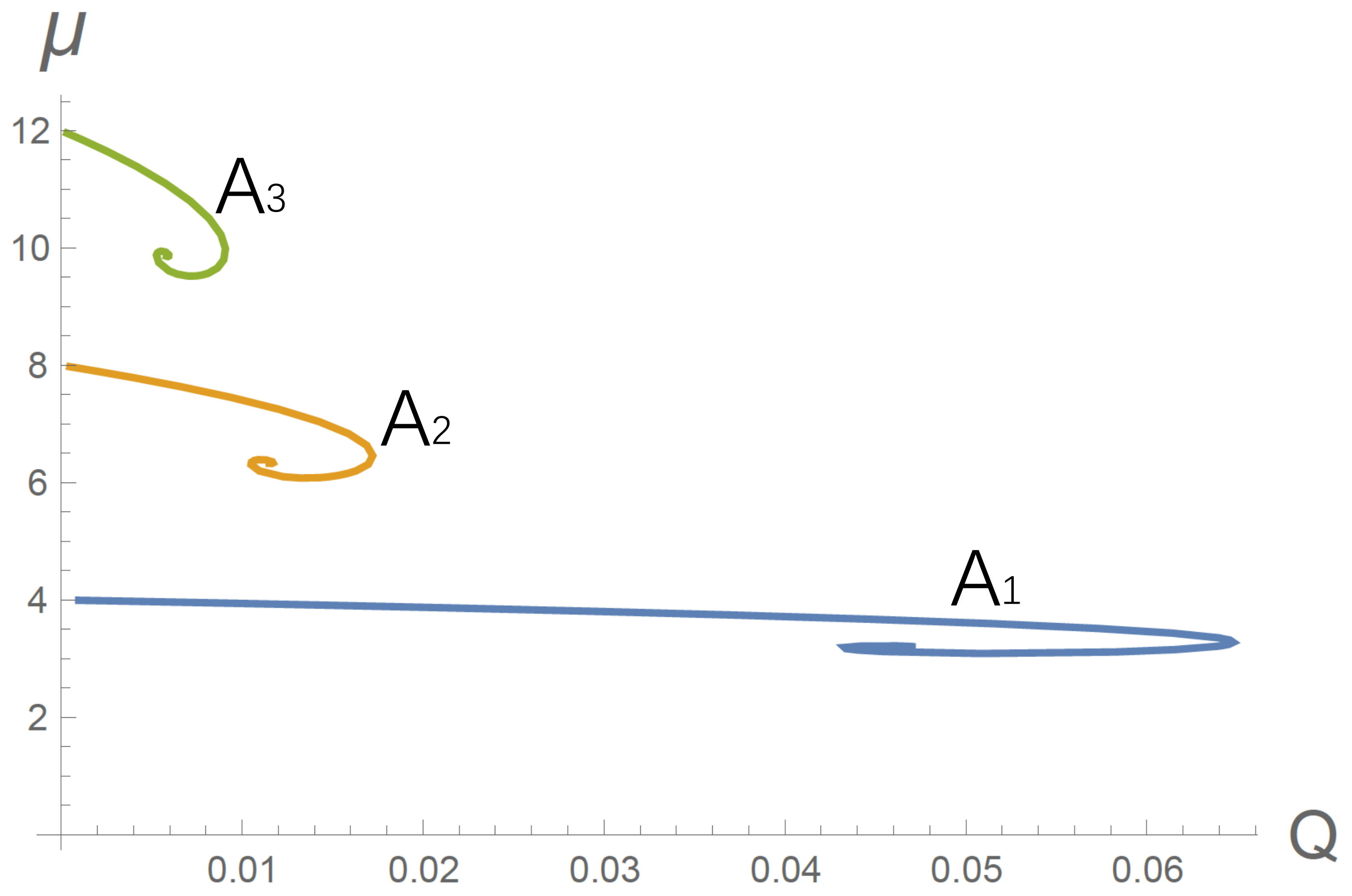}
\includegraphics[width=250pt]{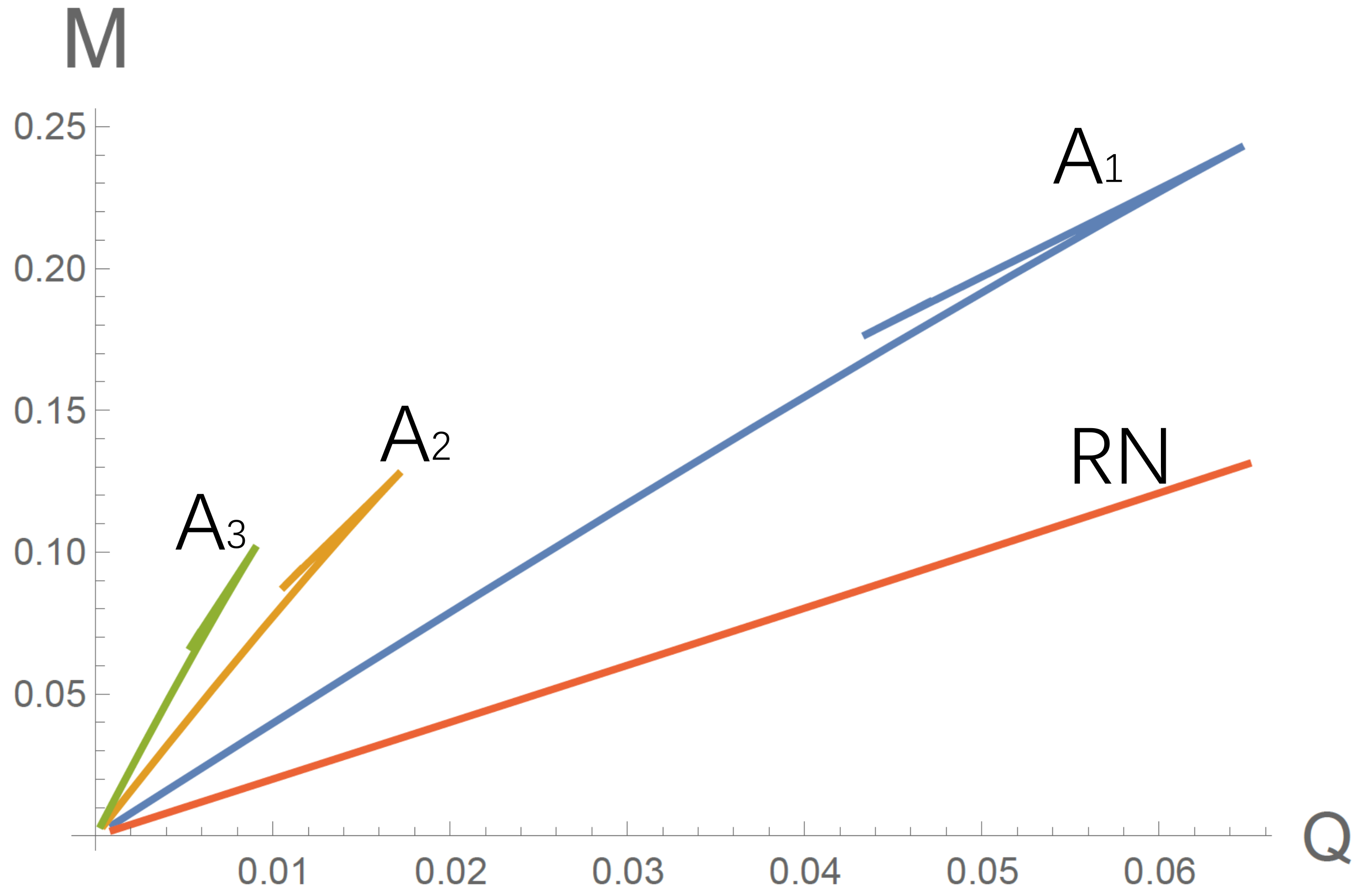}
\end{center}
\caption{\small\it Here are the properties of the $A$ series of AdS boson stars in the $U(1)^4$ supergravity model. For these small AdS boson stars ($Q\ll 1$, the nonlinear effect of the scalar potential is negligible and the solutions are effectively the same as the toy model with $q^2=1/4$. It is also clear that the extremal RN-AdS is the ground state.}
\label{sgrq12A-all}
\end{figure}

\begin{figure}
\begin{center}
\includegraphics[width=220pt]{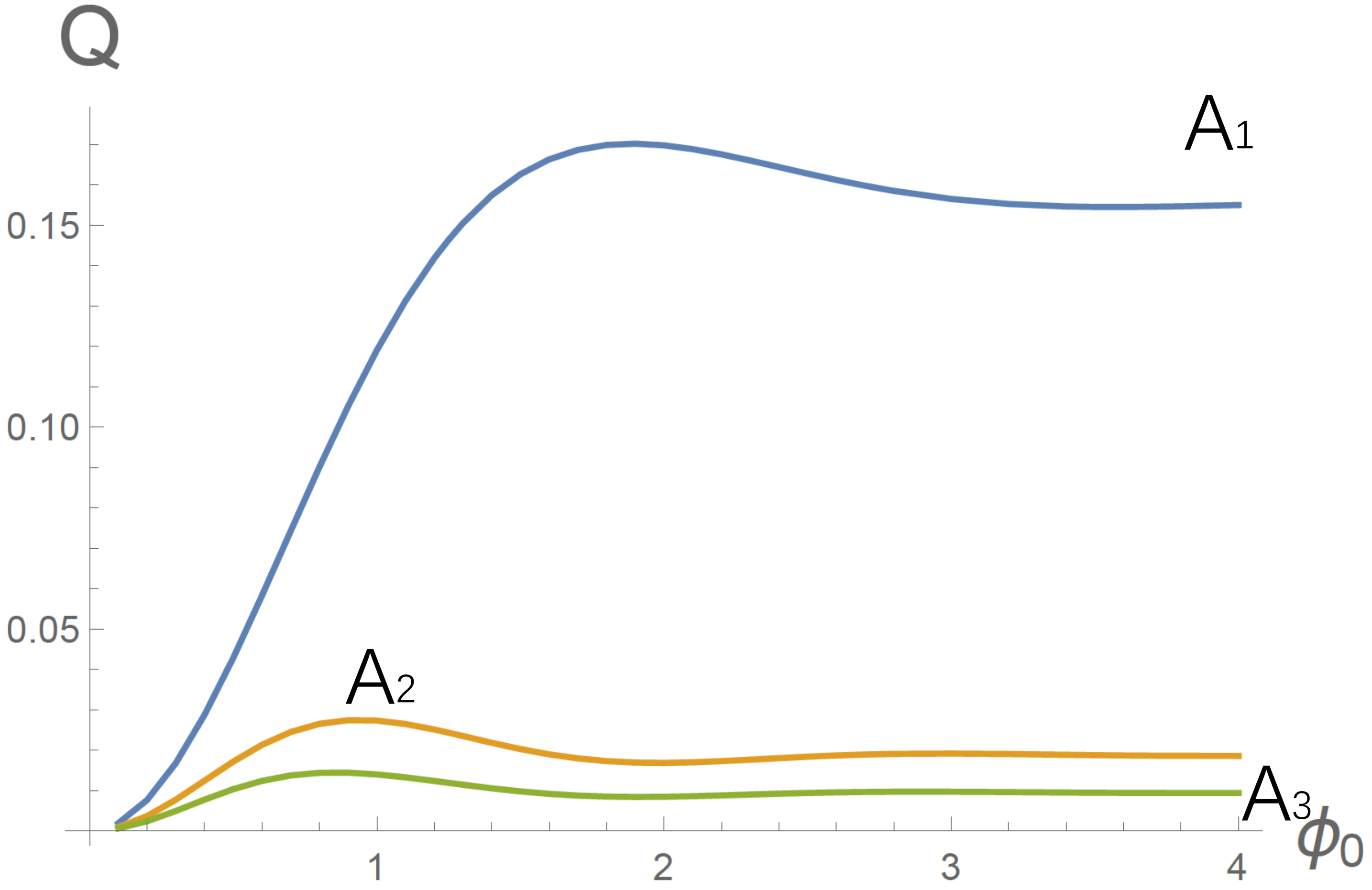}\!
\includegraphics[width=220pt]{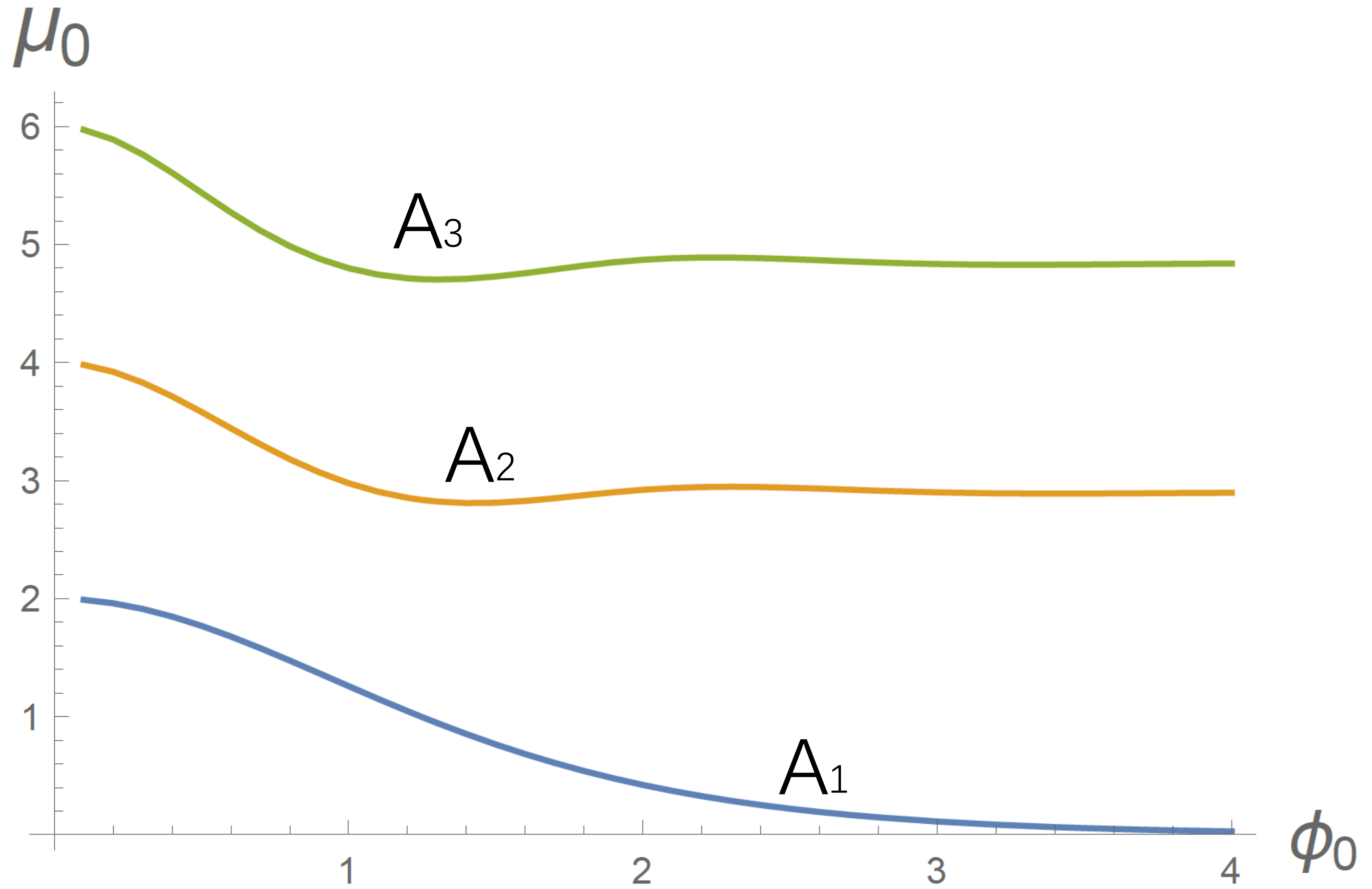}\!
\includegraphics[width=220pt]{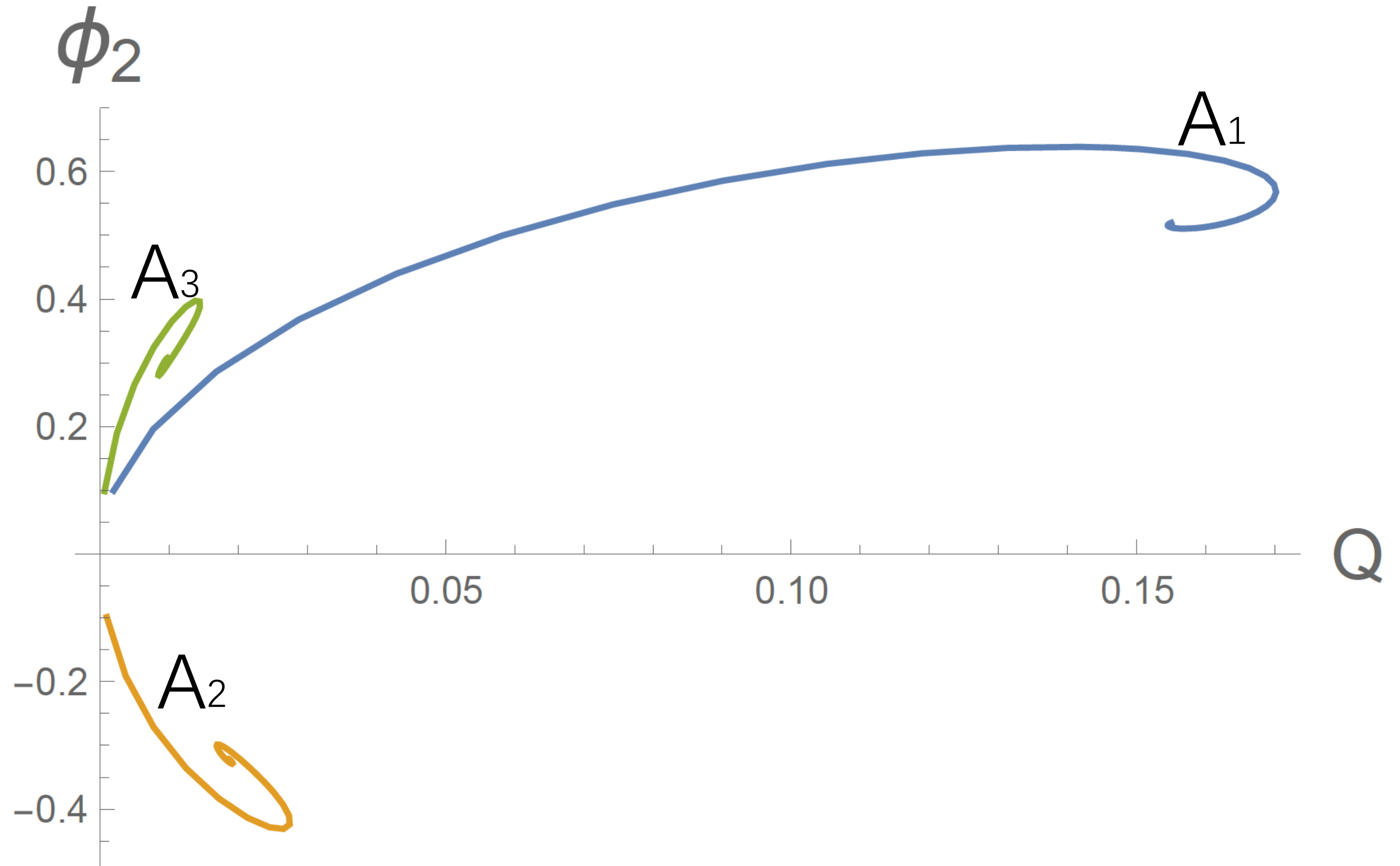}
\includegraphics[width=220pt]{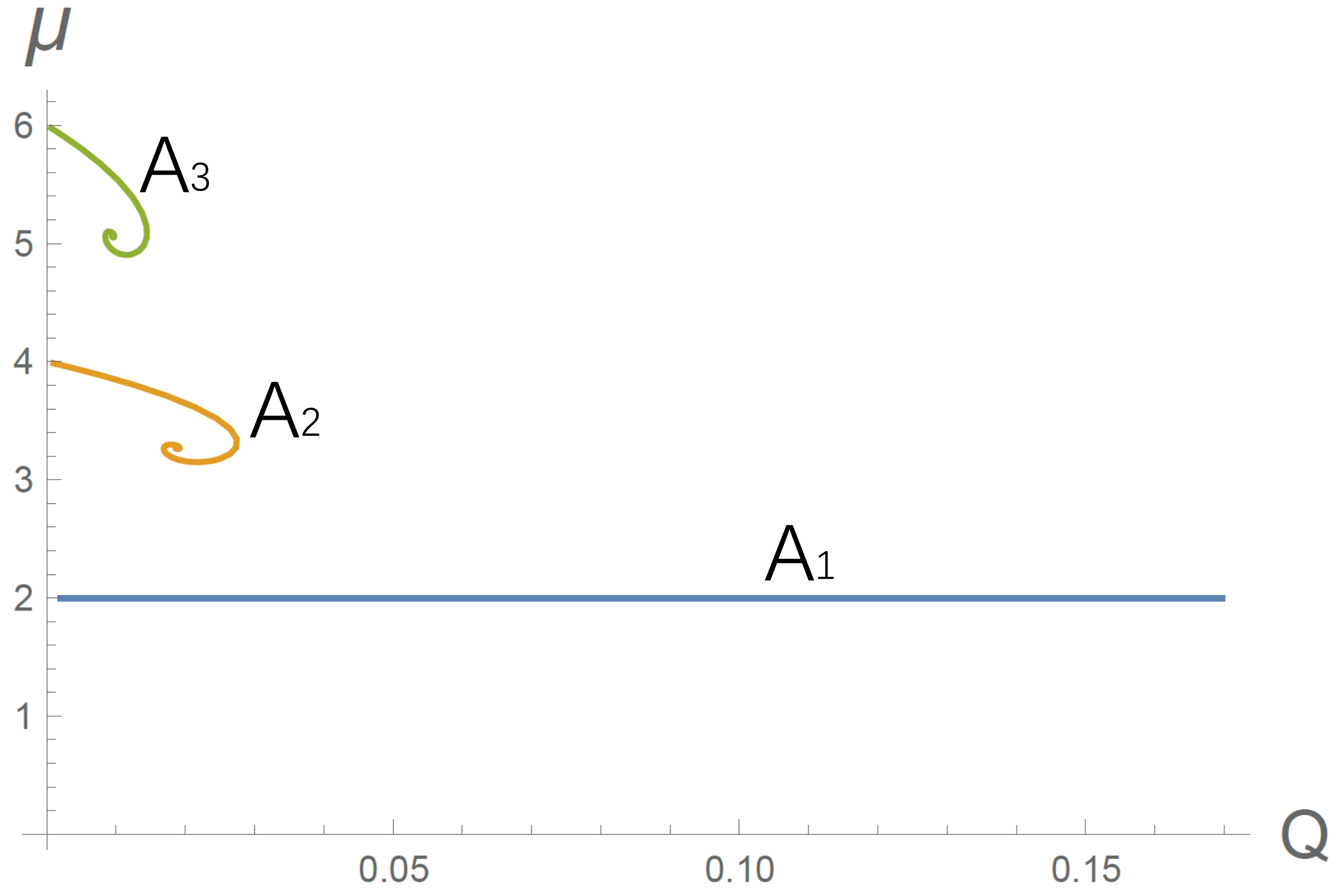}
\includegraphics[width=250pt]{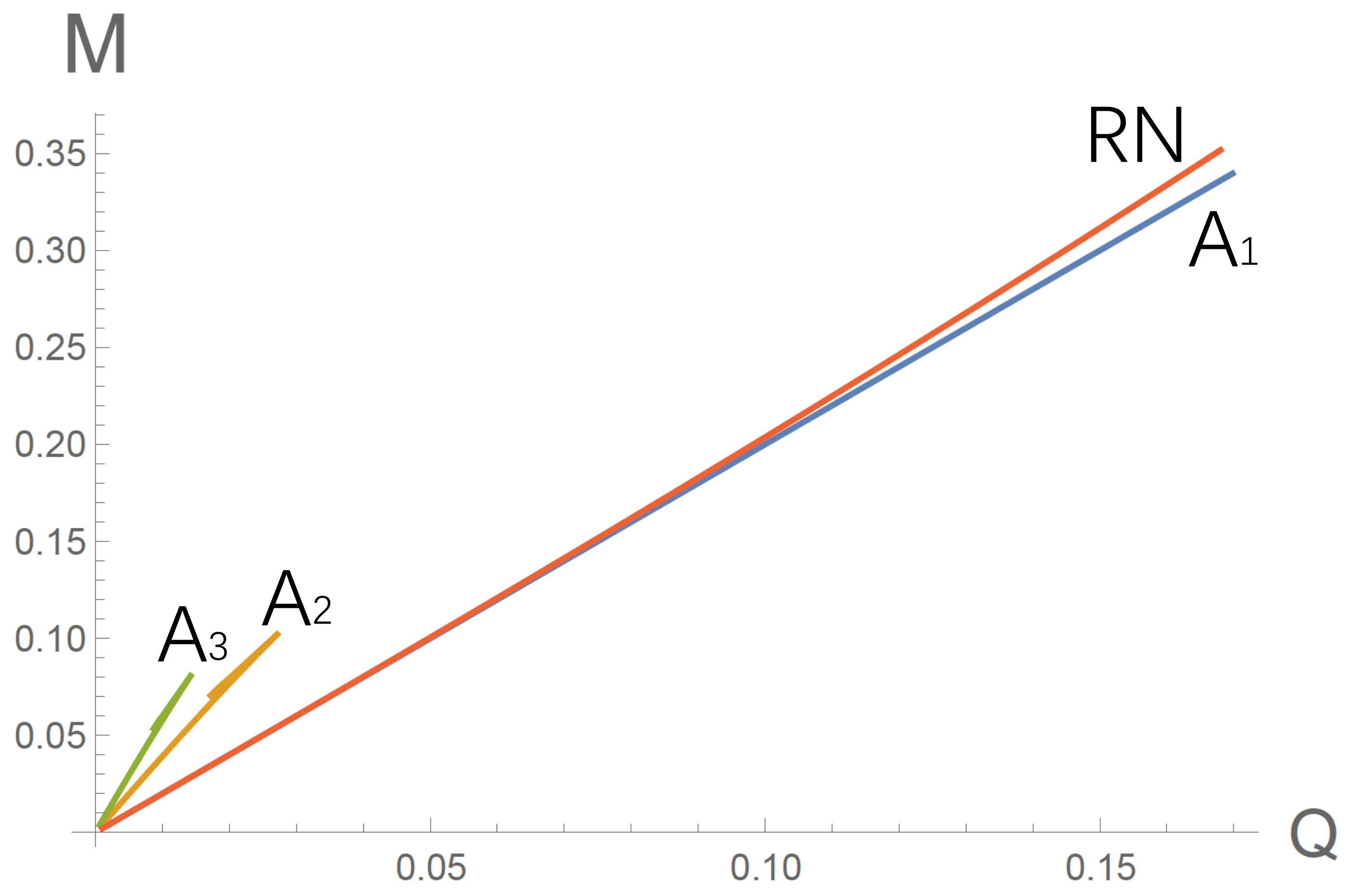}
\end{center}
\caption{\small\it These describe the $A$ series of AdS boson stars in the $SU(3)$ supergravity model.  One unusual feature is that $\mu=2$ for the $A_1$ series.  Its mass-charge relation $M=2Q$ is precisely the same as the extremal RN-Minkowski black hole.  It is clear that the $A_1$ series is the ground state in the appropriate charge region.}
\label{sgrq1A-all}
\end{figure}

\begin{figure}
\begin{center}
\includegraphics[width=220pt]{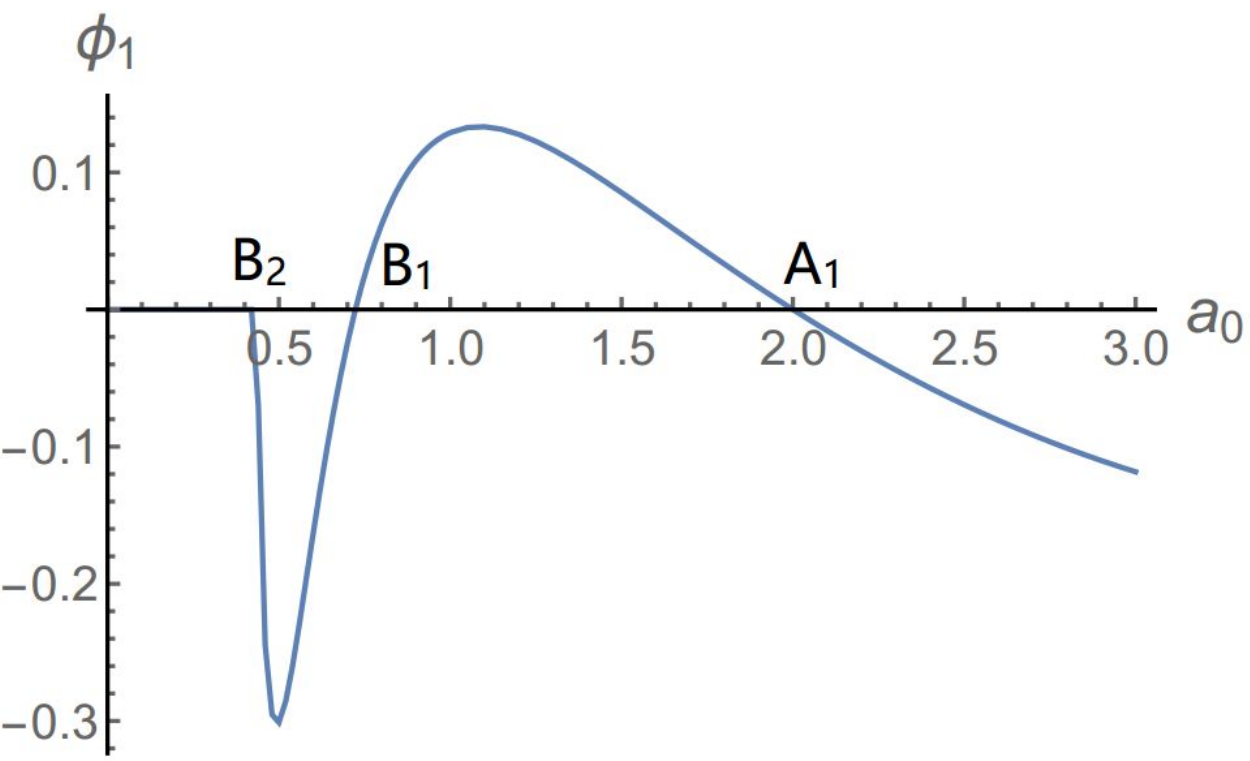}
\includegraphics[width=220pt]{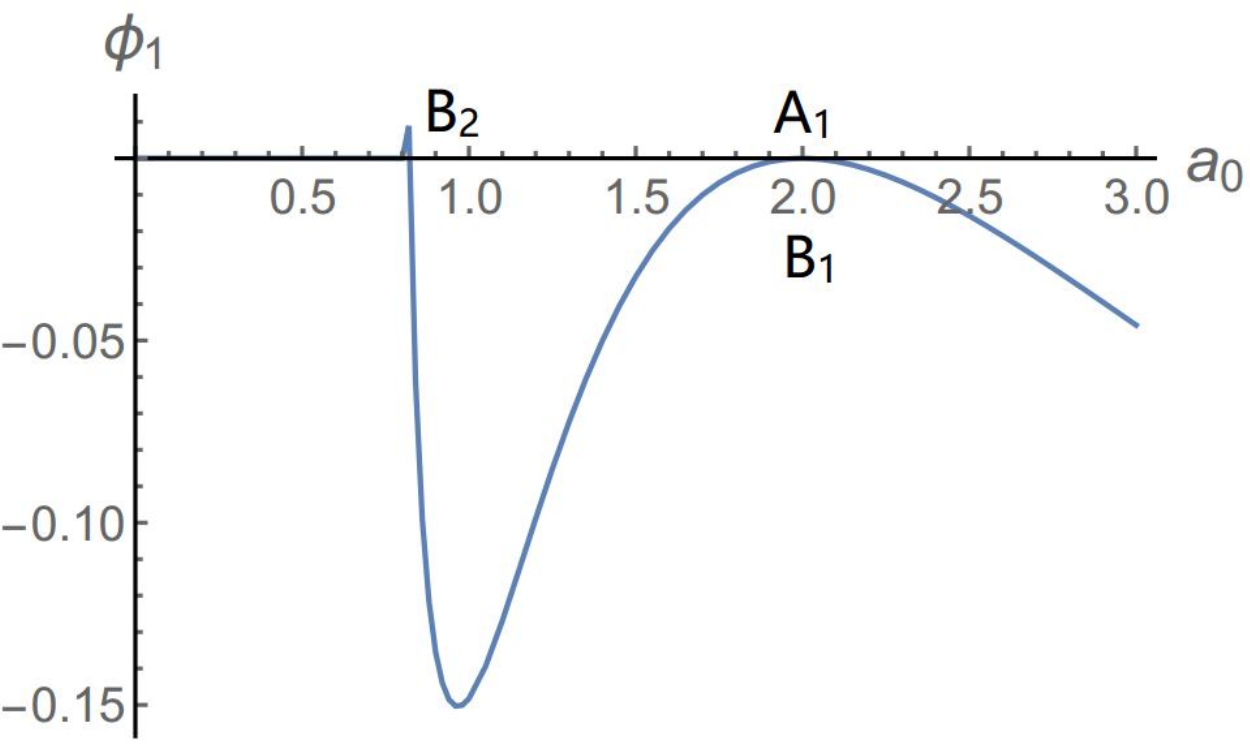}
\includegraphics[width=220pt]{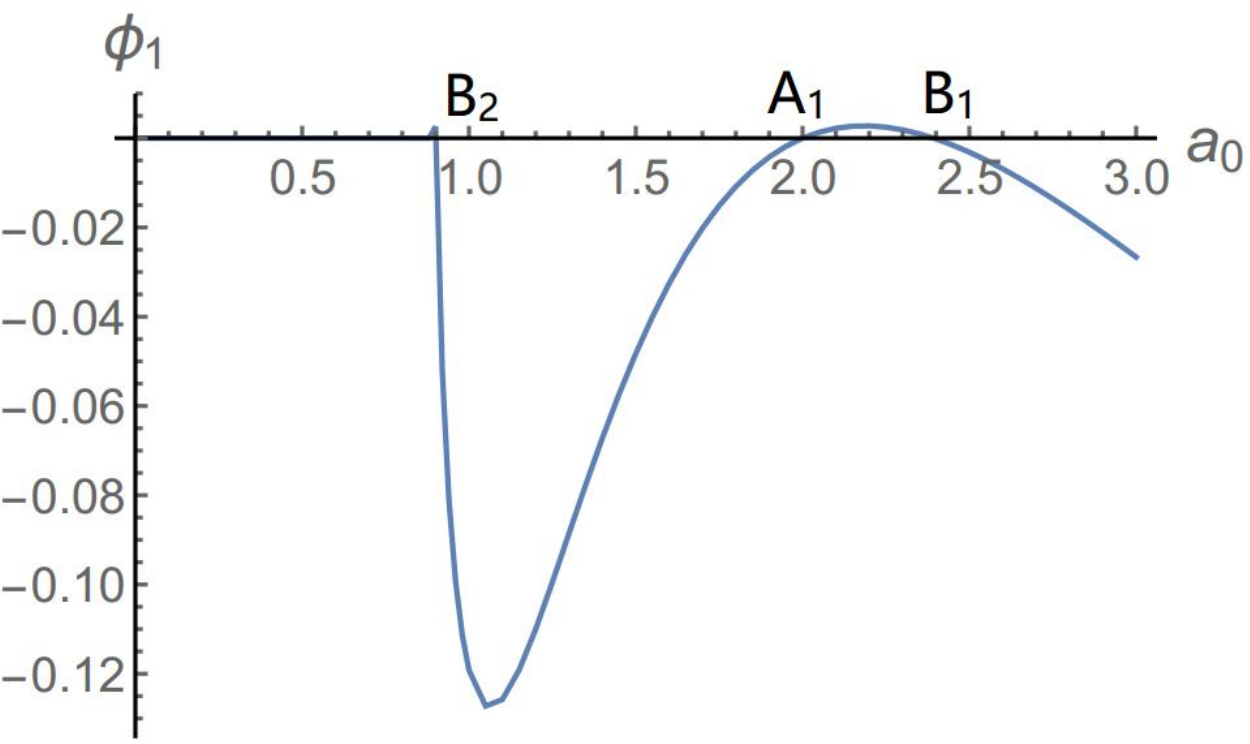}
\end{center}
\caption{\small\it The $B$ roots emerge after certain critical $\phi_0^c$, and they migrate towards right as $\phi_0$ increases.  From left to the right, $\phi_0=1.5, 1.888$ and $2.0$ respectively.  At $\phi_0=1.888$, the $A_1$ and $B_1$ roots join, creating a double root.}
\label{sgrBphi1(a0)}
\end{figure}

\begin{figure}
\begin{center}
\includegraphics[width=220pt]{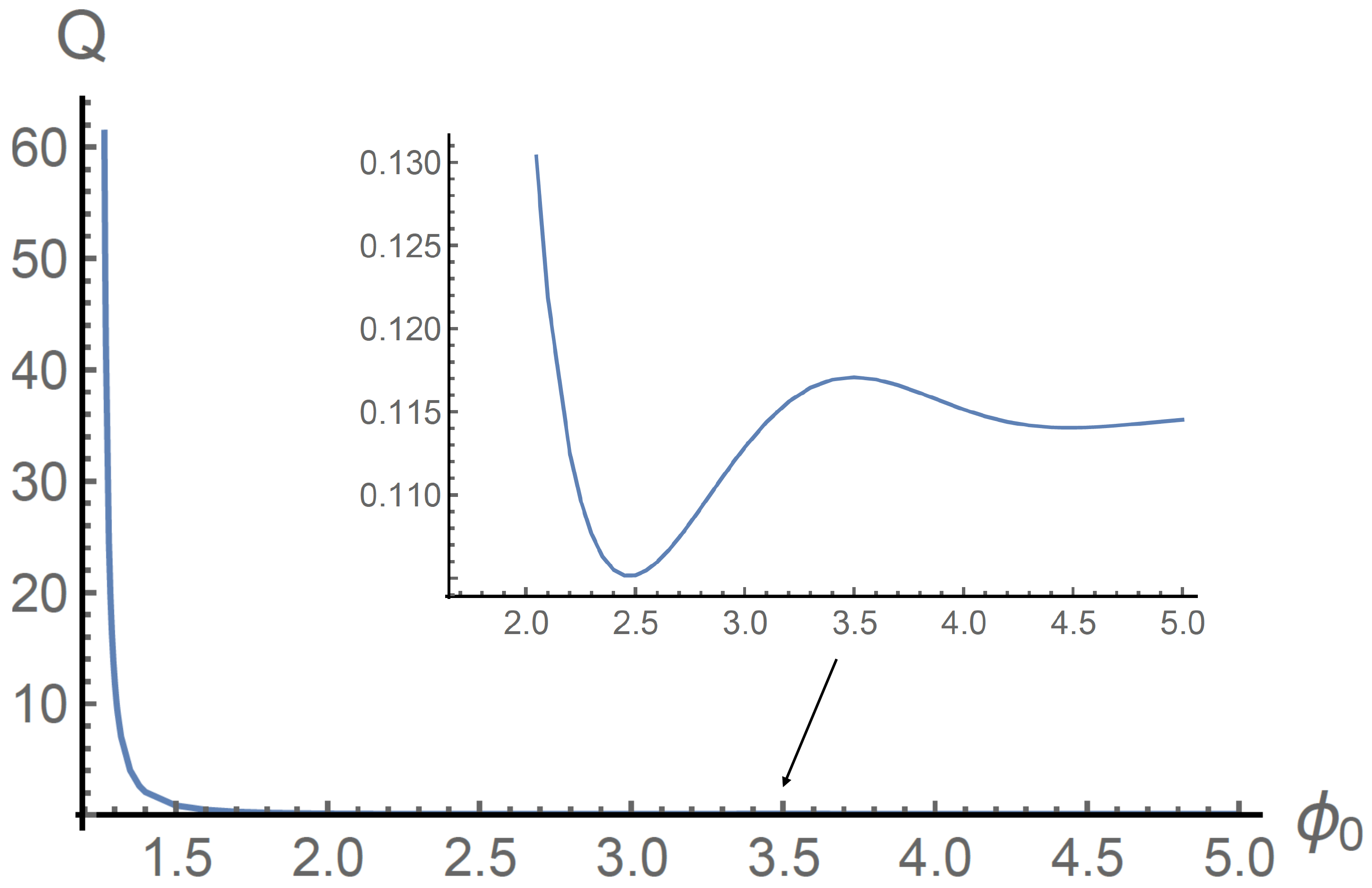}\!
\includegraphics[width=220pt]{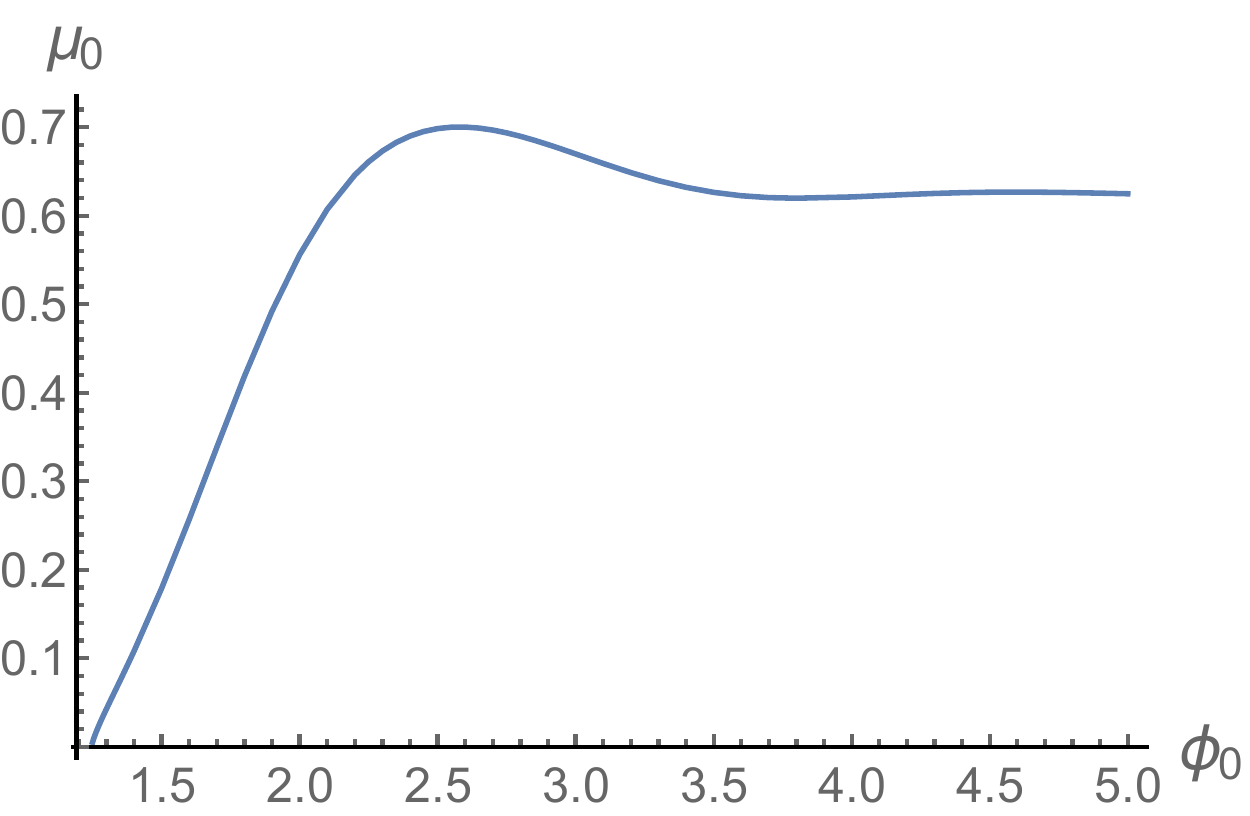}\!
\includegraphics[width=220pt]{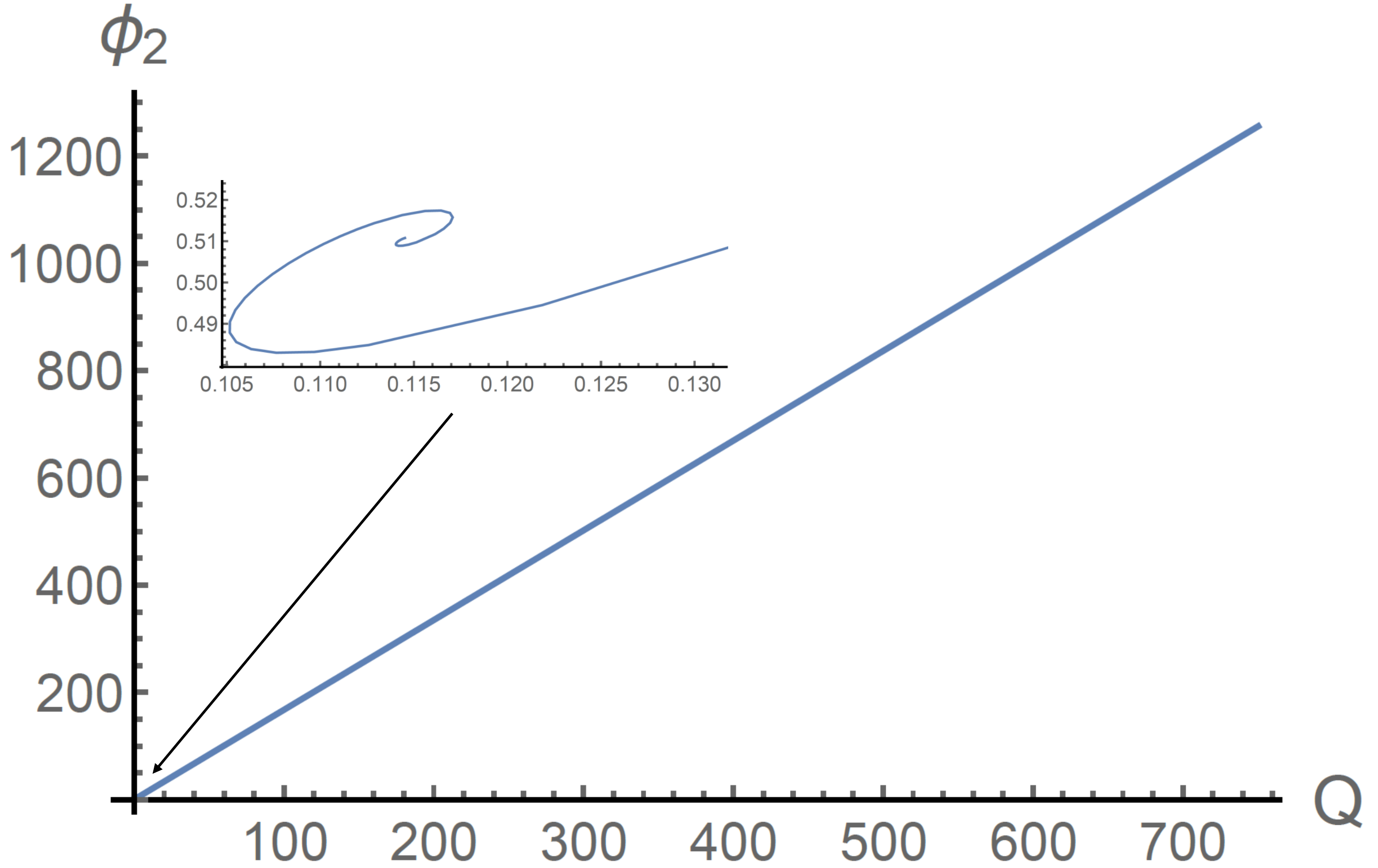}
\includegraphics[width=220pt]{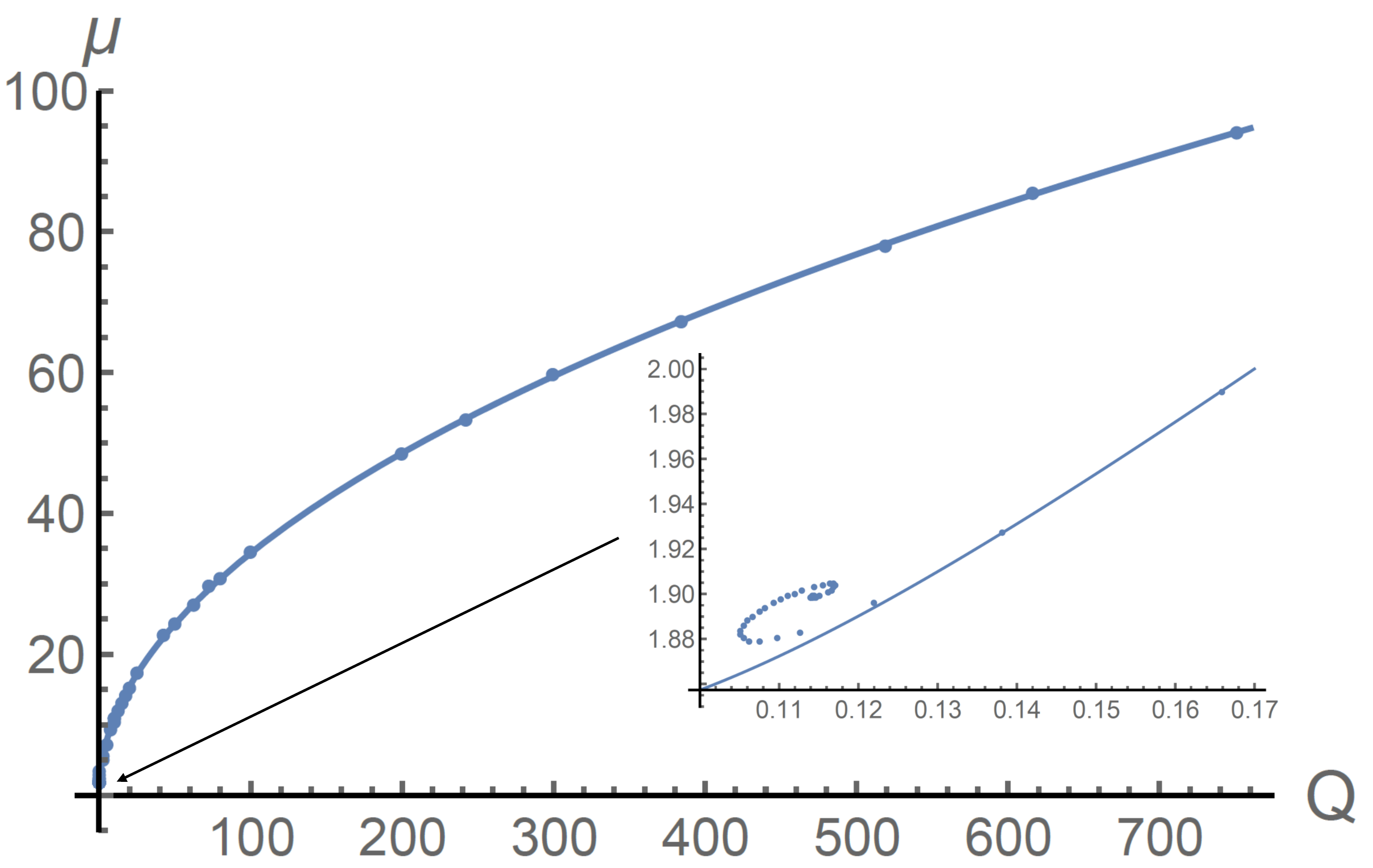}
\includegraphics[width=250pt]{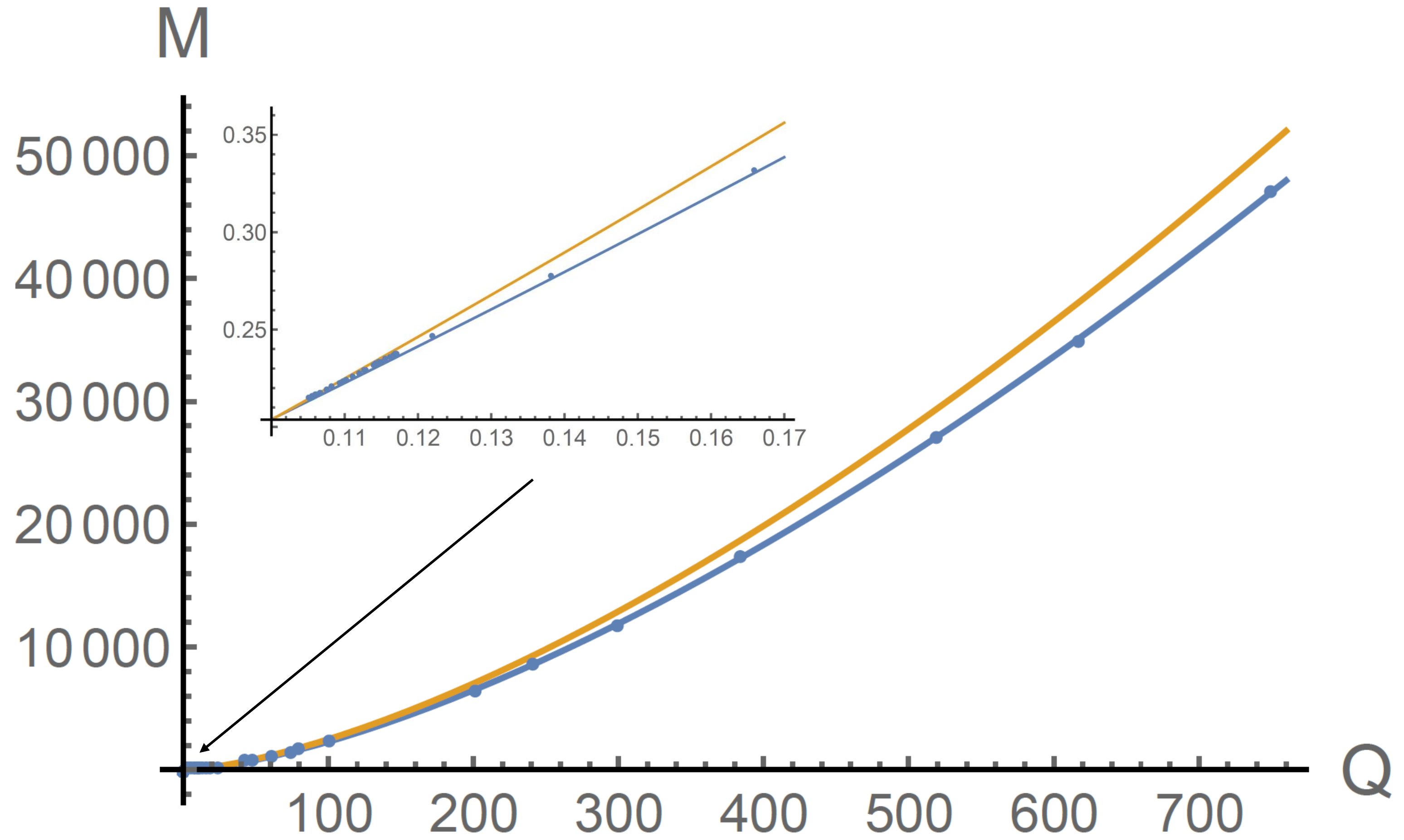}
\end{center}
\caption{\small\it This is the $B_1$ series of AdS boson stars in the $SU(3)$ supergravity model. The extra line in the $M(Q)$ is the mass-charge relation for the extremal RN-AdS black hole and we see that it has larger energy than the boson star.}
\label{sgrq1B-all}
\end{figure}

\begin{figure}
\begin{center}
\includegraphics[width=220pt]{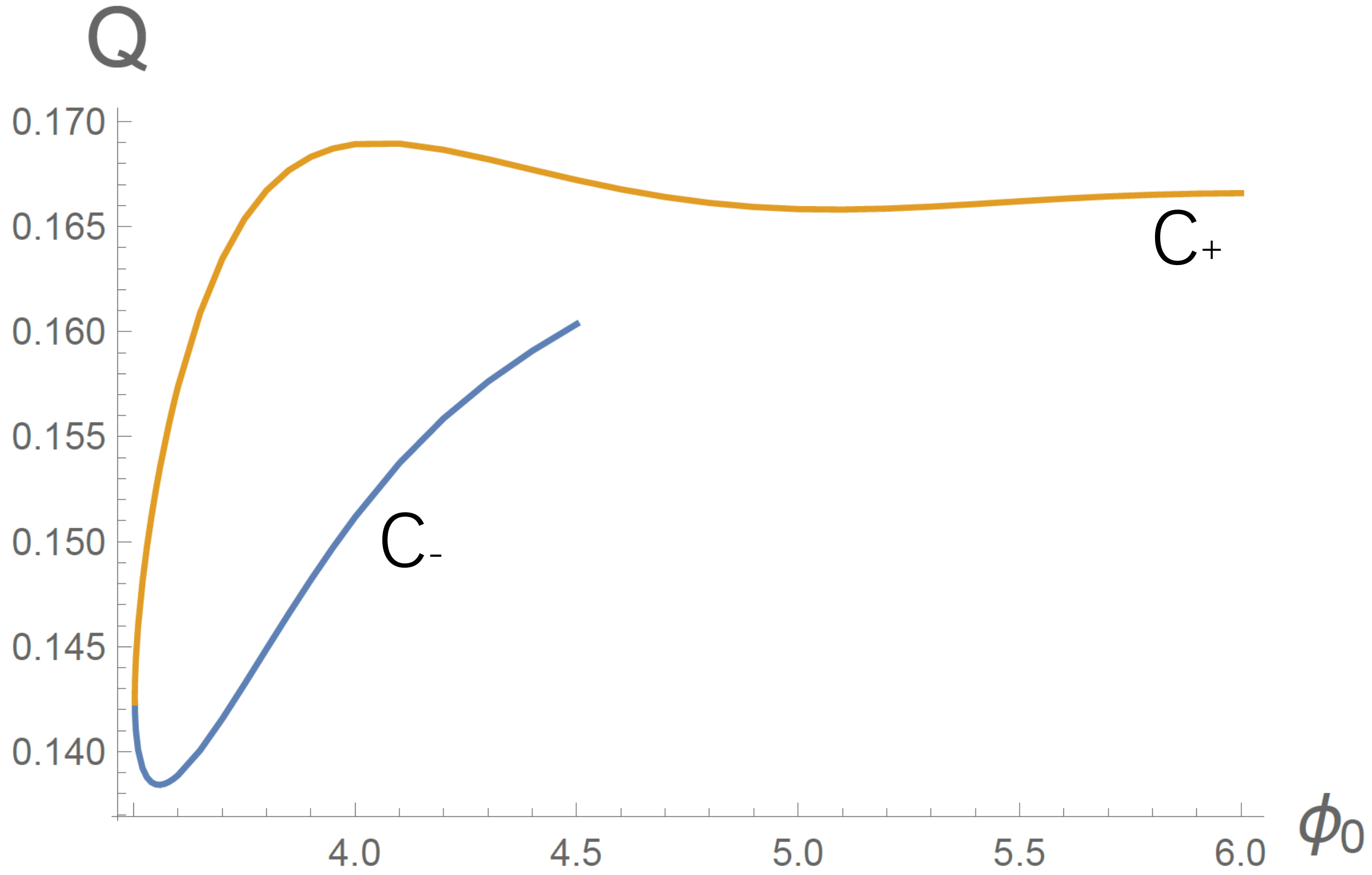}\!
\includegraphics[width=220pt]{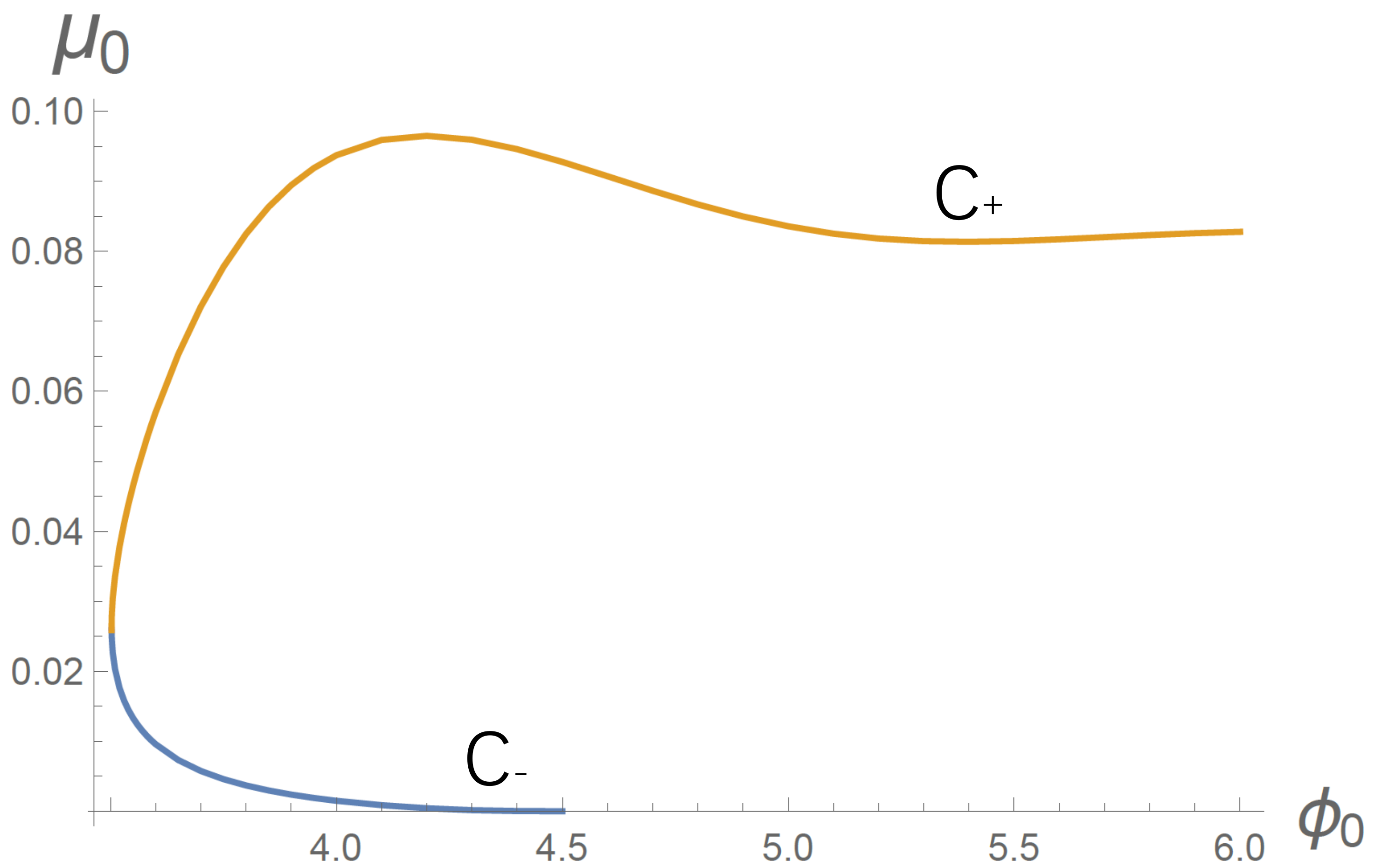}\!
\includegraphics[width=220pt]{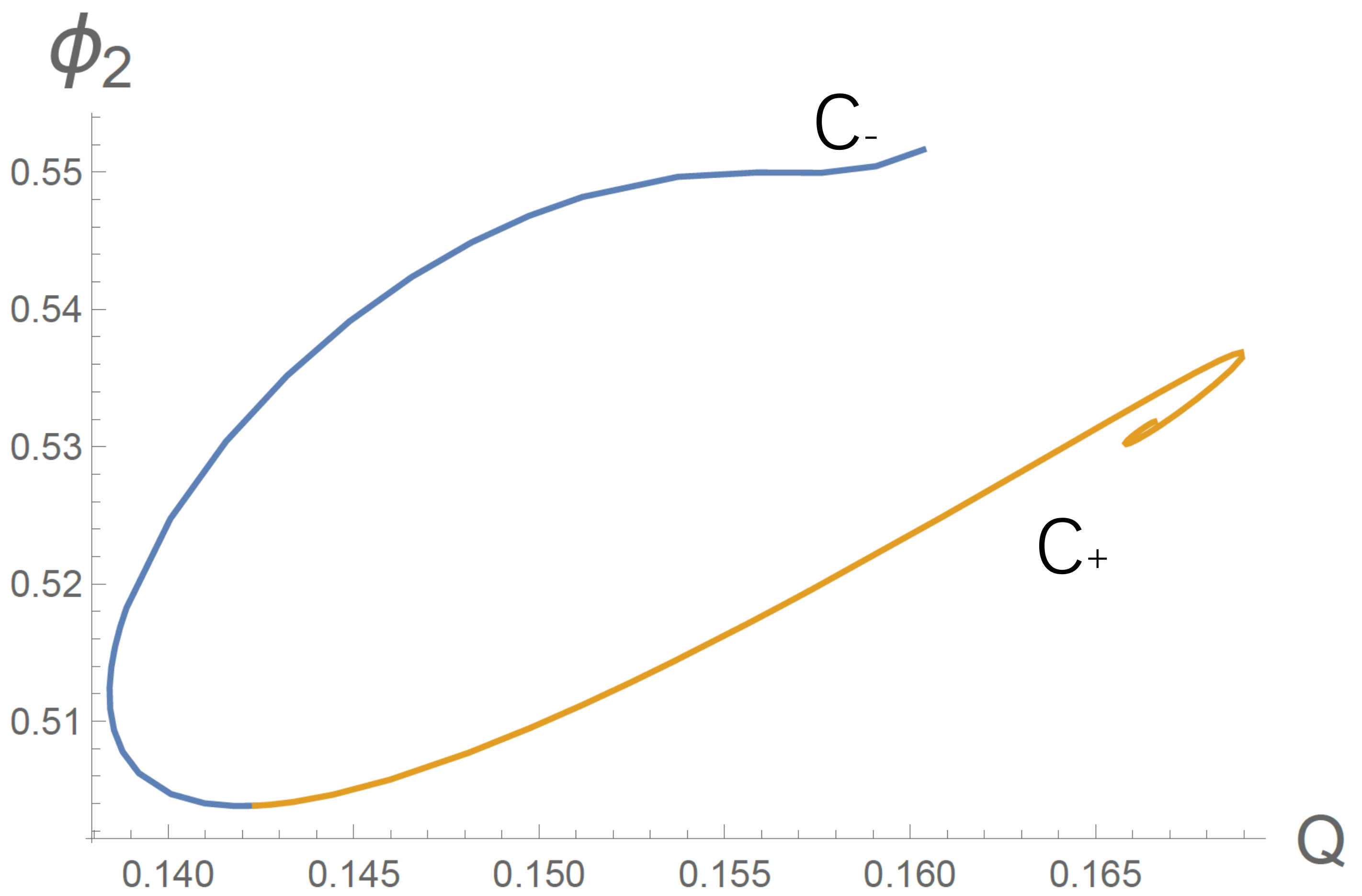}
\includegraphics[width=220pt]{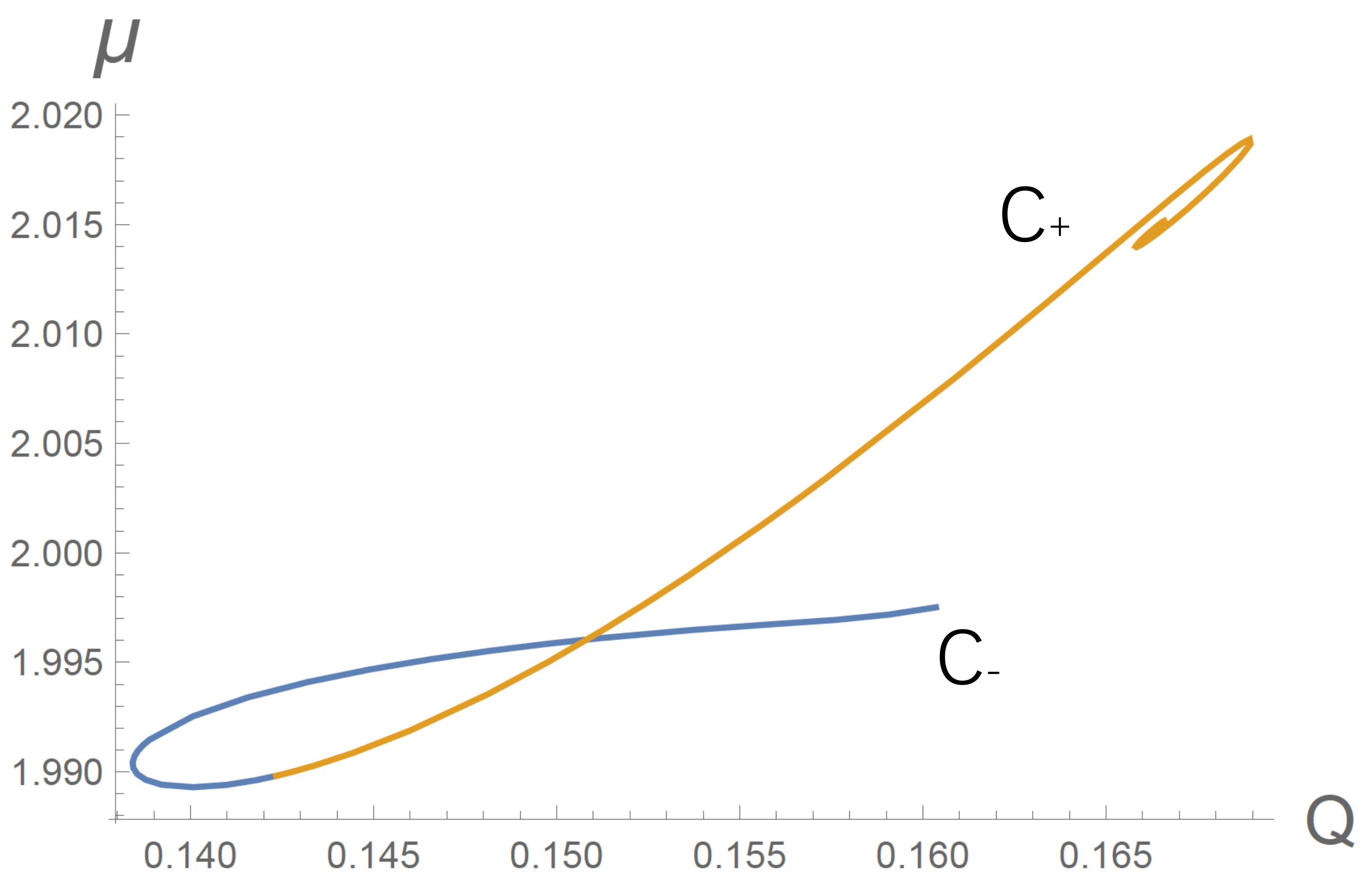}
\includegraphics[width=250pt]{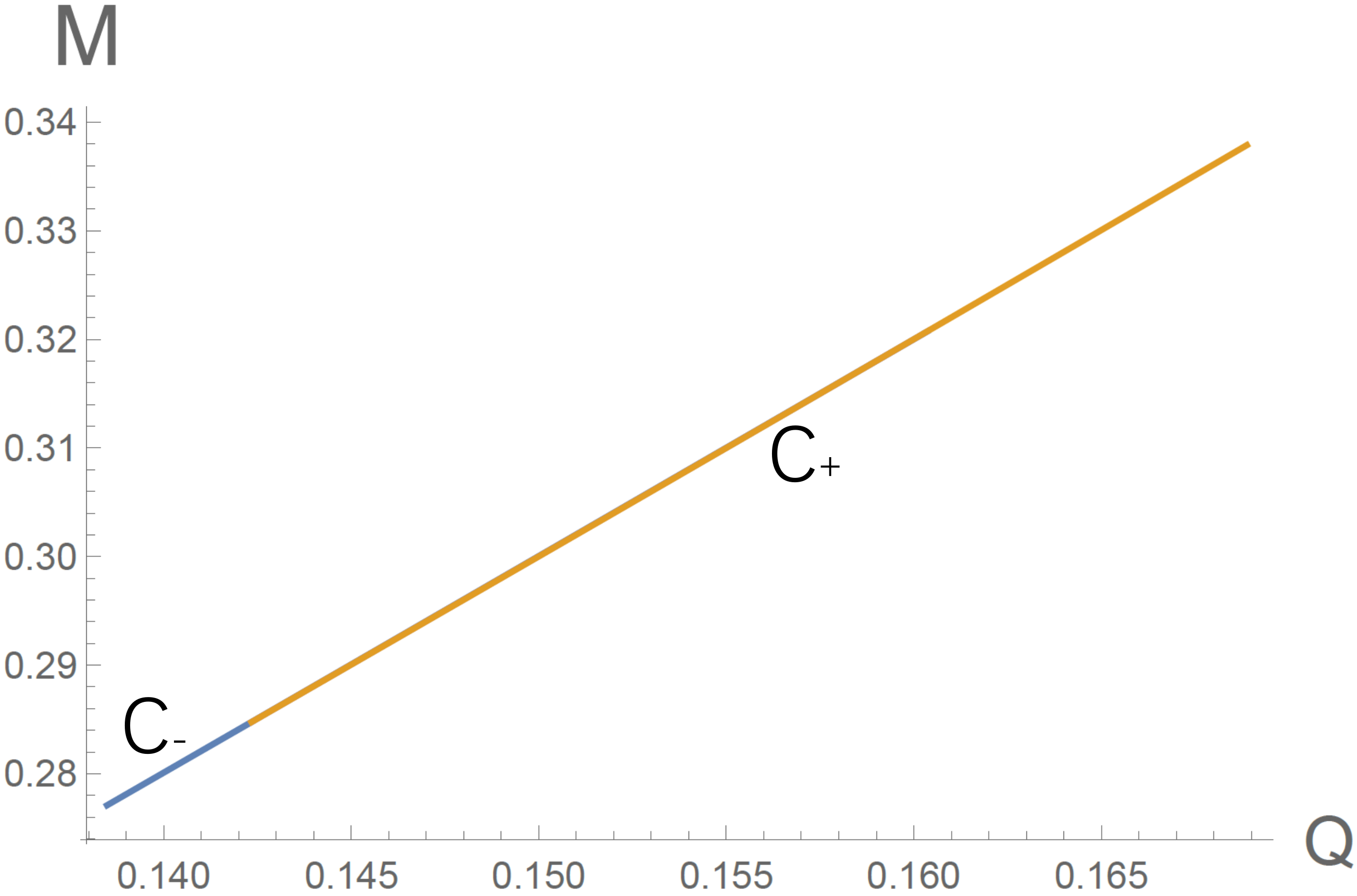}
\end{center}
\caption{\small\it This is the $C_1$ series of AdS boson stars in the $SU(3)$ supergravity model. The mass and charge are bounded both above and below.  The value of $\mu$ lies in a tight region so that the mass-charge relation looks like a straight line.}
\label{sgrq1C-all}
\end{figure}

\end{document}